\documentclass[
draft=false, 
toc=bibliography, 
]{scrreprt}
\usepackage{url}
\usepackage{hyperref}
\usepackage{booktabs} 
\usepackage{graphicx}
\usepackage{verbatim}
\usepackage{amsmath}
\usepackage{color}
\usepackage{makeidx}
\usepackage[normalem]{ulem}
\usepackage{footnote}

\def\lsim{\mathrel{\rlap{\lower4pt\hbox{\hskip1pt$\sim$}}
    \raise1pt\hbox{$<$}}}                
\def\gsim{\mathrel{\rlap{\lower4pt\hbox{\hskip1pt$\sim$}}
    \raise1pt\hbox{$>$}}}                

\makeindex
\begin{document}
\title{CompOSE Reference Manual}
\subtitle{
    Version 3.01\\[5ex]
  	CompStar Online Supernov\ae{} Equations of State
	\\[5ex]
        \emph{``harmonising the concert of nuclear physics and astrophysics''}
	\\[5ex]
        \url{https://compose.obspm.fr}
	\\[8ex] 
        \mbox{}
}
\author{\textbf{CompOSE Core Team}\\
(S. Typel, M. Oertel, T. Kl\"{a}hn, \\
D. Chatterjee, V. Dexheimer, C. Ishizuka, \\
M. Mancini, J. Novak, H. Pais, C. Providencia, \\
A. Raduta, M. Servillat, L. Tolos)}

\date{\today}

\maketitle
\tableofcontents

\begin{abstract}
    \noindent
    \textbf{Abstract}\\
    \mbox{}\\
    CompOSE (CompStar Online Supernovae Equations of State) is an online repository of equations of state (EoS) for use in nuclear physics and astrophysics, e.g., in the description of compact stars or the simulation of core-collapse supernovae and neutron-star mergers, see \texttt{http://compose.obspm.fr}. The main services, offered via the website, are: a collection of data tables in a flexible and easily extendable data format for different EoS types and related physical quantities with extensive documentation and referencing; software for download to extract and to interpolate these data and to calculate additional quantities; webtools to generate EoS tables that are customized to the needs of the users and to illustrate dependencies of various EoS quantities in graphical form.  This manual is an update of previous versions that are available on the CompOSE website, at \texttt{arXiv:1307.5715 [astro-ph.SR]}, and that was originally published in the journal "Physics of Particles and Nuclei' with \texttt{doi:10.1134/S1063779615040061}. It contains a detailed description of the service, containing a general introduction as well as instructions for potential contributors and for users. Short versions of the manual for EoS users and providers will also be available as separate publications.
    \mbox{}\\[8ex]
    \noindent
    \textbf{Preliminary Remark}\\
    \mbox{}\\
    This CompOSE reference manual, version 3.01, is an updated and extended version of previous manuals that appeared as a preprint 	
    \texttt{arXiv:1307.5715 [astro-ph.SR]} on \texttt{arXiv.org}.
    and as a refereed article 
    in the journal "Physics of Particles and Nuclei" \cite{manual:1}.
    The content of this publication was adapted by permission from Springer Nature Customer Service Centre GmbH: Springer Nature, Physics of Particles and Nuclei, Vol.\ 46, No.\ 4, 633-664 ("CompOSE CompStar online supernova equations of state harmonising the concert of nuclear physics and astrophysics compose.obspm.fr", S. Typel, M. Oertel, and T. Kl\"{a}hn), Copyright Pleiades Publishing, Ltd.\ (2015).
    Previous versions are available on the CompOSE website \texttt{compose.obspm.fr}.
\end{abstract}

\part{Introduction}

\chapter{What CompOSE can do and what not}
The online service CompOSE provides information and
data tables for different equations of 
state (EoS) ready for further use
in astrophysical applications,
nuclear physics and beyond.
See the review \cite{Oertel:2016bki} for a general introduction.
CompOSE has three major purposes:
\begin{itemize}
\item CompOSE is a repository of EoS tables in a common format
  for direct usage with
  information on a large number of
  thermodynamic properties, on the chemical composition 
  of dense matter and, if available, on
  microphysical quantities of the constituents.
\item CompOSE allows to interpolate the provided tables  
  using different schemes to obtain the relevant quantities, selected by
  the  user, for grids that are tailored to specific applications.
\item CompOSE can provide information on
  additional thermodynamic quantities, 
  which are not stored in the original data tables, and on further
  quantities,
  which characterize an EoS, such as nuclear matter parameters and compact
  star properties.
\end{itemize} 
The format of the files, as well as the calculational mesh, is mainly determined 
according to the needs of scientific groups performing extensive numerical 
simulations of astrophysical objects.
We try to provide the tables in a large parameter space to cover most
applications.

We cannot offer an online service for all features of CompOSE,
e.g.\ to run all codes online.
This is mostly due to limitations in storage and computation times,
but also gives better control on 
avoiding unphysical input parameters.
However, we offer several computational tools that allow the user
to extract the data from the tables that are relevant for her/him.
These tools can be downloaded from the CompOSE website
\begin{quote}
\url{https://compose.obspm.fr} \: .
\end{quote}
Appendix \ref{citation} describes how to cite CompOSE and the EoS tables it contains.

CompOSE is designed in a modular way, thus allowing to extend the service over 
time.
More and more models and parameter sets will be provided in time.
It is foreseen that additional features will be added in the future, too.

\chapter{How to read this document}
\label{ch:document}
While reading this document please always keep in mind: this document was 
written by physicists for physicists.
It is divided into three major parts.

The first one is relevant both to providers and users of equations of state,
as it serves as a basis for the discussions in the following parts
of the manual and the web site. 
Both, contributors and users, should first of all have a look at the 
introductory chapter \ref{ch:defnot}
where we will discuss general conventions and the notation used throughout
CompOSE. In addition, we give definitions and details on
the system of units that is used within CompOSE and 
on physical constants that should be used 
in order to standardise the generation of new equations of state.

The second part concerns those persons
who wish to contribute to an extension of the CompOSE data base
by the active development of an EoS. CompOSE allows them to make their
favourite EoS available for a broad range of astrophysical and nuclear
physics applications.
In part \ref{part:contrib}, detailed instructions, minimal
requirements and recommendations are 
specified for the preparation of EoS tables
that can be incorporated in future versions of the CompOSE data base. 
If you plan to contribute your EoS, 
you should contact the CompOSE core team (\url{develop.compose@obspm.fr}), 
see appendix \ref{app:team}.
A summary of possible future extensions of the CompOSE data base
are summarized in chapter \ref{ch:extensions}.

The third part concerns the users of EoS data, that are provided by
CompOSE,
who want to test various equations of state in their
simulations of core-collapse supernovae, neutron-star mergers and
other scenarios. In general, they can
safely skip the second part and go directly to the third one. 
This part gives a brief introduction on nuclear
matter properties relevant for the construction of an EoS, as well as a
classification of different types of EoS models 
in chapter \ref{ch:eosmodels}.
The models can be distinguished either by using
different techniques to treat the many-body system 
of strongly interacting particles
or by assuming a different particle composition. The main aim
is to give the relevant information for the
interpretation of the data sheets, provided with each
available EoS table on the web site.
You will find remarks on 
the range of applicability of the various models,
i.e.\ the range of parameters where the code is tested and/or the 
made approximations are still valid.
Characteristic parameters of each model will be specified.
For more detailed information about the physics behind each model, we refer to
the original references. 

In addition, the third part of the manual
explains in chapter \ref{ch:online-service}
how to proceed in order to download an EoS table and
the computational tools. 
The latter allow 
the generation of tables with a
mesh different from the original one via interpolation, in order to adapt
the table to the need of the user. In addition, several thermodynamic
quantities can be calculated which are not contained in the original tables.
The use of the online service and web interface  is described in
chapter \ref{ch:online-service} too.


\section{Recent Changes}

Version 1.00 of this manual was published in 
\cite{manual:1} and is still available on the CompOSE web site. 
Version 2.00 contains some major changes, corrections and additions, in particular concerning the 
structure of the manual and the use of the 
provided software for handling data.
The main changes from version 1.00 to 2.00 in the manual are:
\begin{itemize}
\item the numerical values of physical constants, recommended for the preparation of EoS tables,
  in table \ref{tab:codata} were updated;
\item the discussion of constraints on nuclear matter properties in section
  \ref{sec:nucmatpar} has been modified;
\item sections 6.2 (Overview of models) and 6.3 (Phase transitions) have been removed since a
  comprehensive review \cite{Oertel:2016bki} is available;
\item the title 'User registration' of section \ref{sec:access} has been changed to
  'Access to CompOSE' with an adjusted description of the online sevice;
\item the description of how to use the \texttt{compose} program in section \ref{sec:direct_use}
  was adapted to the new version and the description of the files \texttt{eos.parameters}
 (with a change of the structure)
  and \texttt{eos.quantities} was moved to appendix \ref{app:files};
\item subsection 7.4.3 was moved to appendix  \ref{app:code} and the corresponding tables
  with the (sub-)routines and functions were updated;
\item tables 7.1, 
\ref{tab:ident_compo}, and \ref{tab:hdf5micro} were
  amended and a new table \ref{tab:f_derivatives} was introduced;
\item new appendices \ref{app:files} and \ref{app:code} were introduced;
\item the appendices \ref{app:team} and \ref{app:ack} were updated;
\item the bibliography has been updated.
\end{itemize}
The main changes to the \texttt{compose} program are
\begin{itemize}
\item most vectors and matrices are allocated dynamically with the appropriate size for the
  particular EoS used for generating a user-defined EoS table;
\item the program \texttt{compose} runs by default in a 
  standard version that guides the user
  through the selection of output data and the tabulation scheme;
\item the input files \texttt{eos.parameters} (with modified structure)
  and \texttt{eos.quantities} don't have to be provided
  by the user because they are created automatically by running
  the \texttt{compose} program;
\item an option for generating tables for given entropy per baryon
  instead of the temperature has been added;
\item the number of output quantities has been extended.
\end{itemize}
Major changes in the current version 3.01 are:
\begin{itemize}
    \item update of references to physical constants and nuclear properties in section 3.1;
    \item update of physical constants in table 3.1;
    \item distinguishing of particle and antiparticle densities as well as net number densities for fermions with extended indexing scheme, modified new tables 3.2 and 3.3;
    \item discussion of conversion from nuclear to astrophysical units in section 3.5;
    \item additional discussion of calculation of speed of sound in section 3.6;
    \item addition of new section 3.8 on transport properties;
    \item addition of new section 3.10 on properties of neutron stars;
    \item extended classification of EoS tables in section 4,2;
    \item introduction of new tables with transport properties and information on neutron stars in section 4.2;
    \item discussion of different families of tabulated data and EoS models in section 7.4;
    \item extended description of using EoS tables with software from CompOSE and LORENE;
    \item availability of web tools to calculate EoS tables and visualise the results in section 7.6;
    \item extension of table 7.2;
    \item update of CompOSE team in appendix D;
    \item information on citing CompOSE in appendix E.
\end{itemize}

Examples on using the \texttt{compose}\index{compose} program are given in a new ``quick guide for users''\index{quick guide for users} with some concrete examples.

\chapter{Definitions and notation}
\label{ch:defnot}

The equations of state in the CompOSE data base are provided under
some common assumptions on the physical conditions of the matter
that are specified in this chapter. In order to fix the notation in 
the present document, the definition of all relevant quantities is
given.

\section{Units and conventions}

We use natural units\index{units} 
$\hbar=c=k_{B}=1$\index{$\hbar$}\index{$c$}\index{$k_{B}$}
throughout this document as
is customary in nuclear physics.
Energies\index{energy} and temperatures\index{temperature} 
are measured in MeV, lengths\index{length} in fm.
Units are given in parentheses $[\dots]$ for all quantities
when they are defined within this chapter.
For conversion among your favourite units, the use 
of the  CODATA\index{CODATA} values \cite{CODATA2018}
(\url{http://physics.nist.gov/cuu/Constants/index.html}, \url{www.codata.org})
is recommended, see 
Tab.~\ref{tab:codata}\footnote{The compilation 
on physical constants and review of particle properties 
\cite{ParticleDataGroup:2020ssz}
by the Particle Data Group,
\url{pdg.lbl.gov}, contains the CODATA values together with some additional
constants concerning the properties and 
interaction of elementary particles.}.
They should also be used in the preparation
of new EoS tables to be incorporated in the CompOSE data base.

In astrophysical applications of equations of state it is often customary to change from nuclear units to units that are traditionally used in that field, in particular for temperature, pressure and energy densities. See section \ref{sec:thpot} for examples.

For experimental 
binding energies\index{binding energy} of nuclei, the values of the
2016 Atomic mass evaluation\index{Atomic Mass Evaluation} \cite{Ame2016a,Ame2016b}
(AME2016) 
or the more recent 2020 Atomic mass evaluation\index{Atomic Mass Evaluation}
\cite{Huang:2021nwk,Wang:2021xhn}
(AME2020, \url{http://amdc.impcas.ac.cn})
and corresponding up\-dates are recommended.
Ground state spin\index{spin!ground state} assignments should be taken from
the Nubase2016\index{Nubase} \cite{Nubase2016}
or the Nubase2020\index{Nubase} \cite{Kondev:2021lzi}
evaluation of nuclear properties
(\url{http://amdc.impcas.ac.cn})
and corresponding updates.

\begin{table}[ht]
\begin{center}
\caption{\label{tab:codata}%
Recommended values for physical constants from the 2018
CODATA evaluation and the compilation of the Particle Data Group \cite{ParticleDataGroup:2020ssz}. 
}
\begin{tabular}{llll}
\toprule
Quantity & Symbol & Value & Unit \\
\midrule
speed of light\index{speed of light} 
in vacuum & $c\index{$c$}$ & $299792458$ & m~s$^{-1}$
\\
Planck's constant\index{constant!Planck's} 
&$\hbar\index{$\hbar$}$ & $1.054571817 \times 10^{-34}$ & J~s\\
                 &        & $6.582119569 \times 10^{-22}$ & MeV~s \\
                 &        & $197.3269804$ & MeV~fm~$c^{-1}$ 
\\
Boltzmann's constant\index{constant!Boltzmann's} 
& $k_{B}$\index{$k_{B}$} & $1.380649 \times 10^{-23}$ & J~K$^{-1}$
\\
                     &         & $8.617333262 \times 10^{-11}$ & MeV~K$^{-1}$ 
\\
gravitational constant\index{constant!gravitational} 
& $G$\index{$G$} & $6.67430 \times 10^{-11}$ & m$^3$~kg$^{-1}$~s$^{-2}$
\\
                       &     & $6.70883 \times 10^{-39}$ &
                       GeV$^{-2}c^{-4}$
\\
square of elementary charge & $e^{2}$ &
$1.439964547$ & MeV~fm
\\
 fine structure constant\index{constant!fine structure} 
 & $\alpha=e^{2}/(\hbar c)$\index{$\alpha$}\index{$e^{2}$} &
$1/137.035999084$ & $-$ \\
\\
 neutron mass\index{mass!neutron} & $m_{n}$\index{$m_{n}$} & $939.56542052$ & MeV~$c^{-2}$ \\
 proton mass\index{mass!proton}  & $m_{p}$\index{$m_{p}$} & $938.27208816$
 & MeV~$c^{-2}$ \\
 electron mass\index{mass!electron} 
& $m_{e}$\index{$m_{e}$} & $0.51099895000$ & MeV~$c^{-2}$ \\
 muon mass\index{mass!muon} & $m_{\mu}$\index{$m_{\mu}$} & $105.6583745$ & MeV~$c^{-2}$ 
\\
unified atomic mass unit & $u$\index{$u$} & $931.49410242$ & MeV~$c^{-2}$
\\
solar mass & $M_{\odot}$ & $1.98841 \times 10^{30}$ & kg \\
 & & $1.11542 \times 10^{60}$ & MeV~$c^{-2}$
\\
\bottomrule
\end{tabular}
\end{center}
\end{table}

\section{Physical conditions}
\label{sec:physcon}

Predictions for the 
properties of dense matter can differ considerably depending on the
employed model, the considered constituents and interactions.
In the CompOSE database, the equation of state
is considered to describe dense matter in thermodynamic 
equilibrium\index{equilibrium!thermodynamic},
a short denotation of the fulfillment of several equilibrium conditions. Each
of them corresponds to the existence of an intensive thermodynamic variable that applies to the full
system under consideration. First of all there are the thermal and mechanic equilibrium that allow
to define the temperature $T$ and the pressure $p$. Chemical equilibrium means equilibrium with regard
to strong, electromagnetic and weak reactions. For the latter, one has to distinguish several cases, i.e.,
reactions within a generation of particles, e.g., electronic leptons and muonic leptons, and reactions across flavor generations, e.g., strangeness changing reaction. The equilibrium of each type of reaction is linked to the corresponding chemical potential. These are the baryonic ($\mu_{b}$), charge ($\mu_{q}$), electron lepton ($\mu_{le}$), muon lepton ($\mu_{l\mu}$), and strangeness ($\mu_{s}$) chemical potentials.

It is assumed that all the constituents are in chemical
equilibrium\index{equilibrium!chemical} regardless 
of the time scales\index{time scale} 
and reaction rates\index{reaction!rate}
for the relevant conversion reactions mediated by strong and
electromagnetic interactions. This condition leads to
relations between the 
chemical potentials\index{potential!chemical} 
of all particles.

In contrast, an equilibrium with
respect to weak\index{equilibrium!weak} interaction reactions, in particular
$\beta$-equilibrium\index{equilibrium!$\beta$}, is not supposed in
general. Similarly, 
the chemical potentials of the massive leptons\index{lepton} of all generations,
i.e.\ electrons\index{electron} and muons\index{muon}
(tauons\index{tauon} are irrelevant for the considered conditions)
cannot be assumed to be equal.
There is an option available such that EoS tables can be provided
in particular cases which also take
$\beta$-equilibrium into account, reducing the number of independent
parameters. Assumptions on the relation between the electron and muon chemical potentials are discussed in the description of each model separately. 
In almost all cases they are assumed to be identical.

For EoS models with
strangeness\index{strangeness}\index{quark!strange} 
bearing particles,
e.g.\ hyperons\index{hyperon} or kaons\index{kaon}, 
it is assumed
that the strangeness chemical
potential\index{potential!chemical!strangeness} 
vanishes. This means that we assume the strangeness changing weak
interactions to be in equilibrium. Note that this is not the case in
heavy-ion collisions\index{collision!heavy-ion}, 
where, on the contrary, strangeness is conserved,
i.e., there is no net strangeness.

Except for the tables of pure hadronic\index{matter!hadronic} 
or quark matter\index{matter!quark} without massive
leptons, also local charge neutrality\index{charge!neutrality} 
is assumed to hold.
Neutrinos\index{neutrino} and their contribution to thermodynamic
properties are never included in the present tables. They are
usually treated independently from the EoS in astrophysical
simulations because a thermodynamic
equilibrium\index{equilibrium!thermodynamic} 
can not be assumed in general.

Photons\index{photon} are usually included in equations of state for astrophysical
applications. They only contribute at nonzero temperatures. The treatment of photons 
is discussed in section \ref{sec:photons}.

\section{Particle number densities and particle fractions}

Particle number densities\index{density!number} 
of all particles $i$ are given by
\begin{equation}
 n_{i}\index{$n_{i}$} = \frac{N_{i}}{V} \qquad [\mbox{fm}^{-3}]
\end{equation}
where $N_{i}$\index{$N_{i}$} [dimensionless] is the particle
number\index{particle!number} 
inside the volume\index{volume} 
$V$\index{$V$} [fm$^{3}$].
Note that for particles with half-integer spin\index{spin!half-integer},
$n_{i}$ represents the net particle density\index{density!net}, i.e.\
it is the difference between the particle and antiparticle\index{antiparticle}
density. E.g., for electrons we have $n_{e} = n_{e^{-}}-n_{e^{+}}$.
For particles with integer spin\index{spin!integer}, e.g.\ 
mesons\index{meson}, 
particle and antiparticle\index{antiparticle} 
densities\index{density!meson}\index{density!antiparticle} 
are distinguished and given separately. For particles with half-integer spin\index{spin},
it will be possible to give the particle and antiparticle densities separately in am
upcoming version of the \texttt{CompOSE} program.

From the individual particle number densities several
new \textit{composite} number densities\index{density!number!composite} 
can be deduced that are convenient to
characterize the state of the system.
The baryon number density $n_{b}$\index{density!number!baryon} 
is given by
\begin{equation}
 n_{b}\index{$n_{b}$} = \frac{B}{V}\index{$B$} =
\sum_{i} B_{i} n_{i} \qquad [\mbox{fm}^{-3}]
\end{equation}
with the baryon number\index{baryon!number} 
$B_{i}$\index{$B_{i}$} [dimensionless] for a particle $i$
and the total baryon number
\begin{equation}
    B = \sum_{i} B_{i}  \qquad [\mbox{dimensionless}] 
\end{equation}
inside the volume $V$.
The baryon number of a nucleus
$i$ is just the mass number $A_{i}$\index{$A_{i}$}
[dimensionless] and for a 
quark\index{quark!baryon number} $i$ one has $B_{i}= 1/3$.
\textbf{Warning:} In many astrophysical applications, a mass
density\index{density!mass} $\varrho = m \, n_{b}$\index{$\varrho$} 
[MeV~fm$^{-3}$]
is introduced as a parameter with
$m$ representing, e.g., the neutron mass\index{mass!neutron} or the
atomic mass unit u\index{atomic mass unit}.
But, 
$\varrho$\index{$\varrho$} is not identical
to the total mass\index{density!mass!total} 
($=$ total internal energy\index{energy!internal!total}) density. 
Hence, $\varrho V$ is not a conserved quantity.

The strangeness number density $n_{s}$\index{density!number!strangeness}
\begin{equation}
 n_{s}\index{$n_{s}$} = \frac{S}{V}\index{$S$} =
\sum_{i} S_{i} n_{i} \qquad [\mbox{fm}^{-3}]
\end{equation}
with the total strangeness 
\begin{eqnarray}
 S = \sum_{i} S_{i} \qquad [\mbox{dimensionless}] 
\end{eqnarray}
inside the volume $V$ 
with strangeness numbers\index{strangeness!number}
$S_{i}$\index{$S_{i}$} [dimensionless],
as well as
the lepton number densities $n_{le}$, $n_{l\mu}$\index{density!number!lepton}
and number of leptons $N_{le}$, $N_{l\mu}$\index{lepton!number} 
\begin{eqnarray}
 n_{le}\index{$n_{le}$} = \frac{N_{le}}{V}\index{$N_{le}$}
& = & \sum_{i} L^{e}_{i} n_{i} \qquad [\mbox{fm}^{-3}]
 \\
 n_{l\mu}\index{$n_{l\mu}$} = \frac{N_{l\mu}}{V}\index{$N_{l\mu}$}
& = & \sum_{i} L^{\mu}_{i} n_{i} \qquad [\mbox{fm}^{-3}]
\end{eqnarray}
with 
lepton numbers
 $L^{e}_{i}$\index{$L^{e}_{i}$},
 $L^{\mu}_{i}$\index{$L^{\mu}_{i}$} 
[dimensionless], respectively, 
defined similarly to
the baryon number density. 
Since neutrinos are not included in the CompOSE EoS, $n_{le} =
n_{e}$\index{$n_{e}$} and $n_{l\mu}=n_{\mu}$\index{$n_{\mu}$}. 
We also consider the charge density of strongly
  interacting particles $n_{q}$\index{density!charge}
\begin{equation}
 n_{q}\index{$n_{q}$} = \frac{Q}{V}\index{$Q$} 
= \sum_{i}{}^{\prime}\index{$\sum_{i}{}^{\prime}$} 
 Q_{i} n_{i} \qquad [\mbox{fm}^{-3}]
\end{equation}
with electric charge numbers
$Q_{i}$\index{$Q_{i}$} [dimensionless] 
and the total charge
\begin{equation}
    Q = \sum_{i}{}^{\prime}\index{$\sum_{i}{}^{\prime}$} 
 Q_{i} \qquad [\mbox{dimensionless}] 
\end{equation}
inside the volume $V$.
The prime at the
sum symbols indicates that the summation runs over all particles
(including quarks) except leptons. 
For a nucleus\index{nucleus} 
${}^{A_{i}}Z_{i}$, the baryon number $B_{i}$ and the charge number
$Q_{i}$ are simply given by the mass number $A_{i}$\index{$A_{i}$} 
and atomic number $Z_{i}$\index{$Z_{i}$}, respectively.
Tables \ref{tab:partindex_fermions_sep} and \ref{tab:partindex_bosons} 
summarize baryon, strangeness, charge 
and lepton numbers of the most relevant particles
that appear in the definition of the composite number densities.

Corresponding to the particle number densities $n_{i}$,
the particle number fractions\index{fraction!particle number}
\begin{equation}
\label{eq:Ydef}
 Y_{i}\index{$Y_{i}$} = \frac{n_{i}}{n_{b}} \qquad [\mbox{dimensionless}]
\end{equation}
are defined. Due to this definition with $n_{b}$, we have the
normalization condition
\begin{equation}
 \sum_{i}{} B_{i} Y_{i} = 1
\end{equation}
with a summation over all particles. Note that $Y_{i}$ is not
necessarily
identical to the particle mass number fraction\index{fraction!particle
mass}
\begin{equation}
 X_{i} = B_{i}Y_{i}\index{$X_{i}$} \qquad [\mbox{dimensionless}]
\end{equation}
that is often introduced. We prefer to use $Y_{i}$ and not $X_{i}$ 
since the latter quantity is zero
for all particles with baryon number $B_{i}=0$, e.g.\ mesons.

Because of the imposed physical conditions (see \ref{sec:physcon}), 
the state of the system
is uniquely characterized by the three quantities
temperature $T$\index{$T$} [MeV], baryon number density
$n_{b}$\index{$n_{b}$} [fm$^{-3}$]
and charge density of strongly interaction particles
$n_{q}$\index{$n_{q}$} [fm$^{-3}$]. 
Charge neutrality\index{charge!neutrality} implies 
$n_{q} = n_{le}+n_{l\mu}$\index{$n_{le}$}\index{$n_{l\mu}$} and the strangeness number 
density\index{density!number!strangeness} $n_{s}$\index{$n_{s}$} is fixed by
the condition $\mu_{s}=0$\index{$\mu_{s}$}. 
Instead of $n_{q}$, it is more convenient
to use the charge fraction
of strongly interacting particles\index{fraction!charge}
\begin{equation}
 Y_{q}\index{$Y_{q}$} = \frac{n_{q}}{n_{b}} \qquad [\mbox{dimensionless}]
\end{equation}
as the third independent quantity.
The choice of $Y_{q}$ instead of the electron fraction\index{fraction!electron}
\begin{equation}
 Y_{e}\index{$Y_{e}$} = \frac{n_{e}}{n_{b}} \qquad [\mbox{dimensionless}]
\end{equation}
as a parameter is motivated by the
following facts. 
In pure hadronic (quark) equations of state (i.e.\ without leptons)
only $Y_{q}$ and not $Y_{e}$ is defined.
In EoS models with the condition of charge neutrality\index{charge!neutrality}
that contain electrons as the only considered charged lepton, 
the charge fraction of strongly interacting particles
$Y_{q}$ is identical to the electronic charge fraction $Y_{e}$.
In models with electrons and muons, charge neutrality requires
\begin{equation}
 Y_{q} = Y_{e}+Y_{\mu} = Y_{l}\index{$Y_{l}$}
\end{equation}
with the muon fraction
\begin{equation}
 Y_{\mu}\index{$Y_{\mu}$} 
= \frac{n_{\mu}}{n_{b}} \qquad [\mbox{dimensionless}] 
\end{equation}
and the total lepton fraction $Y_{l}$\index{fraction!lepton!total} [dimensionless].
In this case, the balance between the electron and muon densities
depends on the assumed relation of the electron and muon chemical
potentials.


\section{Particle indexing}

The composition\index{composition} 
of dense matter can rapidly change with 
temperature\index{temperature} $T$ [MeV],
baryon number density\index{density!number!baryon} 
$n_{b}$ [fm${}^{-3}$] and the charge
fraction of strongly interacting particles\index{fraction!charge} $Y_{q}$
[dimensionless]. We introduce
an indexing\index{indexing!particle} scheme for the 
particles\index{particle!index} that allows to identify them
uniquely in order to store only the most
abundant particles in the EoS tables. In tables \ref{tab:partindex_fermions_sep}, \ref{tab:partindex_fermions_net},  
and \ref{tab:partindex_bosons}
an overview of indices for most of the relevant particles is
presented. Indices of missing particles can be added on request.
In the case of fermions, indices of quantities for particles and antiparticles can be given separately as well as net quantities, 
i.e., net particle number densities as differences of particle and antiparticle densities.
Thus in table \ref{tab:partindex_fermions_net} indices for net quantities are given without quantum numbers.

In addition to the various particles that can appear in dense matter,
there is the possibility of strong two- or even three-particle
correlations\index{correlation} such as pairing\index{pairing} of nucleons or quarks 
and phenomena such as superfluidity can arise. Hence, an indexing
scheme for identifying these channels is introduced, too. In table
\ref{tab:corrindex} the notation is given for these correlations.  
Indices of missing particles can be added on request.
In the case of fermions, quantities can be given
for particles and antiparticles separately as well as for the corresponding net
quantities, e.g,, number densities as differences of particle and antiparticle densities.

\begin{table}[ht]
\begin{center}
\caption{\label{tab:partindex_fermions_sep}%
Baryon\index{baryon!number} ($B_{i}$\index{$B_{i}$}), 
strangeness\index{strangeness!number} ($S_{i}$\index{$S_{i}$}), 
charge\index{charge!number} ($Q_{i}$\index{$Q_{i}$}),
lepton\index{lepton!number} ($L^{e}_{i}$\index{$L^{e}_{i}$},
$L^{\mu}_{i}$\index{$L^{\mu}_{i}$})
numbers and indices\index{particle!index} $I_{i}\index{$I_{i}$}$ 
of nuclei and the most relevant fermions $i$
in dense matter. (Not yet implemented in the \texttt{CompOSE} program.)}
{\small 
\begin{tabular}{llrrrrrr}
\toprule
particle class & symbol of particle $i$
 & $B_{i}$ & $S_{i}$ & $Q_{i}$ & $L^{e}_{i}$ & $L^{\mu}_{i}$ & 
 particle index $I_{i}$ \\
\toprule
 nuclei ($A > 1$) & ${}^{A}Z$ & $A$\index{$A$}   
 & $0$   & $Z$\index{$Z$} & $0$ & $0$ & $1000000 + 1000 \cdot A+Z$   \\
         & ${}^{A}\bar{Z}$ & $-A$   
 & $0$   & $-Z$ & $0$ & $0$ & $2000000 + 1000 \cdot A+Z$   \\
\midrule
 leptons & $e^{-}$       & $0$ & $0$ & $-1$ & $+1$ & $0$ & 2 \\
        & $e^{+}$       & $0$ & $0$ & $+1$ & $-1$ & $0$ & 4 \\
        & $\mu^{-}$     & $0$ & $0$ & $-1$ & $0$ & $+1$ & 3 \\
        & $\mu^{+}$     & $0$ & $0$ & $+1$ & $0$ & $-1$ & 5 \\
\midrule
baryons & $n$\index{$n$}          & $+1$ & $0$  & $0$  & 0 & 0 & 12  \\
        & $\overline{n}$\index{$\overline{n}$}    & $-1$ & $0$  & $0$  & 0 & 0 & 14  \\
        & $p$\index{$p$}          & $+1$ & $0$  & $+1$  & 0 & 0 & 13  \\
        & $\overline{p}$\index{$\overline{p}$}    & $-1$ & $0$  & $-1$  & 0 & 0 & 15  \\
        & $\Delta^{-}$\index{$\Delta^{-}$}  & $+1$ & $0$  & $-1$ & 0 & 0 & 24 \\
        & $\overline{\Delta}^{-}$\index{$\overline{\Delta}^{-}$}  & $-1$ & $0$  & $+1$ & 0 & 0 & 28 \\
        & $\Delta^{0}$\index{$\Delta^{0}$}  & $+1$ & $0$  & $0$  & 0 & 0 & 25 \\
        & $\overline{\Delta}^{0}$\index{$\overline{\Delta}^{0}$}  & $-1$ & $0$  & $0$  & 0 & 0 & 29 \\
        & $\Delta^{+}$\index{$\Delta^{+}$}  & $+1$ & $0$  & $+1$ & 0 & 0 & 26 \\
        & $\overline{\Delta}^{+}$\index{$\overline{\Delta}^{+}$}  & $-1$ & $0$  & $-1$ & 0 & 0 & 30 \\
        & $\Delta^{++}$\index{$\Delta^{++}$} & $+1$ & $0$  & $+2$ & 0 & 0 & 27 \\
        & $\overline{\Delta}^{++}$\index{$\overline{\Delta}^{++}$} & $-1$ & $0$  & $-2$ & 0 & 0 & 31 \\
        & $\Lambda$\index{$\Lambda$}    & $+1$ & $-1$  & $0$    & 0 & 0 & 101  \\
        & $\overline{\Lambda}$\index{$\overline{\Lambda}$}    & $-1$ & $+1$  & $0$    & 0 & 0 & 102  \\
        & $\Sigma^{-}$\index{$\Sigma^{-}$}  & $+1$ & $-1$  & $-1$   & 0 & 0 & 113  \\
        & $\overline{\Sigma}^{-}$\index{$\overline{\Sigma}^{-}$}  & $-1$ & $+1$  & $+1$   & 0 & 0 & 116  \\
        & $\Sigma^{0}$\index{$\Sigma^{0}$}  & $+1$ & $-1$  & $0$    & 0 & 0 & 114  \\
        & $\overline{\Sigma}^{0}$\index{$\overline{\Sigma}^{0}$}  & $-1$ & $+1$  & $0$    & 0 & 0 & 117  \\
        & $\Sigma^{+}$\index{$\Sigma^{+}$}  & $+1$ & $-1$  & $+1$   & 0 & 0 & 115  \\
        & $\overline{\Sigma}^{+}$\index{$\overline{\Sigma}^{+}$}  & $-1$ & $+1$  & $+1$   & 0 & 0 & 118  \\
        & $\Xi^{-}$\index{$\Xi^{-}$}     & $+1$ & $-2$  & $-1$   & 0 & 0 & 122  \\
        & $\overline{\Xi}^{-}$\index{$\overline{\Xi}^{-}$}     & $-1$ & $+2$  & $+1$   & 0 & 0 & 124  \\
        & $\Xi^{0}$\index{$\Xi^{0}$}     & $+1$ & $-2$  & $0$    & 0 & 0 & 123  \\
        & $\overline{\Xi}^{0}$\index{$\overline{\Xi}^{0}$}     & $-1$ & $+2$  & $0$    & 0 & 0 & 125  \\
\midrule
quarks  & $u$\index{$u$}       & $1/3$ & $0$   & $+2/3$ & 0 & 0 & 503 \\
        & $\overline{u}$\index{$\overline{u}$} & $-1/3$ & $0$ & $-2/3$ & 0 & 0 & 506 \\
        & $d$\index{$u$}       & $1/3$ & $0$   & $-1/3$ & 0 & 0 & 504 \\
        & $\overline{d}$\index{$\overline{u}$} & $-1/3$ & $0$ & $+1/3$ & 0 & 0 & 507 \\
        & $s$\index{$s$}       & $1/3$ & $-1$  & $-1/3$ & 0 & 0 & 505 \\
        & $\overline{s}$\index{$\overline{s}$} & $-1/3$ & $+1$ & $+1/3$ & 0 & 0 & 508 \\
\bottomrule
\end{tabular}
}
\end{center}
\end{table}

\begin{table}[ht]
\begin{center}
\caption{\label{tab:partindex_fermions_net}%
Indices\index{particle!index} $I_{i}\index{$I_{i}$}$ 
of nuclei and the most relevant fermions $i$
in dense matter for net quantities.}
{\small 
\begin{tabular}{llr}
\toprule
particle class & symbol of particle $i$
 &  particle index $I_{i}$ \\
\toprule
 nuclei ($A > 1$) & ${A}^{Z}$/$\overline{A}^{Z}$ & $1000 \cdot A+Z$   \\
\midrule
leptons & $e^{-}$/$e^{+}$       & 0 \\
        & $\mu^{-}$/$\mu^{+}$     & 1 \\
\midrule
baryons & $n$\index{$n$}/$\overline{n}$\index{$\overline{n}$}          & 10  \\
        & $p$\index{$p$}/$\overline{p}$\index{$\overline{p}$}          & 11  \\
        & $\Delta^{-}$\index{$\Delta^{-}$}/$\overline{\Delta}^{-}$\index{$\overline{\Delta}^{-}$}  & 20 \\
        & $\Delta^{0}$\index{$\Delta^{0}$}/$\overline{\Delta}^{0}$\index{$\overline{\Delta}^{0}$}  & 21 \\
        & $\Delta^{+}$\index{$\Delta^{+}$}/$\overline{\Delta}^{+}$\index{$\overline{\Delta}^{+}$}  & 22 \\
        & $\Delta^{++}$\index{$\Delta^{++}$}/$\overline{\Delta}^{++}$\index{$\overline{\Delta}^{++}$} & 23 \\
        & $\Lambda$\index{$\Lambda$}/$\overline{\Lambda}$\index{$\overline{\Lambda}$}    & 100  \\
        & $\Sigma^{-}$\index{$\Sigma^{-}$}/$\overline{\Sigma}^{-}$\index{$\overline{\Sigma}^{-}$}  & 110  \\
        & $\Sigma^{0}$\index{$\Sigma^{0}$}/$\overline{\Sigma}^{0}$\index{$\overline{\Sigma}^{0}$}  & 111  \\
        & $\Sigma^{+}$\index{$\Sigma^{+}$}/$\overline{\Sigma}^{+}$\index{$\overline{\Sigma}^{+}$}  & 112  \\
        & $\Xi^{-}$\index{$\Xi^{-}$}/$\overline{\Xi}^{-}$\index{$\overline{\Xi}^{-}$}     & 120  \\
        & $\Xi^{0}$\index{$\Xi^{0}$}/$\overline{\Xi}^{0}$\index{$\overline{\Xi}^{0}$}     & 121  \\
\midrule
quarks  & $u$\index{$u$}/$\overline{u}$\index{$\overline{u}$}       & 500 \\
        & $d$\index{$d$}/$\overline{d}$\index{$\overline{d}$}       & 501 \\
        & $s$\index{$s$}/$\overline{s}$\index{$\overline{s}$}       & 502 \\
\bottomrule
\end{tabular}
}
\end{center}
\end{table}

\clearpage

\begin{table}[ht]
\begin{center}
\caption{\label{tab:partindex_bosons}%
Baryon\index{baryon!number} ($B_{i}$\index{$B_{i}$}), 
strangeness\index{strangeness!number} ($S_{i}$\index{$S_{i}$}), 
charge\index{charge!number} ($Q_{i}$\index{$Q_{i}$}),
lepton\index{lepton!number} ($L^{e}_{i}$\index{$L^{e}_{i}$},
$L^{\mu}_{i}$\index{$L^{\mu}_{i}$})
numbers and indices\index{particle!index} $I_{i}\index{$I_{i}$}$ 
of the most relevant bosons $i$
in dense matter.}
{\small 
\begin{tabular}{llrrrrrr}
\toprule
particle class & symbol of particle $i$
 & $B_{i}$ & $S_{i}$ & $Q_{i}$ & $L^{e}_{i}$ & $L^{\mu}_{i}$ & 
 particle index $I_{i}$ \\
\toprule
mesons  & $\omega$\index{$\omega$}      & $0$ & $0$ & $0$ & 0 & 0 & 200 \\
        & $\sigma$\index{$\sigma$}      & $0$ & $0$ & $0$ & 0 & 0 & 210 \\
        & $\eta$\index{$\eta$}        & $0$ & $0$ & $0$ & 0 & 0 & 220 \\
        & $\eta^{\prime}$\index{$\eta^{\prime}$} & $0$ & $0$ & $0$ & 0 & 0 &230 \\
        & $\rho^{-}$\index{$\rho^{-}$}  & $0$ & $0$ & $-1$ & 0 & 0 & 300 \\
        & $\rho^{0}$\index{$\rho^{0}$}  & $0$ & $0$ & $0$  & 0 & 0 & 301 \\
        & $\rho^{+}$\index{$\rho^{+}$}  & $0$ & $0$ & $+1$ & 0 & 0 & 302 \\
        & $\delta^{-}$\index{$\delta^{-}$} & $0$ & $0$ & $-1$ & 0 & 0 & 310 \\
        & $\delta^{0}$\index{$\delta^{0}$} & $0$ & $0$ & $0$ & 0 & 0 & 311 \\
        & $\delta^{+}$\index{$\delta^{+}$} & $0$ & $0$ & $+1$ & 0 & 0 & 312 \\
        & $\pi^{-}$\index{$\pi^{-}$}    & $0$ & $0$ & $-1$ & 0 & 0 & 320 \\
        & $\pi^{0}$\index{$\pi^{0}$}    & $0$ & $0$ & $0$  & 0 & 0 & 321 \\
        & $\pi^{+}$\index{$\pi^{+}$}    & $0$ & $0$ & $+1$ & 0 & 0 & 322 \\
        & $\phi$\index{$\phi$}       & $0$ & $0$ & $0$ & 0 & 0 & 400 \\
        & $\sigma_{s}$\index{$\sigma_{s}$}  & $0$ & $0$ & $0$ & 0 & 0 & 410 \\
        & $K^{-}$\index{$K^{-}$}       & $0$ & $-1$ & $-1$ & 0 & 0 & 420 \\
        & $K^{0}$\index{$K^{0}$}       & $0$ & $+1$ & $0$ & 0 & 0 & 421 \\
        & $\bar{K}^{0}$\index{$\bar{K}^{0}$} & $0$ & $-1$ & $0$ & 0 & 0 & 422 \\
        & $K^{+}$\index{$K^{+}$}       & $0$ & $+1$ & $+1$ & 0 & 0 & 423 \\
\midrule
photon & $\gamma$\index{$\gamma$}   & $0$   & $0$   & $0$    & 0 & 0 & 600 \\
\bottomrule
\end{tabular}
}
\end{center}
\end{table}

\begin{table}[ht]
\begin{center}
\caption{\label{tab:corrindex}%
Baryon\index{baryon!number} ($B_{i}$\index{$B_{i}$}), 
strangeness\index{strangeness!number} ($S_{i}$\index{$S_{i}$}),
charge\index{charge!number} ($Q_{i}$\index{$Q_{i}$})
numbers and indices\index{particle!index} $I_{i}\index{$I_{i}$}$ 
of the most relevant two-particle correlations $i$
in dense matter. The lepton numbers $L^{e}_{i}$\index{$L^{e}_{i}$} 
and $L^{\mu}_{i}$\index{$L^{\mu}_{i}$} are
always zero for these correlations.}
{\small
\begin{tabular}{lccrrrr}
\toprule
correlation class & particles & channel 
 & $B_{i}$\index{$B_{i}$} & $S_{i}$\index{$S_{i}$} & 
 $Q_{i}$\index{$Q_{i}$} & index $I_{i}$ of correlation\\
\toprule
two-body & $nn$ & ${}^{1}S_{0}$\index{${}^{1}S_{0}$} & $2$ & $0$ & $0$ & 700 \\
         & $np$ & ${}^{1}S_{0}$ & $2$ & $0$ & $1$ & 701 \\
         & $pp$ & ${}^{1}S_{0}$ & $2$ & $0$ & $2$ & 702 \\
         & $np$ & ${}^{3}S_{1}$\index{${}^{3}S_{1}$} & $2$ & $0$ & $1$ & 703 \\
\bottomrule
\end{tabular}
}
\end{center}
\end{table}

\section{Thermodynamic potentials and basic quantities}
\label{sec:thpot}

All thermodynamic properties\index{property!thermodynamic} 
of a system are completely determined
if a thermodynamic potential is known as a function of its 
natural variables.
In the general case, this can be formulated as follows.
The properties can be derived
from the thermodynamic potential\index{potential!thermodynamic} 
$\Xi = \Xi(\{x_{i}\},\{\xi_{j}\})$\index{$\Xi$}
depending on $n$ natural variables\index{variable!natural} 
$x_{i}$, $i=1,\dots,n_{1}$\index{$x_{i}$} and $\xi_{j}$, $j = 1\dots,
n_{2}$\index{$\xi_{j}$} with $n_{1} + n_{2} = n$.
The quantities $x_{i}$ and $\xi_{i}$ represent extensive and intensive
variables\index{variable!intensive}\index{variable!extensive}, respectively.
The relations
\begin{equation}
 \xi_{i} = \left. \frac{\partial \Xi}{\partial x_{i}} \right|_{x_{k}, k \neq i; \xi_{j}}
 = \xi_{i}(\{x_{i}\},\{\xi_{j}\})
\end{equation}
and
\begin{equation}
 x_{j} =  -\left. \frac{\partial \Xi}{\partial \xi_{j}} \right|_{\xi_{k}, k \neq j; x_{i}}
 = x_{j}(\{x_{i}\},\{\xi_{j}\})
\end{equation}
are the thermodynamic equations of state\index{equation of state} and
define the variables\index{variable!conjugate} 
$\xi_{i}$ that are conjugate to $x_{i}$
as first partial derivatives of $\Xi$ and vice versa.
Due to Euler's theorem\index{theorem!Euler's} 
on homogeneous functions\index{function!homogeneous}, the
thermodynamic potential is given by the sum
\begin{equation}
\label{eq:xix}
 \Xi(\{x_{i}\},\{\xi_{j}\}) = \sum_{i=1}^{n_{1}} \xi_{i} x_{i}
\end{equation}
running over all 
$i=1,\dots,n_{1}$ extensive variables\index{variable!extensive}
that are primary variables of the thermodynamic potential $\Xi$. 
Thus, the knowledge of all relevant equations of
state or first derivatives is sufficient to recover the thermodynamic
potential $\Xi$ completely.
For the mixed second partial derivatives of a thermodynamic potential $\Xi$,
the result is independent of the order of differentiation
and the Maxwell relations\index{relation!Maxwell}
\begin{equation}
 \left. \frac{\partial \xi_{l}}{\partial x_{k}} 
 \right|_{x_{j}, j\neq k} 
 = \left. \frac{\partial \xi_{k}}{\partial x_{l}} 
 \right|_{x_{j}, j \neq l} 
\end{equation}
are obtained. For a more detailed discussion of these and further
aspects we refer the reader to standard text books on thermodynamics.

In most models for the EoS of dense matter, 
the temperature\index{temperature} $T$\index{$T$} [MeV], 
the volume\index{volume} $V$\index{$V$}
[fm$^{3}$] and the individual 
particle numbers\index{particle!number} $N_{i}$ \index{$N_{i}$} [dimensionless]
are selected as natural variables. This case corresponds to the (Helmholtz) free
energy\index{energy!free}\index{energy!Helmholtz free} 
$F=F(T,V,N_{i})$\index{$F$}
[MeV] as the relevant thermodynamic potential that contains all
information. Note that we assume that the free energy includes contributions 
by the rest masses\index{rest mass} 
of the particles. Keeping the volume $V$ fixed,
it is convenient to define the free energy density\index{density!energy!free}
\begin{equation}
 f(T,n_{i})\index{$f$} = \frac{F}{V} \qquad [\mbox{MeV~fm}^{-3}] 
\end{equation} 
and the entropy density\index{density!entropy}
\begin{equation}
 s(T,n_{i})\index{$s$} =  - \left. \frac{\partial f}{\partial
     T}\right|_{n_{i}}  \qquad [\mbox{fm}^{-3}]
\end{equation}
with the entropy\index{entropy}
\begin{equation}
 S(T,V,N_{i})\index{$S$} = V s(T,n_{i}) = - \left. \frac{\partial F}{\partial
     T}\right|_{V,N_{i}} \qquad [\mbox{dimensionless}] \: .
\end{equation}
The chemical potential\index{potential!chemical} of a particle $i$ is given by
\begin{equation}
 \mu_{i}\index{$\mu_{i}$} = \left. \frac{\partial F}{\partial N_{i}}
 \right|_{T,V,N_{j},j\neq i} =  \left. \frac{\partial f}{\partial n_{i}}
 \right|_{T,n_{j},j\neq i}
 \qquad [\mbox{MeV}] 
\end{equation}
including the rest mass $m_{i}$\index{$m_{i}$} [MeV].
The pressure\index{pressure} is obtained from
\begin{equation}
 p\index{$p$} = - \left. \frac{\partial F}{\partial V}
 \right|_{T,N_{i}} 
 =  n_{b}^{2} \left. \frac{\partial (f/n_{b})}{\partial n_{b}}
 \right|_{T,Y_{q}} 
 = \sum_{i} \mu_{i} n_{i} - f
 \qquad [\mbox{MeV~fm}^{-3}]  \: .
\end{equation}

Each of the composite densities\index{density!composite} 
$n_{b}$, $n_{s}$, $n_{le}$, $n_{l\mu}$, $n_{q}$
\index{$n_{b}$}\index{$n_{s}$}\index{$n_{le}$}\index{$n_{l\mu}$}\index{$n_{q}$}
is accompanied by a 
corresponding chemical potential,
i.e.\ we have the 
baryon number chemical potential\index{potential!chemical!baryon} 
$\mu_{b}$\index{$\mu_{b}$} [MeV],
the strangeness number chemical
potential\index{potential!chemical!strangeness} 
$\mu_{s}$\index{$\mu_{s}$} [MeV],
the electron lepton number chemical potential\index{potential!chemical!electron} 
$\mu_{le}$\index{$\mu_{le}$} [MeV],
the muon lepton number chemical potential\index{potential!chemical!muon} 
$\mu_{l\mu}$\index{$\mu_{l\mu}$} [MeV],
and the charge chemical
potential\index{potential!chemical!charge} 
$\mu_{q}$\index{$\mu_{q}$} [MeV].
The chemical potential
of a particle $i$ is then given by
\begin{equation}
  \mu_{i}\index{$\mu_{i}$} = B_{i} \mu_{b} + Q_{i} \mu_{q} + S_{i}
  \mu_{s} + L^{e}_{i} \mu_{le} + L^{\mu}_{i} \mu_{l\mu}
  \qquad [\mbox{MeV}] \: ,
\end{equation}
e.g.\ $\mu_{n} = \mu_{b}$, $\mu_{p}=\mu_{b}+\mu_{q}$ and $\mu_{e} = \mu_{le}-\mu_{q}$.
Note again that we use the relativistic definition of chemical potentials
including rest masses\index{rest mass}.
It has to be mentioned that $n_{b}$, $n_{s}$, $n_{le}$ and $n_{l\mu}$
are conjugate to $\mu_{b}$, $\mu_{s}$, $\mu_{le}$ and $\mu_{l\mu}$,
respectively. However, this is not true for $n_{q}$ and $\mu_{q}$ if
leptons are included in the EoS.
In general, it is assumed that $\mu_{s}=0$ since
strangeness changing weak interaction reactions are in equilibrium.
Nevertheless, the strangeness density $n_{s}$ can
be non-zero if 
particles with strangeness are considered in the EoS.

In addition to the free energy density, several other thermodynamic
potentials can be defined by applying Legendre
transformations\index{transformation!Legendre}, e.g.\ 
the internal energy\index{energy!internal}
\begin{equation}
 E\index{$E$} = E(S,V,N_{i}) = F+TS  \qquad [\mbox{MeV}] \: ,
\end{equation}
the free enthalpy\index{enthalpy!free} (Gibbs potential\index{potential!Gibbs})
\begin{equation}
 G\index{$G$} 
 = G(T,p,N_{i}) = F+pV = \sum_{i}\mu_{i}N_{i}   \qquad [\mbox{MeV}] \: ,
\end{equation}
the enthalpy\index{enthalpy}
\begin{equation}
 H\index{$H$} = H(S,p,N_{i}) = E+pV   \qquad [\mbox{MeV}] \: ,
\end{equation}
and the grand canonical potential\index{potential!grand canonical}
\begin{equation}
 \Omega\index{$\Omega$} = \Omega(T,V,\mu_{i}) = F-\sum_{i}\mu_{i}N_{i} = -pV 
  \qquad [\mbox{MeV}] 
\end{equation}
with the corresponding densities $e=E/V$\index{$e$}, 
$g=G/V$\index{$g$}, $h=H/V$\index{$h$} and 
$\omega = \Omega/V$\index{$\omega$} [MeV~fm${}^{-3}$], respectively.

It is convenient to define the free energy per baryon
\begin{equation}
 \mathcal{F}\index{$\mathcal{F}$} 
 = \frac{F}{N_{b}} = \frac{f}{n_{b}} \qquad [\mbox{MeV}] \: ,
\end{equation}
the internal energy per baryon
\begin{equation}
 \mathcal{E}\index{$\mathcal{E}$} 
 = \frac{E}{N_{b}} = \frac{e}{n_{b}} \qquad [\mbox{MeV}] \: ,
\end{equation}
the enthalpy per baryon
\begin{equation}
 \mathcal{H}\index{$\mathcal{H}$}
 = \frac{H}{N_{b}} = \frac{h}{n_{b}} \qquad [\mbox{MeV}] \: ,
\end{equation}
and 
the free enthalpy per baryon
\begin{equation}
 \mathcal{G}\index{$\mathcal{G}$}
 = \frac{G}{N_{b}} = \frac{g}{n_{b}} \qquad [\mbox{MeV}] 
\end{equation}
by dividing the corresponding thermodynamic potential by the 
total number of baryons
\begin{equation}
 N_{b}\index{$N_{b}$} = n_{b} V \qquad [\mbox{dimensionless}] \: .
\end{equation}

All equations of state can be obtained from the thermodynamic potentials $E(S,V,N_{i})$, $F(T,V,N_{i})$, $G(T,p,N_{i})$, $H(S,p,N_{i})$ or $\Omega(T,V,\mu_{i})$ by appropriate partial derivatives, in particular
\begin{eqnarray}
 T & = & \left. \frac{\partial E}{\partial S} \right|_{V,N_{i}}
 = \left. \frac{\partial H}{\partial S} \right|_{p,N_{i}} \: ,
 \\
 S & = & -\left. \frac{\partial F}{\partial T} \right|_{V,N_{i}}
 = -\left. \frac{\partial G}{\partial T} \right|_{p,N_{i}}
 = -\left. \frac{\partial \Omega}{\partial T} \right|_{V,\mu_{i}} \: ,
 \\
 -p & = & \left. \frac{\partial E}{\partial V} \right|_{S,N_{i}}
 = \left. \frac{\partial F}{\partial V} \right|_{T,N_{i}} \: ,
 \\
 V & = & \left. \frac{\partial G}{\partial p} \right|_{T,N_{i}}
 = \left. \frac{\partial H}{\partial p} \right|_{S,N_{i}}
 = \left. \frac{\partial \Omega}{\partial p} \right|_{T,\mu_{i}} \: ,
 \\
 \mu_{i} & = & \left. \frac{\partial E}{\partial N_{i}} \right|_{S,V,N_{j\neq i}}
 = \left. \frac{\partial F}{\partial N_{i}} \right|_{T,V,N_{j\neq i}}
 = \left. \frac{\partial G}{\partial N_{i}} \right|_{T,p,N_{j\neq i}}
 = \left. \frac{\partial H}{\partial N_{i}} \right|_{S,p,N_{j\neq i}} \: ,
 \\
 N_{i} & = & -\left. \frac{\partial \Omega}{\partial \mu_{i}} \right|_{S,V,\mu_{j\neq i}} \: .
\end{eqnarray}

The nuclear physics units given above can be converted easily to units that are frequently used in astrophysics. The most relevant cases are the following: 
\begin{enumerate}
    \item The temperature $T$ in MeV can be expressed in K (Kelvin) by a division with the Boltzmann constant $k_{B}$ as given in table \ref{tab:codata}. E.g., a temperature of $1$~MeV corresponds to about $1.16\cdot 10^{10}$~K and, conversely, $10^{10}$~K are about $0.862$~MeV.
    \item Any energy density, e.g, the internal energy density, in MeV~fm${}^{-3}$ can be expressed in the CGS unit erg/cm$^{-3}$ ($1$~erg = $10^{-7}$~J) by a multiplication with the factor $1.602176634\cdot 10^{33}$. 
    \item The pressure $p$ in MeV~fm$^{-3}$ can be expressed in the CGS unit dyn/cm$^{2}$ ($1$~dyn = $1$~erg/cm) by a multiplication with the same numerical factor $1.602176634\cdot 10^{33}$ as for the energy density.
\end{enumerate}

\section{Thermodynamic coefficients}
\label{sec:thermo_coeff}

\index{coefficient!thermodynamic}
In many applications tabulated values of 
the thermodynamic potentials and their first
derivatives are not sufficient and additional quantities that depend
on second derivatives of the thermodynamic potentials are needed.
In general, the relevant thermodynamic potential can depend on
a large number of independent variables such as temperature, volume and  particle
densities or chemical potentials. In applications, however, this
number often reduces due to physical constraints. E.g.,
the condition of weak chemical ($\beta$) equilibrium between certain particle species
reduces the number of independent chemical potentials. 
The condition of charge neutrality relates the particle numbers
of charged constituents. In the following, we consider only systems
that are
completely determined by the temperature $T$, the volume $V$, the
total number of baryons $N_{b}= n_{b} V $ and the charge number 
$N_{q} = Y_{q} N_{b} = Y_{q} n_{b} V$, or in the case of an
equation of state where leptons are present, the number of charged leptons,
which are equal to $N_q$ due to charge neutrality. Then
we have from the free energy per baryon $\mathcal{F}(T,n_{b},Y_{q})$ [MeV]
the specific heat capacity at constant volume\index{heat capacity}
\begin{equation}
 c_{V}\index{$c_{V}$} = \frac{T}{N_{b}} \left. \frac{dS}{dT} \right|_{V,N_{b},N_{q}}
 = -T \left. \frac{\partial^{2} \mathcal{F}}{\partial
     T^{2}}\right|_{n_{b},Y_{q}} \qquad [\mbox{dimensionless}] \: ,
\end{equation}
the tension coefficient at constant volume\index{coefficient!tension}
\begin{equation}
 \beta_{V}\index{$\beta_{V}$} =  \left. \frac{dp}{dT} \right|_{V,N_{b},N_{q}}
 = \left. \frac{dS}{dV} \right|_{T,N_{b},N_{q}}
 = n_{b}^{2} \left. \frac{\partial^{2} \mathcal{F}}{\partial T \partial
     n_{b}}\right|_{Y_{q}} \qquad [\mbox{fm}^{-3}] \: ,
\end{equation}
the isothermal compressibility\index{compressibility!isothermal}
\begin{eqnarray}
\label{eq:kappat}
 \kappa_{T}\index{$\kappa_{T}$}
 & = & -\frac{1}{V} \left. \frac{dV}{dp} \right|_{T,N_{b},N_{q}}
 =   \left( n_{b} \left. \frac{\partial p}{\partial n_{b}}
 \right|_{T,Y_{q}} \right)^{-1}
 \\ \nonumber & = & 
   \left( n_{b}^{2} \left. \frac{\partial^{2} (\mathcal{F}n_{b})}{\partial n_{b}^{2}}
     \right|_{T,Y_{q}} \right)^{-1}
\qquad [\mbox{MeV}^{-1}\mbox{fm}^{3}] \: ,
\end{eqnarray}
the expansion coefficient at constant pressure\index{coefficient!expansion}
\begin{equation}
 \alpha_{p}\index{$\alpha_{p}$} 
 = \frac{1}{V} \left. \frac{dV}{dT} \right|_{p,N_{b},N_{q}}
 = 
  \kappa_{T} \beta_{V}  \qquad [\mbox{MeV}^{-1}] \: ,
\end{equation}
the specific heat capacity at constant pressure\index{heat capacity}
\begin{equation}
 c_{p}\index{$c_{p}$} = \frac{T}{N_{b}} \left. \frac{dS}{dT}
 \right|_{p,N_{b},N_{q}}
 = c_{V} 
 + \frac{T}{n_{b}} \alpha_{p} \beta_{V}
 \qquad [\mbox{dimensionless}] \: ,
\end{equation}
the thermodynamic adiabatic index\index{index!adiabatic}
\begin{equation}
  \label{eq:Gamma}
 \Gamma\index{$\Gamma$} = \frac{c_{p}}{c_{V}} \qquad [\mbox{dimensionless}] 
\end{equation}
and the adiabatic compressibility\index{compressibility!adiabatic}
\begin{equation}
 \kappa_{S}\index{$\kappa_{S}$} 
 = - \frac{1}{V} \left. \frac{dV}{dp} \right|_{S,N_{b},N_{q}}
 = \frac{\kappa_{T}}{\Gamma}  \qquad [\mbox{MeV}^{-1}\mbox{fm}^{3}] \: .
\end{equation}
Different definitions of the adiabatic index are sometimes employed
  in astrophysical contexts. E.g., the form
  \begin{equation}
    \tilde{\Gamma}
    =  \left. \frac{\partial \ln p}{\partial \ln n_{b}}\right|_{S}
    = \frac{1}{p\kappa_{S}}
    \qquad [\mbox{dimensionless}]
  \end{equation}
  is used in applications with polytrope equations of state.
  A tilde is introduced here to distinguish it from (\ref{eq:Gamma}).
  A further definition is given by
  \begin{equation}
    \gamma = \left. \frac{\partial \ln p}{\partial \ln e}\right|_{S}
     \qquad [\mbox{dimensionless}]
  \end{equation}
  with the internal energy density $e$.
The square of the 
speed of sound\index{speed of sound} (isoscalar longitudinal
compression wave)
in the medium for a one-fluid flow at finite temperatures
is given by the relativistic definition
\begin{equation}
 c_{s}^{2}\index{$c_{s}$} 
 =  \left. \frac{dp}{de} \right|_{S,N_{b},N_{q}} 
 = \frac{1}{h\kappa_{S}}
  = \frac{p}{h} \tilde{\Gamma} = \frac{p}{e}\gamma
  \qquad [c^{2}]
\end{equation}
with the enthalpy density $h=e+p$ [MeV~fm$^{-3}$].
Alternatively, the expression
\begin{equation}
c_{s}^{2} =  
  \left. \frac{n_{b}}{h} \frac{\partial p}{\partial n_{b}}
   \right|_{Y_{q},e} 
 +  \left. \frac{\partial p}{\partial e} 
 \right|_{n_{b},n_{q}} 
  \qquad [c^{2}]
\end{equation}
can be used for constant hadronic charge fraction $Y_{q}$. 
In the application to cold compact stars, i.e., matter at zero temperature in $\beta$
equilibrium, the EoS depends only on a single variable, e.g., the baryon
density $n_{b}$ or the baryon chemical potential $\mu_{b}$. 
Then the square of the speed of sound can be calculated from
\begin{equation}
    c_{s}^{2} = \left. \frac{dp}{de} \right|_{T=0,\beta~eq.}
    =  \frac{n_{b}}{\mu_{b}} \left. \frac{d \mu_{b}}{d n_{b}} \right|_{T=0,\beta~eq.}
    \qquad [c^{2}]
\end{equation}
in the unit of the squared speed of light.
All of the above quantities can be calculated with the help
of the second derivatives of $\mathcal{F}$ or $f$ with respect to the
parameters $T$, $n_{b}$, and $n_{q}$.

\section{Microscopic quantities}

Most EoS models are based on 
microscopic models that
can give information on microscopic particle properties
in addition to the thermodynamic and compositional
quantities. Only few 
models are just simple parametrizations of thermodynamic quantities
without recurrence to the underlying microphysics. The
CompOSE database allows to store these microscopic values, too, 
such that they can be used in applications. 
In the present version, 
we consider a small set of quantities that are discussed in the
following. It can be easily extended if required.

In particular, we consider the effective masses of the particles 
in which case
one has to distinguish between different definitions.
The effective Landau mass\index{effective mass!Landau}
\begin{equation}
 m^{L}_{i}\index{$m^{L}_{i}$} \qquad [\mbox{MeV}]
\end{equation}
of a particle $i$ is related to the single-particle density of
states\index{density of states} and defined as
\begin{equation}
\frac{1}{m^{L}_{i}} = \frac{1}{p^{F}_{i}} \left. \frac{d E_{i}}{d
    p}\right|_{p = p_{i}^{F}}
 \: ,
\end{equation}
where $p^{F}_{i}$ denotes the respective Fermi-momentum and $E_{i}$ is the
single-particle energy. In non-relativistic Skyrme-Hartree-Fock models it
appears in the 
kinetic energy\index{energy!kinetic} contribution 
\begin{equation}
 T_{i}\index{$T_{i}$} = \frac{p_{i}^{2}}{2m^{L}_{i}} \qquad [\mbox{MeV}]
\end{equation}
to the total single-particle energy.
In addition to the kinetic single-particle energy $T_{i}$, a single-particle
potential\index{potential!single-particle}
\begin{equation}
 U_{i}\index{$U_{i}$} \qquad [\mbox{MeV}]
\end{equation}
contributes to the single-particle energy\index{energy!single-particle}
\begin{equation}
E_{i}\index{$E_{i}$} = T_{i} + U_{i} + m_{i}
\end{equation}
in nonrelativistic models. 
The effective Dirac mass\index{effective mass!Dirac}
\begin{equation}
 m^{D}_{i}\index{$m^{D}_{i}$} = m_{i} - S_{i} \qquad [\mbox{MeV}]
\end{equation}
is found in the single-particle Hamiltonian of relativistic models,
such as relativistic mean-field approaches. This quantity does not
reflect directly the density of single-particle states. At zero
temperature the corresponding effective Landau mass can be obtained
from
\begin{equation}
 m^{L}_{i} = \sqrt{(m^{D}_{i})^{2}+(p^{F}_{i})^{2}} \: .
\end{equation}
The effective Dirac mass depends on the
scalar self-energy\index{self-energy!scalar} 
\begin{equation}
 S_{i}\index{$S_{i}$} = \Sigma_{i}\index{$\Sigma_{i}$} \qquad [\mbox{MeV}]
\end{equation}
that enters the relativistic single-particle Hamiltonian together with the
vector self-energy\index{self-energy!vector}
\begin{equation}
 V_{i}\index{$V_{i}$} = \Sigma^{0}_{i}\qquad [\mbox{MeV}] 
\end{equation}
that is the time-component of the general four-vector self-energy
$\Sigma^{\mu}_{i}$.\index{$\Sigma^{\mu}_{i}$}

In certain parameter regions of an EoS, the phenomena of
superconductivity\index{superconductivity} 
or superfluidity\index{superfluidity} can be found. In this case,
it is worthwhile to know the size of the pairing gaps\index{pairing!gap}
\begin{equation}
 \Delta_{i}\index{$\Delta_{i}$} \qquad [\mbox{MeV}] \: ,
\end{equation}
e.g. in the $i=nn({}^{1}S_{0})$ channel in case of 
neutron pairing\index{pairing!neutron}. In the general case, it can be
energy and momentum dependent and cannot be easily tabulated in the present
format. In many cases, it is given for specific conditions defined in the 
EoS data sheet.

\section{Transport properties}

Besides the thermodynamic quantities and coefficients as described in sections (\ref{sec:thpot}) and (\ref{sec:thermo_coeff}), there are transport coefficients that are required in astrophysical simulations of dynamically evolving systems
in non-ideal hydrodynamics that consider dissipative effects.
There is no general agreement on the choice of units for the transport coefficients. These have to be specified in the data sheet of each EoS. 

\subsection{Viscosities}
\label{sec:visc}

Corrections to the equations of ideal hydrodynamics \cite{Strickland:2014pga} can be introduced by considering the energy-momentum tensor (in relativistic notation neglecting heat flow)
\begin{equation}
    T^{\mu\nu} = T^{\mu\nu}_\mathrm{ideal} + \Pi^{\mu\nu} 
    \qquad [\mbox{MeV~fm${}^{-3}$}]
\end{equation}
with the ideal contribution
\begin{equation}
    T^{\mu\nu}_\mathrm{ideal} = (e+p)u^{\mu}u^{\nu} - p g^{\mu\nu}
    = e u^{\mu}u^{\nu} -p \Delta^{\mu\nu}
    \qquad [\mbox{MeV~fm${}^{-3}$}]
\end{equation}
where
\begin{equation}
    \Delta^{\mu\nu} = g^{\mu\nu} - u^{\mu}u^{\nu}
    \qquad [\mbox{dimensionless}]
\end{equation}
and the correction $\Pi^{\mu\nu}$. The ideal part
contains the energy density $e$, the pressure $p$, the four-velocity $u^{\mu}$ and the metric tensor $g^{\mu\nu}$. The correction can be written in lowest order as
\begin{equation}
    \Pi^{\mu\nu} = \pi^{\mu\nu} - \Phi \Delta^{\mu\nu}
    \qquad [\mbox{MeV~fm${}^{-3}$}]
\end{equation}
with a symmetric traceless contribution $\pi^{\mu\nu}$ and a traceful part. These can be expressed in first order as
\begin{equation}
 \pi^{\mu\nu}  =  \eta \: \left(
  \nabla^{\mu} u^{\nu} + \nabla^{\nu} u^{\mu}
 - \frac{2}{3} \Delta^{\mu\nu} \nabla_{\alpha} u^{\alpha}
 \right)
 \qquad [\mbox{MeV~fm${}^{-3}$}]
\end{equation}
and
\begin{equation} 
 \Phi = \zeta \: \nabla_{\alpha} u^{\alpha} 
 \qquad [\mbox{MeV~fm${}^{-3}$}] 
\end{equation}
with the coefficients $\eta$ [MeV~fm${}^{-2}$] and $\zeta$ [MeV~fm${}^{-2}$] that are called the shear and bulk viscosity, respectively. 

\subsection{Conductivities}
\label{sec:cond}

The thermal conductivity $\kappa$ [fm${}^{-2}$] relates the heat flux density,
i.e., the transported heat energy per area and time,
\begin{equation}
 \vec{q} = - \kappa \: \vec{\nabla} T \qquad [\mbox{MeV~fm${}^{-3}$}]
\end{equation}
to the gradient of the temperature $T$.
The electrical conductivity $\sigma$ [fm${}^{-1}$] connects the electrical current density
\begin{equation}
    \vec{\jmath} = \sigma \: \vec{E} \qquad [\mbox{$e$~fm${}^{-3}$}]
\end{equation}
and the electrical field strength $\vec{E}=-\vec{\nabla} \varphi$ [$e$~fm${}^{-2}$] with the electric potential $\varphi$ [$e$~fm${}^{-1}$] ($e$ is the elementary charge).
In general, the thermal and electrical conductivities can be tensors if the direction of the induced currents is not in the direction of the field gradients.

\section{Contribution of photons}
\label{sec:photons}

If photons\index{photon} are considered in an EoS, they are treated as a gas of
massless bosons with simple analytical expressions for the relevant basic
thermodynamic quantities. The contribution of the photon can be
simply added to the
thermodynamic quantities of the other constituents. We have
the photon free energy density
\begin{equation}
 f_{\gamma}(T)\index{$f_{\gamma}$} 
 = -\frac{\pi^{2}}{45} T^{4} \qquad [\mbox{MeV~fm}^{-3}]
\end{equation}
the photon entropy density
\begin{equation}
 s_{\gamma}(T)\index{$s_{\gamma}$} 
 = \frac{4\pi^{2}}{45} T^{3} \qquad [\mbox{fm}^{-3}]
\end{equation}
and the photon pressure
\begin{equation}
 p_{\gamma}(T)\index{$p_{\gamma}$} 
 = \frac{\pi^{2}}{45} T^{4} \qquad [\mbox{MeV~fm}^{-3}]
\end{equation}
depending only on the temperature.  The chemical potential of the photon
vanishes
\begin{equation}
 \mu_{\gamma}\index{$\mu_{\gamma}$} = 0 \qquad [\mbox{MeV}]
\end{equation}
since it is its own antiparticle.
From these quantities, all other relevant photon thermodynamic
quantities can be derived.

\section{Properties of compact stars}

If an EoS model can provide the thermodynamic properties of charge-neutral matter in $\beta$ equilibrium at zero temperature, e.g., 
a  compact-star EoS, it is possible to calculate 
the radius\index{radius!neutron star} $R$\index{$R$}, 
mass\index{mass!neutron star} $M$\index{$M$}, 
and tidal deformability\index{tidal deformability} $\Lambda$\index{$\Lambda$} of a sequence of compact stars. For non-rotating, spherically symmetric stars, the mass-radius relation is obtained by solving the 
Tolman-Oppenheimer-Volkoff (TOV)\index{TOV} equation
\cite{Tolman:1939,Oppenheimer:1939}
\begin{equation}
\label{eq:TOV}
    \frac{dp}{dr} = - G \frac{M(r)e(r)}{
    r^{2}}
    \left[1 + \frac{p(r)}{e(r)}\right]
    \left[1 + \frac{4\pi r^{3} p(r)}{M(r)
    } \right]
    \left[ 1 - \frac{2GM(r)}{
    r}\right]^{-1}
\end{equation}
with the pressure $p$ and the mass inside the radius $r$ [km]
\begin{equation}
 M(r) = 4\pi 
 \int_{0}^{r} dr^{\prime} \:
 \left( r^{\prime}\right)^{2} \: e(r^{\prime})
 \qquad [\mbox{M}_{\odot}]
\end{equation}
that depends on the (internal) energy density $e$. 
Equation (\ref{eq:TOV}) is a first-order differential equation that can be solved by assuming a particular central density $n_{c}$\index{density!central}\index{$n_c$} with corresponding pressure and energy density at the center of the star and then integrating outward until the pressure vanishes (or is sufficiently small) which defines the surface of the star at a radius $R$ [km]. 
Thus the radius\index{radius!compact star} $R$\index{$R$} and 
mass $M(R)$\index{mass!compact star} of a particular compact star are obtained. 
Varying the central density $n_{c}$, the mass-radius relation of a sequence of stars is found.

A compact star will change its shape
under the influence of an external
gravitational field. The major effect will be a quadrupole deformation
that can be quantified with the 
tidal deformability\index{tidal deformability}
\cite{Hinderer:2007mb,Hinderer:2009ca}.
This quantity is defined as
\begin{equation}
    \Lambda\index{$\Lambda$} = \frac{2}{3} k_{2} 
    C^{-5} 
    \qquad [\mbox{dimensionless}]
\end{equation}
with the compactness
\begin{equation}
    C = \frac{GM(R)}{
    R} \qquad [\mbox{dimensionless}]
\end{equation}
and the $l=2$ Love number $k_{2}$. It can be found as
\begin{eqnarray}
 k_{2} & = & \frac{8}{5} C^{5} \left( 1 - 2 C \right)^{2} 
 \left[ 2 + 2 C \left( y-1\right) -y\right]
 \\ \nonumber & & \times
 \left\{ 2C \left[ 6 - 3 y + 3 C \left( 5 y - 8 \right)\right]
 + 4 C^{3} \left[ 13 - 11 y + C \left( 3 y - 2 \right) +
 2 C^{2} \left( 1 + y \right) \right] 
 \right. \\ \nonumber  & & \left.
  + 3 \left( 1 - 2 C \right)^{2} 
 \left[ 2 - y + 2 C \left( y - 1 \right) \right]
 \ln \left( 1 - 2 C \right)
 \right\}^{-1} \qquad [\mbox{dimensionless}]
\end{eqnarray}
when the logarithmic derivative
\begin{equation}
    y = \frac{d \ln H}{d \ln R} \: , \qquad [\mbox{dimensionless}]
\end{equation}
of a function $H(R)$ is known for a compact star configuration with radius $R$. The quantity $H$ for a particular star is obtained by solving the
second-order differential equation
\begin{eqnarray}
 0 & = & \frac{d^{2}H}{dr^{2}} + \frac{dH}{dr}
 \left( \frac{2}{r} + \frac{d\Phi}{dr} - \frac{d\lambda}{dr} \right)
 \\ \nonumber & &
  + H \left[ - \frac{6}{r^{2}} e^{2\lambda}
  - 2 \left( \frac{d\Phi}{dr} \right)^{2}
  + 2 \frac{d\Phi^{2}}{dr^{2}}
  + \frac{3}{r} \frac{d\lambda}{dr}
  + \frac{7}{r} \frac{d\Phi}{dr}
  + 2 \frac{d\Phi}{dr}\frac{d\lambda}{dr}
  + \frac{f}{r} \left( \frac{d\Phi}{dr} + \frac{d\lambda}{dr}\right)
  \right]
\end{eqnarray}
simultaneously with (\ref{eq:TOV}) for $dp/dr$ using 
the potential
\begin{equation}
    e^{2\lambda}  = \left[ 1 -  \frac{2GM(r)}{
 r}\right]^{-1} 
 \qquad [\mbox{dimensionless}] 
\end{equation}
of the Schwarzschild metric and 
the auxiliary quantities
\begin{eqnarray}
 \Phi & = & - \frac{1}{\varepsilon(r)+p(r)} \frac{dp}{dr} 
 \qquad [\mbox{km}^{-1}] \:, 
\\ 
    f & = & \frac{de}{dp} = 
    1/c_s^2\qquad [\mbox{dimensionless}]
\end{eqnarray}
that are obtained from the compact-star EoS. 
The boundary condition for $H(r)$
is given by the condition that $H(r) \to a_{0}r^{2}$ for $r \to 0$ with an arbitrary constant $a_{0}$ which has no influence on the value of the tidal deformability. 

Publicly available numerical codes exist to solve Einstein equations, which can in particular be used to obtain stationary models of rotating axisymmetric stars for a given EoS, see, e.g., the \textsc{RNS}\index{RNS} code \cite{Stergioulas:1994ea,Nozawa:1998ak,RNS} and the \textsc{LORENE}\index{LORENE} library \cite{LORENE}. The latter has an interface to directly employ EoS data from CompOSE, see 
subsection \ref{sec:direct_use}. It can compute initial data for binary neutron stars, too. 

Another interesting quantity for neutron stars with respect to cooling is the threshold density for the operation of the direct Urca process\index{Urca}. For a neutron-star EoS with only electrons as leptonic degrees of freedom, the threshold charge fraction is given by
\begin{equation}
    Y_{q}^{\mathrm{dUrca}} = \frac{1}{9}
\end{equation}
or
\begin{equation}
    Y_{q}^{\mathrm{dUrca}} = \left\{1+\left[1+\left(\frac{Y_{e}}{Y_{e}+Y_{\mu}}\right)^{1/3}\right]^{3} \right\}^{-1}
\end{equation}
if both electrons and muons are accounted for
\cite{Lattimer:1991ib,Alvarez-Castillo:2016yma}. The direct Urca process can only operate if $Y_{q}$ is larger than these threshold
values. Otherwise the neutrino generating direct Urca reactions such as for example
\begin{equation}
    n \to p + e^{-} + \bar{\nu}_{e}
\end{equation}
are kinematically blocked; energy and momentum conservation cannot be fulfilled simultaneously. From the solution of the TOV equation and the resulting NS masses as function on the central density -- together with the range of baryon densities allowing for direct Urca processes -- threshold NS masses for the processes can be obtained. Since in general $Y_{q}$ increases with $n_{b}$ in the core, this gives a lower limit on the NS mass for the direct Urca process 
to operate for a given EoS.

\part{For Contributors}
\label{part:contrib}
The success of the CompOSE data base depends on the support
of nuclear physicists providing tables with their favourite EoS.
Some well known EoS used extensively in astrophysical applications
are already incorporated in the CompOSE data base and are ready for use. 
However, a larger collection of EoS from different models
is highly desirable.
In order to be suitable for a simple usage, contributors should follow
the guidelines specified in this part of the manual.

\chapter{How to prepare EoS  tables}

\index{table!EoS}
In general, tables with EoS data contain a wealth of information on the
thermodynamic properties, the composition of dense matter and on the
microphysical properties of the constituents.
In order to minimize the memory size, only the essential
thermodynamic quantities should be stored in the tables that are
used in the CompOSE data base. These quantities are selected such that
they are sufficient to calculate a particular thermodynamic
potential, here the free energy\index{energy!free} $F$\index{$F$} [MeV], 
and its first derivatives with respect to the
parameters that correspond to the natural
variables\index{variable!natural} 
of $F$, i.e.\
temperature\index{temperature} $T$\index{$T$} [MeV], 
volume\index{volume} $V$\index{$V$} [fm${}^{3}$] 
and particle numbers\index{particle!number} 
$N_{i}$\index{$N_{i}$} [dimensionless].
However, it is worthwhile to have redundant information in order
to check the consistency of the EoS tables or to simplify the
determination of additional quantities. 
Further thermodynamic quantities relevant for astrophysical
applications can, if not directly available, be generated with the help of
thermodynamic identities and, for quantities depending on
higher derivatives, by using appropriate
interpolation\index{interpolation} schemes.
The organization of the data 
concerning composition etc.\ depends on the 
particular EoS. Thus it is required that up to three separate tables are
provided with unique formats\index{table!format} that  contain the 
thermodynamic quantities (required), the 
information on the composition of matter (depending
on the choice of the constituents of the model, optional) and the microscopic
properties of the particles (optional), respectively.
Details of the employed discretization\index{discretization} mesh of the
parameters have to be supplied in additional tables.

Additional information that is not contained in the Eos tables has to
be provided by the contributor. It will appear in the data
sheet\index{data sheet} that
accompanies each EoS and is available on the CompOSE web
site\index{web site}. Predictions for properties
of cold, spherical compact stars will be made available in a separate file if it is possible to calculate these quantities in the model. See the new ``quick guide for providers''\index{quick guide for providers} for a summary.

\section{Tabulation of quantities}

\subsection{Parameters and parameter ranges}
\label{ssec:para}

\index{parameter!range}
Because of the imposed physical conditions, see section \ref{sec:physcon}, 
the state of the system
is uniquely characterized by only three quantities,
see section \ref{sec:thpot}, that are used as parameters in the
EoS tables with the properties of dense matter:

\begin{enumerate}
\item \textbf{temperature} $T$\index{$T$} [MeV],
\item \textbf{baryon number density} $n_{b}$\index{$n_{b}$} [fm$^{-3}$],
\item \textbf{charge fraction of strongly interacting particles} 
$Y_{q}$\index{$Y_{q}$} [dimensionless].
\end{enumerate}

In order to be useful in many applications, the 
parameter ranges\index{parameter!range} should be chosen as wide as possible
and the mesh spacing as fine as possible, i.e.\ the number of
grid points as large as possible 
within reason. The recommended parameter\index{parameter!recommended} 
values for the definition of the grid are
given in the tables in the end of this chapter. In some cases it might be
useful to adapt the resolution\index{resolution} 
of the tables, in particular for the
baryon density, to the
physical situation and occurring phenomena. E.g.\ a rapid change
in some quantities or a phase transition\index{phase!transition} can occur.
For this purpose, several tables for different blocks in density
with different resolution
can be supplied for a single EoS model.

The recommended discretization\index{discretization} scheme 
for a general purpose EoS table depends on the parameters:
\begin{enumerate}
\item \textbf{temperature}\index{temperature}\index{$T$}\\
In this case, two standards are suggested:\\
a. linear mesh in $\ln T$, i.e.\
\begin{equation}
 T(i_{T}\index{$i_{T}$}) = T^{\textrm{ref}} (f_{T})^{i_{T}-1} \quad
 f_{T} = 10^{1/M_{T}} \quad 
 i_{T}=N_{T}^{\textrm{min}},N_{T}^{\textrm{min}}+1, \dots, N_{T}^{\textrm{max}} 
\end{equation}
with parameters $T^{\textrm{ref}}\index{$T^{\textrm{ref}}$} > 0$~MeV, 
$N_{T}^{\textrm{min}}\index{$N_{T}^{\textrm{min}}$} <
N_{T}^{\textrm{max}}\index{$N_{T}^{\textrm{max}}$}$
and $M_{T}\index{$M_{T}$}>1$ points per decade in temperature;\\
b. a mesh that is linear in $T$ at low temperatures and
linear in $\ln T$ at high temperatures, i.e.
\begin{equation}
 T(i_{T}) = T^{\textrm{ref}} \frac{\sinh (f_{T} i_{T})}{\sinh (f_{T})} \quad
 i_{T}=N_{T}^{\textrm{min}},N_{T}^{\textrm{min}}+1, \dots, N_{T}^{\textrm{max}} 
\end{equation}
 with parameters
$T^{\textrm{ref}}>0$~MeV, 
$N_{T}^{\textrm{min}} < N_{T}^{\textrm{max}}$ and $f_{T}>0$. 
The resolution\index{resolution} 
parameter $M_{T}>0$ determines $f_{T}\index{$f_{T}$} = \ln (10)/M_{T}$;\\
In both cases the index $i_{T}=0$ is reserved for $T=0$~MeV;
\item \textbf{baryon number density}\index{density!number!baryon}\index{$n_{b}$}\\
  linear mesh in $\ln n_{b}$, i.e.\
  \begin{equation}
    n_{b}(i_{n_{b}}\index{$i_{n_{b}}$}) = n_{b}^{\textrm{ref}} (f_{n_{b}})^{i_{n_{b}}-1} \quad
    f_{n_{b}} = 10^{1/M_{n_{b}}} \quad
    i_{n_{b}} =N_{n_{b}}^{\textrm{min}},N_{n_{b}}^{\textrm{min}}+1, \dots,
    N_{n_{b}}^{\textrm{max}} 
  \end{equation} 
  with parameters $n_{b}^{\textrm{ref}}\index{$n_{b}^{\textrm{ref}}$}>0$~fm${}^{-3}$, 
  $N_{n_{b}}^{\textrm{min}}\index{$N_{n_{b}}^{\textrm{min}}$} < 
    N_{n_{b}}^{\textrm{max}}\index{$N_{n_{b}}^{\textrm{max}}$}$
  and $M_{n_{b}}\index{$M_{n_{b}}$}>1$ 
  points per decade in the baryon number density;
\item \textbf{charge fraction of strongly interacting particles}
  \index{fraction!charge}\index{$Y_{q}$}\\
  linear mesh in $Y_{q}$, i.e.\
  \begin{equation}
    Y_{q}(i_{Y_{q}}\index{$i_{Y_{q}}$}) = \frac{i_{Y_{q}}}{M_{Y_{q}}} \quad
    i_{Y_{q}}=N_{Y_{q}}^{\textrm{min}},N_{Y_{q}}^{\textrm{min}}+1, \dots,
    N_{Y_{q}}^{\textrm{max}} 
  \end{equation} 
  with
  parameters 
  $N_{Y_{q}}^{\textrm{min}}\index{$N_{Y_{q}}^{\textrm{min}}$} 
    < N_{Y_{q}}^{\textrm{max}}\index{$N_{Y_{q}}^{\textrm{max}}$}$
  and $M_{Y_{q}}\index{$M_{Y_{q}}$}>N_{Y_{q}}^{\textrm{max}}$
  with a resolution in $Y_{q}$ of
  $1/M_{Y_{q}}$.
\end{enumerate}
Each grid point\index{grid point} in the table is identified with the triple
$(i_{T},i_{n_{b}},i_{Y_{q}})$ of indices.
The recommended values for the parameters are given in section
\ref{ssec:tot}.
Actually used meshes 
for particular EoS are specified
in the online documentation for each individual EoS table.

\subsection{Thermodynamic consistency}

The equations of state included in the CompOSE data base
are required to fullfill some basic
thermodynamic consistency\index{consistency!thermodynamic}
relations. 
Of course, the consistency can only hold up to some numerical level.
Measures of the thermodynamic consistency are included in the
characterisation of the EoS tables on the web site. 

Because the free energy density is considered in the present case
as the basic thermodynamic potential,
the homogeneity\index{homogeneity} condition, cf.\ eq. (\ref{eq:xix}),
\begin{equation}
\label{eq:f_con}
 f(T,n_{b},Y_{q})\index{$f$} = 
 -p+\mu_{b} n_{b}+\mu_{q} \left( n_{q}-n_{le} - n_{l\mu}\right)  
 +\mu_{le} n_{le} +\mu_{l\mu} n_{l\mu} 
\end{equation}
with the pressure $p$, the chemical potentials $\mu_{b}$,
$\mu_{q}$, 
$\mu_{le}$, $\mu_{l\mu}$ and corresponding densities
$n_{b}$, $n_{q}$, 
$n_{le}$, $n_{l\mu}$ should hold.
Note that 
$n_{q}$ contains only the  contribution
of strongly interacting particles
to the charge density, that 
a fixed relation between $\mu_{le}$ and $\mu_{l\mu}$ is
assumed and that $n_{le}+n_{l\mu}=n_{q} = Y_{q}n_{b}$ due to local
charge neutrality 
if leptons are considered in the equation
of state. Additionally, in eq.\ (\ref{eq:f_con}) it was assumed
that the strangeness chemical potential $\mu_{s}$ is
zero.

It is convenient to define the effective lepton
density\index{density!lepton!effective} as 
\begin{equation}
 n_{l}\index{$n_{l}$} = n_{le} + n_{l\mu} \qquad [\mbox{fm}^{-3}]
\end{equation}
and the effective lepton chemical 
potential\index{chemical potential!lepton!effective}
\begin{equation}
 \mu_{l}\index{$\mu_{l}$} = \frac{\mu_{le} n_{le} +\mu_{l\mu}
   n_{l\mu}}{n_{l}} \qquad [\mbox{MeV}]
\end{equation}
such that
\begin{equation}
 f(T,n_{b},Y_{q})\index{$f$} =
  -p + \left( \mu_{b} + Y_{q} \mu_{q} \right) n_{b}
\end{equation}
in case of an EoS without leptons and
\begin{equation}
 f(T,n_{b},Y_{q})\index{$f$} =
  -p + \left( \mu_{b} + Y_{l} \mu_{l} \right) n_{b}
\end{equation}
in the case with leptons and charge neutrality\index{charge!neutrality}
($Y_{l} =  Y_{q}$), respectively. 
The tabulated values for 
the entropy density, pressure and chemical potentials
should be given by the first partial derivatives as
\begin{eqnarray}
 s\index{$s$} & = &
 - \left. \frac{\partial f}{\partial T} \right|_{n_{b},Y_{q}} \: ,
 \\
 p\index{$p$} & = &
 n_{b}^{2} \left. \frac{\partial (f/n_{b}) }{\partial n_{b}}
  \right|_{T,Y_{q}} \: ,
 \\
  \mu_{b}\index{$\mu_{b}$} 
  + Y_{i} \mu_{i} 
  & = &
 \left. \frac{\partial f}{\partial n_{b}} \right|_{T,Y_{q}} \: ,
 \\ 
   \mu_{i}  & = & \frac{1}{n_{b}}
 \left. \frac{\partial f}{\partial Y_{i}} \right|_{T,n_{b}} 
\end{eqnarray}
where $i=l(q)$ if leptons are (not) included in the EoS. 
These relations 
will be used to derive the first derivatives of $f$ from
the tabulated thermodynamic quantities $s$, $p$, $\mu_{b}$,
$\mu_{q}$, and $\mu_{l}$ 
in some particular interpolation schemes.

For a mixed second partial derivative, 
the Maxwell relations\index{relation!Maxwell} are
\begin{eqnarray}
 - n_{b}^{2} \left. \frac{\partial (s/n_{b})}{\partial n_{b}}
 \right|_{T,Y_{q}} 
 & = & \left. \frac{\partial p}{\partial T} \right|_{n_{b},Y_{q}} \: ,
 \\
 - \left. \frac{\partial s}{\partial n_{x}} \right|_{T,n_{y},y \neq x}
 & = &  \left. \frac{\partial \mu_{x}}{\partial T} \right|_{n_{y}} \: ,
 \\
  \left. \frac{\partial p}{\partial n_{x}} \right|_{T,n_{y},y \neq x}
 & = &   n_{b} \left. \frac{\partial \mu_{x}}{\partial n_{b}}
 \right|_{T,n_{y}, y \neq b} \: , 
  \\
  \left. \frac{\partial \mu_{x}}{\partial n_{y}} \right|_{T,n_{z},z \neq y}
 & = &  \left. \frac{\partial \mu_{y}}{\partial n_{x}}
 \right|_{T,n_{z}, z \neq x}
\end{eqnarray}
where $x,y,z \in \left\{ b,q,le,l\mu \right\}$ should hold.

The knowlegde of three functions $p=p(T,n_{b},Y_{q})$,
$\mu_{b}\index{$\mu_{b}$}=\mu_{b}(T,n_{b},Y_{q})$, and
$\mu_{q}\index{$\mu_{q}$}=\mu_{q}(T,n_{b},Y_{q})$ or
$\mu_{l}\index{$\mu_{l}$}=\mu_{l}(T,n_{b},Y_{q})$
is sufficient to recover the free energy density
$f\index{$f$}=f(T,n_{b},Y_{q})$
for the particular physical conditions of section \ref{sec:physcon}
($\mu_{s} = 0$ and $n_{q}=n_{l}$).
Further quantities can be derived by
partial derivatives with respect to the parameters
$T$, $n_{b}$ and $Y_{q}$.

Thus it would be convenient
to store only these three quantities in the EoS table\index{table!EoS}
for the thermodynamic properties in order to reduce the memory size.
However, we require that the pressure, the entropy, the three chemical potentials 
and also the free and internal energy densities should be
provided independently in the EoS tables for checking purposes. 
It is also allowed to
store more than these seven basic quantities. 

\section{Structure of tables}
\label{sec:tab_structure}

The CompOSE data base contains for each EoS model data tables
that specify the used \emph{parameter grid} and tables that contain
the \emph{thermodynamic properties, the compositional information}
and \emph{the microscopic information}, respectively. Properties of
a sequence of cold, spherical \emph{compact stars} will be collected in a separate file if caclculable in the EoS model.

\subsection{Types of EoS tables}
\label{ssec:tot}

In general, three types of EoS input tables\index{table!input} 
should be available for each EoS
model in order to address different applications:

\begin{enumerate}

\item \textbf{Three-dimensional tables}
\index{equation of state!three-dimensional}

These tables depend on all three indepedent parameters $T$, $n_{b}$
and $Y_{q}$. There is only one case:

\begin{itemize}
\item \textbf{General purpose EoS table}
\index{equation of state!general purpose}

The recommended parameter\index{parameter!recommended} 
values for the definition of the grid are
given in table \ref{tab:eos_GP}.
This choice of discretation 
corresponds to a table with 
$N_{T}^{\textrm{max}} \times N_{n_{b}}^{\textrm{max}} 
\times N_{Y_{q}}^{\textrm{max}} = 1119720 (1462860) $ data points for
the case b(a) in the $T$ grid.
The table includes neither points with $T=0$~MeV nor
with $Y_{q}=0$.

\end{itemize}

\item \textbf{Two-dimensional tables}
\index{equation of state!two-dimensional}

These tables depend on two of the three indepedent parameters $T$, $n_{b}$
and $Y_{q}$. Four cases are offered here:

\begin{itemize}

\item \textbf{Zero-temperature EoS table}
\index{equation of state!zero temperature}

The recommended parameter\index{parameter!recommended}
values for the definition of the grid are
given in table \ref{tab:eos_T0}.
This choice of discretization 
corresponds to a table with 
$N_{n_{b}}^{\textrm{max}} \times (N_{Y_{q}}^{\textrm{max}}+1) = 18361$ data points.
Only $T=0$~MeV is considered. Data points for 
$Y_{q}=0$, i.e.\ pure neutron matter,
are included in this EoS table.

\item \textbf{Symmetric matter EoS table}
\index{equation of state!symmetric matter}

The recommended parameter\index{parameter!recommended}
values for the definition of the grid are
given in table \ref{tab:eos_NM}.
This choice of discretization 
corresponds to a table with 
$N_{T}^{\textrm{max}} \times N_{n_{b}}^{\textrm{max}} = 18963 (24381) $ data points
for the case b(a) in the $T$ grid.
Only $Y_{q}=0.5$ is considered. $T=0$~MeV, i.e.\ the 
zero-temperature case, is (not) included in this EoS table
for case b(a).

\item \textbf{Neutron matter EoS table}
\index{equation of state!neutron matter}

The recommended parameter\index{parameter!recommended}
values for the definition of the grid are
given in table \ref{tab:eos_NM}.
This choice of discretization 
corresponds to a table with 
$N_{T}^{\textrm{max}} \times N_{n_{b}}^{\textrm{max}} = 18963 (24381) $ data points
for case b(a).
Only $Y_{q}=0$ is considered. $T=0$~MeV, i.e.\ the 
zero-temperature case, is (not) included in this EoS table
for case b(a).

\item \textbf{EoS table of $\beta$-equilibrated matter}
\index{equation of state!$\beta$ equilibrium}

The recommended parameter\index{parameter!recommended}
values for the definition of the grid are
given in table \ref{tab:eos_NM}.
This choice of discretization 
corresponds to a table with 
$N_{T}^{\textrm{max}} \times N_{n_{b}}^{\textrm{max}} = 18963 (24381) $ data points
for the case b(a) in the $T$ grid.
The charge fraction of strongly interacting particles $Y_{q}$ is determined by the
conditions of charge neutrality and $\beta$ equilibrium with leptons. 
$T=0$~MeV, i.e.\ the 
zero-temperature case, is (not) included in this EoS table for
case b(a).

\end{itemize}

\item \textbf{One-dimensional tables}
\index{equation of state!one-dimensional}

These tables depend only on the parameter $n_{b}$. 
Three cases are offered here:

\begin{itemize}

\item \textbf{EoS table of cold symmetric matter}
\index{equation of state!symmetric matter}

The recommended parameter\index{parameter!recommended}
values for the definition of the grid are
given in table \ref{tab:eos_1dim}.
This choice of discretization 
corresponds to a table with 
$N_{n_{b}}^{\textrm{max}} = 301$ data points.
Only $Y_{q}=0.5$ and $T=0$~MeV is considered. 

\item \textbf{EoS table of cold neutron matter}
\index{equation of state!neutron matter}

The recommended parameter\index{parameter!recommended}
values for the definition of the grid are
given in table \ref{tab:eos_1dim}.
This choice of discretization 
corresponds to a table with 
$N_{n_{b}}^{\textrm{max}} = 301$ data points.
Only $Y_{q}=0.0$ and $T=0$ is considered. 

\item \textbf{EoS table of cold $\beta$-equilibrated matter}
\index{equation of state!$\beta$ equilibrium}

The recommended parameter\index{parameter!recommended}
values for the definition of the grid are
given in table \ref{tab:eos_1dim}.
This choice of discretization 
corresponds to a table with 
$N_{n_{b}}^{\textrm{max}} = 301$ data points.
Only $T=0$~MeV is considered and the charge fraction 
of strongly interacting particles $Y_{q}$
is determined by the conditions of $\beta$-equilibrium and local
charge neutrality with leptons.

\end{itemize}

\end{enumerate}

Every set of EoS tables\index{table!EoS} contains up to three tables: the first
that specifies the
thermodynamic\index{table!thermodynamic} 
state of the system (always required), the second that
contains information on the composition\index{table!composition} of the matter
(optional), the third that gives microscopic 
information\index{table!microscopic} (optional).

The format\index{table!format} of the thermodynamic, the
compositional
and the microscopic
table, respectively, has to follow the following prescriptions.

\subsection{Table with thermodynamic quantities (required)}
\label{ssec:tabthermoquant}
\index{table!thermodynamic}

The first row of this table contains 
three entries: the mass of the neutron $m_{n}$\index{$m_{n}$} 
and that of the proton
$m_{p}$\index{$m_{p}$} in MeV of the particular EoS model and then an integer
$I_{l}$\index{$I_{l}$} that
indicates whether the EoS contains leptons or not. If electrons (and/or
muons) are considered in the EoS, one sets $I_{l}=1$. For other values
of $I_{l}$, it is assumed that there are no electrons (and/or muons)
and the lepton chemical potential $\mu_{l}$ should be set to zero.
The neutron mass $m_{n}$ is used for scaling
certain quantities as specified below.
The following rows in this table contain information
on the particular grid point in the table and the relevant
thermodynamic quantities.
The actually tabulated quantities are given by
\begin{enumerate}
\item \textbf{pressure divided by the baryon number 
density $Q_{1}= p/n_{b}$} [MeV], 
\item \textbf{entropy per baryon (or entropy density per baryon density) $Q_{2} = s/n_{b}$} [dimensionless],
\item \textbf{scaled and shifted baryon chemical potential
    $Q_{3} =\mu_{b}/m_{n}-1$} [dimensionless],
\item \textbf{scaled charge chemical potential
    $Q_{4} = \mu_{q}/m_{n}$} [dimensionless],
\item \textbf{scaled effective lepton chemical potential
    $Q_{5} = \mu_{l}/m_{n}$} [dimensionless],
\item \textbf{scaled free energy per baryon
    $Q_{6} = f/(n_{b}m_{n})-1$} [dimensionless],
\item \textbf{scaled internal energy per baryon
    $Q_{7} = e/(n_{b}m_{n})-1$} [dimensionless].
\end{enumerate}
For an ideal gas $Q_{1}$\index{$Q_{i}$} is just the temperature $T$.
The traditional scaling of the chemical potentials and the energies
per baryon
with the neutron mass $m_{n}$\index{$m_{n}$} has been introduced. 

Each line in the thermodynamic table 
(except the first) has to contain the following quantities
\begin{equation}
\label{eq:eos.thermo}
 i_{T}\index{$i_{T}$} \:\: i_{n_{b}}\index{$i_{n_{b}}$} 
 \:\: i_{Y_{q}}\index{$i_{Y_{q}}$} \:\:
 Q_{1} \:\: Q_{2} \:\: Q_{3} \:\: Q_{4} \:\: Q_{5} \:\: Q_{6} \:\:
 Q_{7} \:\:
 N_{\textrm{add}} \:\: 
 \underbrace{q_{1} \:\: 
   q_{2} \:\: \dots
 \:\:  q_{N_{\textrm{add}}}}_{N_{\textrm{add}} \: {\textrm{quantities}}}
\end{equation}
with seven required quantities $Q_{i}$ and 
$N_{\textrm{add}}$\index{$N_{\textrm{add}}$} additional
optional thermodynamic quantities $q_{i}$\index{$q_{i}$}. 
The meaning of the additional quantities\index{quantity!additional}
is specified in the data
sheet\index{data sheet} for each EoS table.
It is assumed that the order of all quantities $q_{i}$ does not change
within an EoS table.
Since the triple of table indices 
$(i_{T},i_{n_{b}},i_{Y_{q}})$
is explicitly specified, the order of tabulation\index{table!order} with
respect to the grid points is not relevant. If the table contains
rows with identical indices $i_{T}$, $i_{n_{b}}$, $i_{Y_{q}}$ then
table entries for the quantities $Q_{i}$, $N_{\textrm{add}}$, 
and $q_{i}$ are used from the row that is read last.

\subsection{Table with the composition of matter (optional)}
\label{ssec:compo}
\index{table!composition}

The predicted chemical composition of matter depends strongly
on the employed theoretical model. In principle, the
particle content and hence the data that are required to be
stored can depend on the particular point in the parameter space, too.
Thus the format of the table
has to be adapted to this situation. The essential information is
contained in the particle fractions\index{particle!fraction} 
$Y_{i}$\index{$Y_{i}$} but in some cases
additional information, e.g.\ on the type of phase\index{phase},
might be useful. 

The tabulated quantities in the EoS composition table
are typically given by
\begin{enumerate}
\item \textbf{particle fractions} 
  $Y_{i}$\index{$Y_{i}$} [dimensionless],
\item \textbf{average mass number of a representative nucleus} 
  $A^{\textrm{av}}$\index{$A^{\textrm{av}}$} [dimensionless],
\item \textbf{average charge number of a representative nucleus} 
  $Z^{\textrm{av}}$\index{$Z^{\textrm{av}}$} [dimensionless],
\item \textbf{index encoding the type of phase} 
  $I_{\textrm{phase}}$\index{$I_{\textrm{phase}}$} [dimensionless]\\
(The correspondence of the phase index $I_{\textrm{phase}}$ with the 
actual structure of the phase is defined for each EoS model
on the corresponding web page.).
\end{enumerate}


In the following a standard format for the tabulation of compositional
information is defined. It is suitable
for most of the presently available EoS tables of hadronic and quark 
EoS models.
It allows to change 
from point to point the particle fractions
that are given in the table. Thus, it is possible
to select only those particles that are most abundant under the
considered local conditions. Each row of the table  contains the
following entries
\begin{eqnarray}
  \lefteqn{i_{T} \:\: i_{n_{b}} \:\: i_{Y_{q}} \:\: 
 I_{\textrm{phase}} \:\: N_{\textrm{pairs}}
 \underbrace{I_{1} \:\: Y_{I_{1}} \:\: \dots \:\: 
 I_{N_{\textrm{pairs}}} \:\: Y_{I_{N_{\textrm{pairs}}}}}_{N_{\textrm{pairs}} \: {\textrm{pairs}}}}
 \\ \nonumber & &
 N_{\textrm{quad}} \:\:
 \underbrace{I_{1} \:\:  A^{\textrm{av}}_{I_{1}}\:\:
   Z^{\textrm{av}}_{I_{1}} \:\: Y_{I_{1}} \:\: \dots \:\: 
   I_{N_{\textrm{quad}}} \:\: A^{\textrm{av}}_{I_{N_{\textrm{quad}}}}\:\:
   Z^{\textrm{av}}_{I_{N_{\textrm{quad}}}} \:\: Y_{I_{N_{\textrm{quad}}}}}_{N_{\textrm{quad}} \: {\textrm{quadruples}}}
\end{eqnarray}
with $N_{\textrm{pairs}}$\index{$N_{\textrm{pair}}$} 
pairs that combine the particle index $I_{i}$\index{$I_{i}$}
as defined in tables \ref{tab:partindex_fermions_sep}, 
\ref{tab:partindex_fermions_net}, \ref{tab:partindex_bosons} and 
\ref{tab:corrindex} and the corresponding
particle fraction $Y_{i}$\index{$Y_{i}$}. 
(Note the definition of particle fractions in equation
\ref{eq:Ydef}.) In addition, there are $N_{\textrm{quad}}$\index{$N_{\textrm{quad}}$}
quadruples that contain an index $I_{i}$  that specifies a group of
nuclei $\mathcal{M}_{I_{i}}$ with average mass number 
\begin{equation}
 A^{\textrm{av}}_{I_{i}}\index{$A^{\textrm{av}}_{I_{i}}$} = \frac{\sum_{j\in \mathcal{M}_{I_{i}}} A_{j}
   Y_{j}}{\sum_{j\in \mathcal{M}_{I_{i}}} Y_{j}} \: , 
\end{equation}
average charge number $Z^{\textrm{av}}_{I_{i}}$\index{$Z^{\textrm{av}}_{I_{i}}$}
\begin{equation}
 Z^{\textrm{av}}_{I_{i}} = \frac{\sum_{j\in \mathcal{M}_{I_{i}}} Z_{j}
   Y_{j}}{\sum_{j\in \mathcal{M}_{I_{i}}} Y_{j}} \: , 
\end{equation}
and combined fraction
\begin{equation}
  Y_{I_{i}}\index{$Y_{I_{i}}$} = \frac{\sum_{j\in \mathcal{M}_{I_{i}}}
    A_{j}Y_{j}}{A^{\textrm{av}}_{I_{i}}}
 = \sum_{j\in \mathcal{M}_{I_{i}}} Y_{j} \: .
\end{equation} 
The average neutron number\index{$N^{\textrm{av}}_{I_{i}}$}
\begin{equation}
 N^{\textrm{av}}_{I_{i}} = \frac{\sum_{j\in \mathcal{M}_{I_{i}}} N_{j}
   Y_{j}}{\sum_{j\in \mathcal{M}_{I_{i}}} Y_{j}} \: , 
\end{equation}
is not stored because it is easily found from 
$N^{\textrm{av}}_{I_{i}} = A^{\textrm{av}}_{I_{i}} - Z^{\textrm{av}}_{I_{i}}$.
In case that there are no pairs (quadruples) to be stored, the number
$N_{\textrm{pairs}}$ ($N_{\textrm{quad}}$) is set to zero.
The average mass and charge numbers correspond to those
of a representative heavy nucleus if there is only one group of nuclei
except the lightest that are considered explicitly. 
However, it is also possible to define several subsets of nuclei
with corresponding average mass numbers, charge numbers and fractions.
The correlation between the index $I_{i}$ and the set of
nuclei $\mathcal{M}_{I_{i}}$
is defined for each EoS model separately and given in the
data sheet\index{data sheet} on the CompOSE web pages.
Since the composition of matter can vary rapidly with a change of the
parameters, it is not required that the fractions of all particles or
particle sets have to given in reach row. It is sufficient to list
only those that are relevant, e.g.\ with
nonzero $Y_{i}$. Others can be omitted.

\subsection{Table with information on microscopic quantities (optional)}
\label{ssec:micro}
\index{table!microscopic}

Besides the information on thermodynamic and compositional
quantities, most EoS models can provide additional information
on microscopic quantities\index{quantity!microscopic} 
of an individual particle $i$, e.g.\
\begin{enumerate}
\item \textbf{Landau effective mass divided by the particle mass} 
  $m^{L}_{i}/m_{i}$\index{$m^{L}_{i}$} [dimensionless],
\item \textbf{Dirac effective mass divided by the particle mass} 
  $m^{D}_{i}/m_{i}$\index{$m^{D}_{i}$} [dimensionless],
\item \textbf{non-relativistic single-particle potential} 
  $U_{i}$\index{$U_{i}$} [MeV],
\item \textbf{relativistic vector self-energy} 
  $V_{i}$\index{$V_{i}$} [MeV],
\item \textbf{relativistic scalar self-energy} 
  $S_{i}$\index{$S_{i}$} [MeV],
\item \textbf{pairing gap}
  $\Delta_{i}$\index{$\Delta_{i}$} [MeV].
\end{enumerate}
These quantities are stored in an additional table 
if available.
In this case, each line has the format
\begin{equation}
 i_{T}\index{$i_{T}$} \:\: i_{n_{b}}\index{$i_{n_{b}}$} 
 \:\: i_{Y_{q}}\index{$i_{Y_{q}}$} \:\:
 N_{\textrm{qty}} \:\: 
 \underbrace{K_{1} \:\: q_{K_{1}} \:\: K_{2} \:\:
   q_{K_{2}} \:\: \dots
 \:\: K_{N_{\textrm{qty}}} \:\: q_{K_{N_{\textrm{qty}}}}}_{N_{\textrm{qty}} \: {\textrm{pairs}}}
\end{equation}
where $N_{\textrm{qty}}$\index{$N_{\textrm{qty}}$} 
is the number of stored quantities $q_{K_{i}}$.
The composite index\index{index!composite} 
$K_{i}$\index{$K_{i}$} identifies uniquely both the particle or
correlation and the physical quantity. It is formed as
\begin{equation}
 K_{i} = 1000 \: I_{i} + J_{i}
\end{equation}
with the particle or correlation
index\index{index!particle}\index{index!correlation} 
$I_{i}$\index{$I_{i}$} from tables
\ref{tab:partindex_fermions_sep}, \ref{tab:partindex_fermions_net},
\ref{tab:partindex_bosons} or \ref{tab:corrindex}, respectively, and
the quantity index $J_{i}$\index{$J_{i}$} from table \ref{tab:ident_micro}.

\subsection{Table with information on transport quantities (optional)}
\label{ssec:transport}
\index{table!transport}

The structure of this table will be defined after the \texttt{CompOSE} code has been 
updated to handle transport quantities.

\subsection{Table with information on properties of compact start (optional)}
\label{ssec:mass-radius}
\index{table!mass-radius}

If an EoS model can predict the properties of cold, spherical compact stars, e.g., the mass-radius relation and corresponding tidal deformabilities, these data will be provided in a separate file. The first line of this file identifies the given quantities and their units, followed by their actual values in the subsequent lines. The exact structure depends on the available data. Most relevant are
\begin{itemize}
    \item \textbf{radius}\index{radius!compact star} $R$\index{$R$} [km],
    \item \textbf{mass}\index{mass!compact star} $M$\index{$M$} [M${}_{\odot}$],
    \item \textbf{tidal deformability}\index{tidal deformability} $\Lambda$\index{$\Lambda$} [dimensionless],
    \item \textbf{central density}\index{density!central} $n_{c}$\index{$n_{c}$} [fm${}^{-3}$].
\end{itemize}
It is required that the radius $R$ and the mass $M$ using the units
km and solar masses, M${}_{\odot}$,
are given in the first and second column of the file, respectively.

\subsection{Tables with parameters}
\label{ssec:paratables}
\index{table!parameter}

The mesh points for the general purpose EoS 
are stored in three individual data files that are
used as an input in addition to the three tables containing
the thermodynamic, compositional and microphysical information.
It is assumed that the discretization scheme 
for the relevant parameters in the zero-temperature
and in the neutron-matter table is identical 
to the grid for the general purpose table.
Parameter values should be given with at least eight significant
digits. For the parameter temperature, the corresponding
table contains a single
entry in each row. In the first row, 
$N_{T}^{\textrm{min}}$\index{$N_{T}^{\textrm{min}}$} is given,
and $N_{T}^{\textrm{max}}$\index{$N_{T}^{\textrm{max}}$} 
in the second row. Then the parameter values
$T(N_{T}^{\textrm{min}})$, $T(N_{T}^{\textrm{min}}+1)$, \dots, $T(N_{T}^{\textrm{max}})$, 
are given in every following row. In total, the table contains
$N_{T}^{\textrm{max}}-N_{T}^{\textrm{min}}+3$ rows. The same storage scheme
applies for the parameters $n_{b}$ and $Y_{q}$.

\subsection{Identification of tables}
\label{ssec:identtables}
\index{table!identification}

Each Eos table is identified by a unique name and a unique extension
that correspond to the model and type of table, respectively. 
Using the generic name \texttt{eos} for a particular EoS model, there are 
always three parameter tables
\begin{itemize}
 \item \texttt{eos.t}\index{eos.t} (required)
 \item \texttt{eos.nb}\index{eos.nb} (required)
 \item \texttt{eos.yq}\index{eos.yq} (required)
\end{itemize}
that define the discretization mesh for the three parameters
$T$, $n_{b}$ and $Y_{q}$ as discussed in the previous section.
All three tables are required even for EoS tables with less than three
dimensions.

The thermodynamic, compositional and microscopic quantities are
stored in three tables with the generic names
\begin{itemize}
 \item \texttt{eos.thermo}\index{eos.thermo} (required)
 \item \texttt{eos.compo}\index{eos.compo} (optional)
 \item \texttt{eos.micro}\index{eos.micro} (optional)
\end{itemize}
where only the first one is required for every EoS model since
information
on the composition or microscopic details are not always available. 

If the EoS model allows to calculate properties of cold, spherical compact stars, these data will be stored in a file with the name
\begin{itemize}
    \item \texttt{eos.mr}\index{eos.mr} (optional).
\end{itemize}

\subsection{Precision of tabulated quantities}

EoS quantities of a particular model have a limited
precision\index{precision} in practical calculations. Machine precision is hardly ever
reached. The format\index{table!format} 
of the tabulated data should reflect this fact and
only the significant number of digits should be given to avoid excessive
memory size. Usually eight significant digits are sufficient since
the error in the interpolation can be considerably larger.

\section{Data sheet}

Information on an EoS that is not provided directly in the tables but relevant
for the characterisation of the particular model can be found in a
data sheet\index{data sheet} that comes along with the tables on the
CompOSE web pages\index{web page}. Some quantities are extracted
automatically from the tabulated data, others need to be specified by
the contributor. E.g., characteristic nuclear matter parameters (see
section \ref{sec:nucmatpar}) are obtained by using the program
\texttt{compose.f90} (see section \ref{sec:data}) if an EoS table for pure
nuclear matter without electrons or muons is available. 
Similarly, information on the parameter grid, type and dimension of
the EoS table, considered particles and additionally stored quantities
are extracted by the program. The EoS of
stellar matter in 
$\beta$-equilibrium\index{equation of state!$\beta$ equilibrium} 
is generated by using
\texttt{compose.f90}\index{compose.f90} 
if the EoS table for matter with electrons or muons
is provided by the contributor. From this extracted EoS table,
characteristic properties of neutron stars, e.g.\ 
maximum masses\index{maximum mass} 
or the threshold of the direct Urca process\index{Urca},
are calculated and given in the data sheet. It also contains information
on the considered constituent particles in the model. The contributor
has to supply a short characterisation of the EoS model and relevant
references\index{references}. He/she also needs to define the meaning of the index
$I_{\textrm{phase}}$\index{$I_{\textrm{phase}}$} for the phases\index{phase} considered in the EoS.


\newpage

\begin{table}[t]
\begin{center}
\caption{\label{tab:eos_GP}%
Parameter values for the recommended\index{parameter!recommended} 
discretization in
$T$, $n_{b}$ and $Y_{q}$ for a general purpose EoS table.}
\begin{tabular}{lcccc}
 \toprule
 quantity & reference & minimum & maximum  & resolution \\
                      & value(s) & index & index & parameter \\
 \midrule
 $T$, case a & $T^{\textrm{ref}} = 0.1$~MeV & 
 $N_{T}^{\textrm{min}} = 1$ & $N_{T}^{\textrm{max}} = 81$ & $M_{T}
 = 25$ \\
 $T$, case b & $T^{\textrm{ref}} = 0.1$~MeV & 
 $N_{T}^{\textrm{min}} = 1$ & $N_{T}^{\textrm{max}} = 62$ & $M_{T}=25$ \\
 $n_{b}$ & $n_{b}^{\textrm{ref}} = 10^{-12}$~fm${}^{-3}$ & 
 $N_{n_{b}}^{\textrm{min}} = 1$ & $N_{n_{b}}^{\textrm{max}} = 301$ & $M_{n_{b}}
 = 25$ \\
 $Y_{q}$ & N/A &
 $N_{Y_{q}}^{\textrm{min}} = 1$ & $N_{Y_{q}}^{\textrm{max}} = 60$ & $M_{Y_{q}} = 100$ \\
 \bottomrule
\end{tabular}
\end{center}
\end{table}

\begin{table}[t]
\begin{center}
\caption{\label{tab:eos_T0}%
Parameter values for the recommended\index{parameter!recommended} 
discretization in
$n_{b}$ and $Y_{q}$ for a zero-temperature EoS table.}
\begin{tabular}{lcccc}
 \toprule
 quantity & reference & minimum & maximum  & resolution \\
                      & value & index & index & parameter \\
 \midrule
 $n_{b}$ & $n_{b}^{\textrm{ref}} = 10^{-12}$~fm${}^{-3}$ & 
 $N_{n_{b}}^{\textrm{min}} = 1$ & $N_{n_{b}}^{\textrm{max}} = 301$ & $M_{n_{b}}
 = 25$ \\
 $Y_{q}$ & N/A &
 $N_{Y_{q}}^{\textrm{min}} = 0$ & $N_{Y_{q}}^{\textrm{max}} = 60$ & $M_{Y_{q}} = 100$ \\
 \bottomrule
\end{tabular}
\end{center}
\end{table}

\begin{table}[t]
\begin{center}
\caption{\label{tab:eos_NM}%
Parameter values for the recommended\index{parameter!recommended} 
discretization in
$T$ and $n_{b}$ for EoS tables of symmetric matter, neutron matter or
$\beta$-equilibrated matter.}
\begin{tabular}{lcccc}
 \toprule
 quantity & reference & minimum & maximum  & resolution \\
                      & value & index & index & parameter \\
 \midrule
 $T$, case a & $T^{\textrm{ref}} = 0.1$~MeV & 
 $N_{T}^{\textrm{min}} = 1$ & $N_{T}^{\textrm{max}} = 81$ & $M_{T}
 = 25$ \\
 $T$, case b & $T^{\textrm{ref}} = 0.1$~MeV & 
 $N_{T}^{\textrm{min}} = 0$ & $N_{T}^{\textrm{max}} = 62$ & $M_{T} = 25$ \\
 $n_{b}$ & $n_{b}^{\textrm{ref}} = 10^{-12}$~fm${}^{-3}$ & 
 $N_{n_{b}}^{\textrm{min}} = 1$ & $N_{n_{b}}^{\textrm{max}} = 301$ & $M_{n_{b}}
 = 25$ \\
 \bottomrule
\end{tabular}
\end{center}
\end{table}

\begin{table}[t]
\begin{center}
\caption{\label{tab:eos_1dim}%
Parameter values for the recommended\index{parameter!recommended} 
discretization in
$n_{b}$ for one-dimensional EoS tables.}
\begin{tabular}{lcccc}
 \toprule
 quantity & reference & minimum & maximum  & resolution \\
                      & value & index & index & parameter \\
 \midrule
 $n_{b}$ & $n_{b}^{\textrm{ref}} = 10^{-12}$~fm${}^{-3}$ & 
 $N_{n_{b}}^{\textrm{min}} = 1$ & $N_{n_{b}}^{\textrm{max}} = 301$ & $M_{n_{b}}
 = 25$ \\
 \bottomrule
\end{tabular}
\end{center}
\end{table}

\chapter{Extending CompOSE}
\label{ch:extensions}

CompOSE is meant to be under constant development.
The main aim is to enlarge the data base by adding
EoS tables of more and more models. 
The most simple way is to convert tables of existing model
calculations to the generic CompOSE scheme of tabulation in
the data files.
A more involved task is to develop a new EoS model and to provide the
results in the appropriate format. In either case, we ask
you to contact the CompOSE core team by sending an email to
\begin{quote}
\texttt{develop.compose@obspm.fr}. 
\end{quote}
All questions will be clarified by them
on how your results can be incorporated in the CompOSE data base
and made accessible to the public.

For the future, it is foreseen that the functionality of the CompOSE
database will be extended in order to make even more results
of EoS model calculations available that are not included in
the present sets of data. It is also envisioned to add more quantities
to the data tables that might be relevant for astrophysical
applications and beyond, in particular transport quantities.

\part{For Users}
\label{part:user}

\chapter{Models for the equation of state}
\label{ch:eosmodels}
There is a large number of model equations of state. They fall into
different classes 
depending on the theoretical approach.
The main difficulty in generating an equation of state is the
description of the strongly interacting constituents, hadrons and/or
quarks. See reference \cite{Oertel:2016bki} for a in-depth review.
The EoS tables on the CompOSE website 
are not classified according to the model but to 
their structure of tabulation and possible application.

Details and characteristic parameters of each particular EoS
can be found in a data sheet that accompanies the data tables
on the CompOSE web site 
\begin{quote}
 \url{https://compose.obspm.fr}.
\end{quote}

For electrons\index{electron} and muons\index{muon}, 
a uniform Fermi gas\index{Fermi gas} model is
employed in many cases. This assumption simplifies the calculation significantly
in case of spatially inhomogeneous charge distributions.
Details on the treatment of electrons and muons 
are given for each EoS separately on the data sheet.

At finite temperatures, photons\index{photon} contribute to the thermodynamic
properties of the system. Their treatment is discussed in section
\ref{sec:photons}. For each EoS in the database, a remark is given
whether photons are considered or not.
 
The definition of commonly used
nuclear matter parameters\index{parameter!nuclear matter} 
is given in section \ref{sec:nucmatpar}.
They characterize the essential
properties of nuclear matter and can serve
in a first comparison of the various models.
Be, however,
careful since these parameters are only valid in the vicinity of
saturation for symmetric, i.e.\ same number of protons and neutrons, matter.
Specific values of these parameters are given in the data sheet
for the individual models and in a summary table on the CompOSE website.
Collections of nuclear matter parameters are available
for widely used Skyrme Hartree-Fock models \cite{Dutra:2012mb}
and  for nonrelativistic Hartree-Fock models
with Gogny-type interactions \cite{Sellahewa:2014nia}. 
Results for relativistic mean-field models
can be found in \cite{Dutra:2014qga}. See the review \cite{Oertel:2016bki} for
recent experimental constraints.

In the end of this section, further remarks of caution are given.

\section{Nuclear matter properties}
\label{sec:nucmatpar}
Realistic models of nuclear, i.e.\ purely hadronic,
matter\index{matter!hadronic} 
without leptons show
some general features that are related to the occurence of the
saturation\index{saturation} 
phenomenon. Due to the symmetries of the strong interaction,
uniform nuclear matter\index{matter!uniform} 
at zero temperature reaches a state with the
largest binding energy per nucleon
at a finite saturation density\index{density!saturation} 
$n_{b}^{\textrm{sat}}$\index{$n_{b}^{\textrm{sat}}$} with equal concentrations
of neutrons and protons\footnote{This is exactly true only if the mass
  difference between neutrons and protons is neglected.}. 
For $T=0$~MeV, the energy per nucleon\index{energy!per nucleon} 
$\mathcal{E}$\index{$\mathcal{E}$}
can be considerd as a function of the baryon number density $n_{b}$
and the asymmetry\index{asymmetry}
\begin{equation}
 \alpha\index{$\alpha$} = \frac{n_{n}-n_{p}}{n_{n}+n_{p}} = 1-2Y_{q} \qquad
 [\mbox{dimensionless}]
\end{equation}
that vanishes for symmetric nuclear matter, i.e.\  $\alpha = 0$ and
$Y_{q} = 0.5$. It can be
expanded in a power series around the saturation point as
\begin{eqnarray}
\label{eq:Eseries}
 \mathcal{E}(n_{b},\alpha) & = &
 -B_{\textrm{sat}} + \frac{1}{2} K x^{2} + \frac{1}{6}
 Q 
 x^{3} + \dots 
 \\ \nonumber & & 
 + \alpha^{2} \left( J +  L x + \frac{1}{2} K_{\textrm{sym}}
   x^{2} + \dots \right) + \dots 
 \qquad [\mbox{MeV}]
\end{eqnarray}
with the relative deviation of the actual density from the saturation
density
\begin{equation}
\label{eq:def_x}
 x\index{$x$} = 
 \frac{n_{b}-n_{b}^{\textrm{sat}}}{3n_{b}^{\textrm{sat}}}
  \qquad [\mbox{dimensionless}]
\end{equation}
and the asymmetry $\alpha$. The coefficients 
$B_{\textrm{sat}}$\index{$B_{\textrm{sat}}$},
$K$\index{$K$}, $Q=-K^{\prime}$\index{$Q$}\index{$K^{\prime}$}, 
$J$\index{$J$}, $L$\index{$L$}, $K_{\textrm{sym}}$\index{$K_{\textrm{sym}}$} [MeV], 
\dots and $n_{b}^{\textrm{sat}}$\index{$n_{b}^{\textrm{sat}}$} [fm${}^{-3}$]
characterize the behaviour of the EoS.
They can, of course, only give an indication of the general features
of the EoS, since these
quantities are relevant essentially in the vicinity of the saturation
density\index{density!saturation} at zero
temperature and close to symmetric nuclear matter (equal number
of protons and neutrons).  Extrapolations based on this polynomial
expansion might be dangerous. 
The numerical factor in
equation (\ref{eq:def_x}) 
is of historical origin since 
the Fermi momentum\index{momentum!Fermi} 
$k_{F}$\index{$k_{F}$} [fm$^{-1}$] instead of the density
$n_{b} = 2k_{F}^{3}/(3\pi^{2})$ of symmetric nuclear matter at zero
temperature was originally
used as the expansion parameter. 
Note that contributions linear in
$\alpha$ and linear in $x$ for $\alpha=0$ are absent in the expansion
due to the condition of minimum energy per baryon.

In order to study the dependence on the asymmetry $\alpha$, it is also
convenient to introduce the symmetry energy\index{energy!symmetry}
\begin{equation}
 E_{\textrm{sym}}\index{$E_{\textrm{sym}}$}(n_{b}) = \left.
 \frac{1}{2} \frac{\partial^{2} \mathcal{E}(n_{b},\alpha)}{\partial
   \alpha^{2}} \right|_{\alpha = 0} \qquad [\mbox{MeV}]
\end{equation}
that is a function of the baryon number density $n_{b}$ only.
In many cases the quadratic approximation\index{approximation!quadratic} 
in $\alpha$ in equation 
(\ref{eq:Eseries}) is sufficient. Then the symmetry energy can also
be calculated from the finite difference approximation
\begin{equation}
 E_{\textrm{sym}}(n_{b}) = \frac{1}{2}
 \left[ \mathcal{E}(n_{b},-1)  - 2 \mathcal{E}(n_{b},0) 
 + \mathcal{E}(n_{b},1)\right]  \qquad [\mbox{MeV}]
\end{equation}
comparing symmetric matter with pure neutron and pure proton matter.

In the following, the coefficients appearing in equation (\ref{eq:Eseries})
are discussed in more detail.

\begin{itemize}
\item The \textbf{saturation density}\index{density!saturation} 
$n_{b}^{\textrm{sat}}$\index{$n_{b}^{\textrm{sat}}$} [fm${}^{-3}$]
  of symmetric nuclear
  matter is defined by the condition that the pressure\index{pressure} 
  vanishes, i.e.\
  \begin{equation}
    p\index{$p$} = \left. n_{b}^{2} \frac{d\mathcal{E}(n_{b},0)}{dn_{b}}
    \right|_{n_{b}=n_{b}^{\textrm{sat}}} = 0 \qquad [\mbox{MeV~fm}^{-3}]
  \end{equation}
  and the energy per baryon becomes minimal.
\item The \textbf{binding energy at saturation}\index{energy!binding} 
  $B_{\textrm{sat}}$\index{$B_{\textrm{sat}}$} [MeV]
  of symmetric nuclear matter can be obtained from Bethe-Weizs\"{a}cker mass
  formulas\index{mass formula!Bethe-Weizs\"{a}cker} 
  for nuclei ${}^{A}{Z}$ by an extrapolation of $A=2Z$
  to infinity.
\item The \textbf{incompressibility of bulk nuclear matter}\index{incompressibility}
  $K$\index{$K$} [MeV] quantifies the curvature\index{curvature} 
  of the binding energy per baryon
  with respect to the density variation at saturation
  since it is defined by
  \begin{equation}
    K\index{$K$} =  \left. 9 n_{b}^{2} \frac{\partial^{2} 
        \mathcal{E}(n_{b},0)}{\partial n_{b}^{2}} 
    \right|_{n_{b}=n_{b}^{\textrm{sat}}} 
    =  \left. 9 n_{b} \frac{\partial (p/n_b)}{\partial n_{b}}
\right|_{n_{b}=n_{b}^{\textrm{sat}}} 
\qquad [\mbox{MeV}] \: .
  \end{equation}
  It is related to the isothermal
  compressibility\index{compressibility!isothermal} 
  $\kappa_{T}$\index{$\kappa_{T}$} (\ref{eq:kappat}) at zero temperature, 
  zero asymmetry and saturation density by
  \begin{equation}
   K = \frac{9}{\kappa_{T} n_{b}^{\textrm{sat}}} \qquad [\mbox{MeV}] \: .
  \end{equation}
  Thus $K$ would be better called compression
  modulus\index{compression modulus}.
  It can be derived, e.g.\ from studies of 
  isoscalar giant monopole excitations of nuclei. 
  The extraction from data on isoscalar giant monopole 
  resonances\index{giant resonance!isoscalar monopole} 
  depends on the density dependence of the
  nuclear symmetry energy\index{energy!symmetry}, 
  a quantity under intensive debate in recent
  years. See also a recent comprehensive study of giant monopole resonances
    \cite{Stone:2014wza}. 
\item The \textbf{skewness coefficient of bulk nuclear matter}\index{skewness}
  $Q=-K^{\prime}$ [MeV] 
  is defined by the third derivative of the energy per baryon as
  \begin{equation}
    Q\index{$Q$}
    = 27 n_{b}^{3} \left. \frac{\partial^{3} 
        \mathcal{E}(n_{b},0)}{\partial n_{b}^{3}} \right|_{n_{b} =
      n_{b}^{\textrm{sat}}} 
 = 27 n_{b}^{2} \left. \frac{\partial^{2} 
        (p/n_{b})}{\partial n_{b}^{2}} \right|_{n_{b} = n_{b}^{\textrm{sat}}}  - 6 K
 \qquad [\mbox{MeV}] \: .
  \end{equation}
  The value of $Q$ in combination with $K$ determines the surface
  properties of nuclei, e.g.\ the ratio of the surface tension to the
  surface thickness. Thus, in order to fix $Q$
  in a particular model,
  it is important to include quantities like radii and the surface thickness
  in the procedure to fit the model parameters.
\item The \textbf{symmetry energy at saturation}\index{energy!symmetry} $J$ [MeV]
  is just given by
 \begin{equation}
   J\index{$J$} = E_{\textrm{sym}}(n_{b}^{\textrm{sat}}) \qquad [\mbox{MeV}].
 \end{equation}
  This coefficient mainly determines the isospin\index{isospin} 
  dependence of the 
  binding energy of nuclei. It is important in predicting
  masses of exotic nuclei far away from the valley of 
  stability\index{valley of stability}
  in the chart of nuclei.
\item The \textbf{symmetry energy slope
    coefficient}\index{energy!symmetry!slope} 
  $L$ [MeV] that
  is obtained from
  \begin{equation}
    L\index{$L$} = 3 n_{b} \left. \frac{dE_{\textrm{sym}}(n_{b})}{dn_{b}} \right|_{n_{b} =
      n_{b}^{\textrm{sat}}} \qquad [\mbox{MeV}].
  \end{equation}
  This coefficient is related to the density dependence of the neutron
  matter EoS near the saturation density. 
  It is strongly correlated with the neutron skin
  thickness\index{skin thickness!neutron} 
  of heavy nuclei like ${}^{208}$Pb. Unfortunately,
  experimentally determined values of the neutron skin thickness are not very
  precise so far and a large range of values for $L$ is found by comparing
  different EoS models.
\item The \textbf{symmetry incompressibility}\index{incompressibility!symmetry}
  $K_{\textrm{sym}}$ [MeV] 
  quantifies the curvature of the symmetry energy
  with respect to the density variation at saturation.
  It is defined by
  \begin{equation}
    K_{\textrm{sym}}\index{$K_{\textrm{sym}}$} = 9 n_{b}^{2} \left. \frac{d^{2} 
        E_{\textrm{sym}}}{d n_{b}^{2}} \right|_{n_{b} =
      n_{b}^{\textrm{sat}}} \qquad [\mbox{MeV}] \: .
  \end{equation}
\end{itemize}
For each EoS in the CompOSE data base, the actual values of the nuclear
matter coefficients are given if available.

\section{Remarks of caution}
\label{sec:remarks}
We provide equations of state tables
in the CompOSE data base that cover large ranges
in temperature, baryon number density and charge fraction of strongly
interacting particles.
However, we cannot guarantee that the physical description of 
matter under these conditions is close to reality. The user
has to judge whether a particular EoS is suitable for her or his
application.

The user also has to keep in mind that the thermodynamic and
compositional properties of nuclear matter\index{matter!nuclear} without electrons (and
muons) and dense stellar matter\index{matter!stellar} 
with electrons (and muons) are rather
different because in the former case the Coulomb interaction is
artificially neglected. This can have drastic consequences for
the occurence and type of phase transitions. Thus it is not
recommended to generate a dense matter EoS from a nuclear matter EoS
simply by adding the contribution from electrons (and muons).

In many cases it is desirable to extend a particular EoS to densities
and temperatures 
below or above the available or recommended
parameter ranges\index{parameter!range}. 
In these regimes, an
equation of state depending only on temperature, charge
fraction of strongly interacting particles and baryonic density to describe matter 
might not be sufficient, e.g.\ at very low densities and temperatures
where the time scales of 
thermodynamic equilibration\index{equilibrium!thermodynamic} 
reaction rates become large
and full reaction networks have to be considered. 
However, for many purposes a detailed description of matter
in these regime is not necessary, such that we decided to furnish
tables for these conditions, too. 

Our knowledge about the QCD\index{QCD} 
phase diagram\index{phase!diagram} suggests that there could be
a transition from a
hadronic phase\index{phase!hadronic} to a
quark-gluon plasma\index{quark-gluon plasma} 
within the range of densities and temperatures
reachable in core-collapse supernovae\index{supernova!core-collapse}, 
hence within the range of our
tables. Of course, there are lots of uncertainties about this phase
transition\index{phase!transition}, 
so that its occurence cannot be affirmed, but the
possibility has to be kept in mind when employing a purely hadronic
EoS\index{equation of state!hadronic}
up to densities well above nuclear matter 
saturation density\index{density!saturation} and
temperatures as high as several tens of MeV. Within CompOSE there will be
tables available including this phase transition.
Even without thinking about a
QCD phase transition, other forms of (exotic) matter shall appear at
high densities and temperatures. Already for cold matter\index{matter!cold} EoS
used for neutron star\index{star!neutron} 
modelisation for a long time, hyperons\index{hyperon}, pions\index{pion} and
kaons\index{kaon} 
have been considered. At temperatures above about 20~MeV, this point
becomes even more crucial. Thus using a purely nuclear EoS in this
regime can be a first approximation, technically very appealing, but
again it has to be kept in mind that probably the 
considered physics is too poorly described.

\chapter{Online services and data handling}
\label{ch:online-service}
\section{Access to CompOSE}
\label{sec:access}
The EoS data and routines
provided by CompOSE can be downloaded from the web site 
\begin{quote}
\texttt{https://compose.obspm.fr} \: .
\end{quote}
The tables and codes are offered free of charge but come without warranty.
Please acknowledge and give proper reference to CompOSE if you use our service
for your applications, presentations and publications (see appendix \ref{citation}). 
If you encounter any difficulties, it is possible to contact 
the CompOSE team by sending an email to
\begin{quote}
\texttt{develop.compose@obspm.fr} \: . 
\end{quote}

To subscribe to the CompOSE newsletter, distributed via the 
email list ``compose.info'', where changes and updates will be 
announced, send an e-mail with subject ``Subscribe'' 
to \texttt{develop.compose@obspm.fr}.

\section{EoS data sheets}

Each EoS of the CompOSE data base is accompanied with a data
sheet\index{data sheet}
available for download from the web site. The data sheet provides
essential information that allows the user to decide whether the EoS 
is suitable for her/his application. It contains information on
the origin and creation of the EoS table, a short abstract of the
physical model, references, the parameter ranges and considered
particle species, a summary of the available data files and 
fundamental quantities
that characterize the EoS, such as nuclear matter and
neutron star properties, if available.

\section{Options for using EoS data}
\label{sec:eos}
There are different options for
downloading and/or generating EoS data and 
tables\index{equation of state!table}. 
It is possible to obtain EoS tables
for different models and for different purposes.   
The original EoS tables are given as plain ASCII files that allow
every user to read the data without the need for further codes.
The general idea is that
a user chooses a particular model from those available on
the CompOSE web site\index{web site}.
Then, there are three different ways of accessing EoS data:
\begin{enumerate}
\item The user downloads data tables of an available EoS in the original
form and uses her or his own routines to handle the data.
For every model EoS 
there are three tables that contain the details of the discretization
of the parameters: \texttt{eos.t}\index{eos.t}, \texttt{ eos.nb}\index{eos.nb} 
and \texttt{eos.yq}\index{eos.yq}.
The original EoS data are stored as functions of 
temperature\index{temperature} $T$\index{$T$},
baryon number density\index{density!number!baryon}
$n_{b}$\index{$n_{b}$} 
and charge fraction of strongly interacting particles\index{fraction!charge} 
$Y_{q}$\index{$Y_{q}$}.
The last quantity is identical to the electron
fraction\index{fraction!electron} 
$Y_{e}$\index{$Y_{e}$} if electrons are
the only charged leptons considered in the EoS model.
The actual data on the thermodynamic,
compositional and microscopic properties are stored in the files
\texttt{eos.thermo}\index{eos.thermo}, 
\texttt{eos.compo}\index{eos.compo} and 
\texttt{eos.micro}\index{eos.micro}, respectively,
as far as available for that particular model. See section \ref{sec:tab_structure}
for the format of all the tables. If available for an EoS, the file 
\texttt{eos.mr} contains information of the properties of compact
stars and can be read directly without the need to use the CompOSE
software.
\item The user downloads the original data tables for a particular
EoS as described above and, in addition, 
several files 
that contain routines for reading, testing, interpolating and transforming
the data. These codes serve four major purposes:
\begin{itemize}
\item to interpolate the original EoS data tables in order to obtain
the quantities at parameter values different from the original tabulation,
\item to calculate additional quantities that are not given in the
original data files,
\item to select those quantities that are relevant for a particular
application and to store them in separate data files in a format
more convenient for the user,
\item to provide EoS data tables in the advanced 
 HDF5 format\index{format!HDF5} (see\\
\url{http://www.hdfgroup.org/HDF5/}) that is widely used in the
astrophysics community.
\end{itemize}
Details on how to work with the subroutines are described in 
the section~\ref{sec:data}.
\item Using the tools provided on the webpage to manipulate and visualise EoS data. In contrast to downloading data and the CompOSE software, the access to running the web tools is password restricted. Please contact \texttt{develop.compose(at)obspm.fr} if you wish an account to be created for you.
\end{enumerate}

\section{Different families of EoS data}
\label{sec:eosfamilies}

There is now a large number of different EoS models for which data
  is available on the web page for free download. In order to more easily find the data for a particular application, the different models have been ordered into different categories:
\begin{enumerate}
\item \textbf{Cold Neutron Star EoS}: this category contains
    models designed to describe cold (zero temperature) matter in
    weak ($\beta$) equilibrium. They all contain the contribution of
    electrons (and for some of them muons, too) and are directly
    applicable to construct cold neutron star models, {e.g.} by
    using the \textsc{eos\_compose} class with the \textsc{LORENE}
    library\footnote{\url{https://lorene.obspm.fr}}.
\item \textbf{Neutron Star Crust EoS}: category for EoS that describe the
    non-homogeneous matter of the outer crust and or inner crust,
    e.g. the Baym-Pethick-Sutherland (BPS) EoS.
    The set of crust EoS is divided
  into   \textbf{outer} crust and  \textbf{inner} crust and may be
  used, together with a core EoS, to construct a complete EoS. 
  Some EoS account for the appearance of pasta phases\index{pasta phase} in the transition region from the inner crust to the homogeneous core.
  In order for radii of low mass NS not to be affected by the crust EoS,
  the lowest considered density in the table should
  be sufficiently small, i.e., it has to reach small densities of the order
  of $10^{-9}$~fm$^{-3}$, where iron group nuclei appear in the outer crust. This is not the case for all models. If the model provides at least an inner crust, any outer crust model can be matched, the exact choice of the outer crust EoS does
  not much affect global compact star properties such as radius or mass. On the contrary, if the model only contains a core EoS, then some attention should be paid to the choice of the inner crust EoS for matching and the matching conditions, since in particular radii and tidal deformabilities can depend on that choice.
  The choice is optimal if the symmetry energy properties are
  similar to the ones of the core EoS.
  The crust thickness is sensitive to the crust-core transition density which in turn is sensitive to the medium modifications of the cluster properties and the treatment of the pasta phases.
  A  small routine that allows 
  for a  thermodynamically consistent matching of the crust EoS to
  the core EoS will be made available in the near future on the \texttt{CompOSE} web page, 
  e.g. it checks that $p_{\mathrm{crust}} = p_{\mathrm{
    core}}$ together with $\mu_{\mathrm{crust}} = \mu_{\mathrm{core}}$ at the transition.
\item \textbf{Cold Matter EoS}: this category contains models
    at zero temperature for different charge fractions. It is
    divided into different entrances according to the underlying
    physical approach.
\item \textbf{Neutron Matter EoS}: this category contains models
    for neutron matter, {i.e.} for $Y_q$ = 0 (the net
    abundances of charged particles and antiparticles is zero) at
    both $T=0$ and finite temperature.
\item \textbf{General Purpose EoS}: this category contains models which cover a large range of different temperatures, charge fractions and baryon number densities, as required in particular for simulations of core-collapse supernovae and binary neutron star mergers. For convenience, most models are provided in two versions, one with the contribution of electrons, positrons and photons included and one containing only the baryonic part.
\end{enumerate}
In addition to this very general distinction between EoS models,
  based mainly on distinguishing the different possible applications
  of the EoS data, there are sub-families.
These sub-families first describe the particular assumptions for the particle content of matter:
\begin{enumerate}
\item \textbf{Nucleonic models}, EoS models that assume matter to be composed of nucleons, charged leptons and potentially nuclear clusters,
\item \textbf{Hybrid (quark-hadron) models}, category for EoS models with a transition from hadronic matter at low densities to quark matter at high densities,
\item \textbf{Models with hyperons}, category for EoS models
    allowing for hyperons (strange baryons) to be populated,
\item \textbf{Models with hyperons and $\Delta$-resonances}, category for EoS models
    allowing for hyperons (strange baryons) and $\Delta$-resonances to be populated,
\item \textbf{Models with condensates}, category for EoS models
    allowing for kaon or pion condensates to set in,
    \item \textbf{Quark models}, category for EoS models containing quark matter.
  \end{enumerate}   
  Other sub-families might be added in the future.
  Then, there are 
  sub-families distinguishing a particular framework to determine the EoS (see
  the review \cite{Oertel:2016bki}).
This concerns first the description of homogeneous matter, i.e. the interaction of nucleons, hyperons etc, and the corresponding many-body treatment:
\begin{enumerate}
  \item \textbf{Non-relativistic density functional models}, 
  EoS, where the interaction of baryons is based on non-relativistic  
  density functional theory, in particular this concerns Skyrme\index{Skyrme} like interactions,
  e.g., SLy4, or Gogny-type\index{Gogny} interactions. 
  In these models,  the total energy density of homogeneous baryonic
  matter is given in terms of  the sum of 
a  kinetic and a potential energy density  functional in the space defined by the different particle species number densities: proton and neutron densities for nucleonic matter and proton, neutron and hyperons for hyperonic matter.
\item \textbf{Relativistic density functional models}, EoS based on
    covariant density functional theory, e.g. the STOS EoS (H Shen et
    al). These models are generally applied in the mean-field
    approximation. Baryons, nucleons and, when included, hyperons and
    or Delta-resonances, are described by fermionic fields that interact with
    purely phenomenological mesonic fields introduced in order to
    describe the nuclear interaction. The quantum numbers of the
    interaction channel give the name to the mesons. The most frequently used are
    the scalar and vector-isoscalar mesons named, respectively,
    $\sigma$ and $\omega$-mesons, and the vector-isovector
    $\rho$-meson. Some models also include the scalar-isovector
    $\delta$-meson, and when hyperons are included also the scalar and
    vector-isoscalar  $\sigma^*$ and $\phi$-mesons with hidden
    strangeness ($\bar s s$ mesons) may be included. These models are divided into two categories: models with constant couplings and non-linear mesonic terms (TM1 and SFHo are two examples) and models with density dependent couplings and only quadratic mesonic terms (such as DD2 and DDME2). 
    If the
    actually used values of the couplings differ from those of the original publications, these new values are provided in the data sheet.
\item \textbf{Microscopic calculations}, EoS based on ab initio
    calculations of nuclear matter, {e.g.} the BHF technique or
    variational calculations or Monte Carlo methods. Within these approaches different  many-body methods are used to include correlations in the many-body system starting from a few-body interaction that has been fitted to  observables in nucleon-nucleon scattering in vacuum and properties of bound few-nucleon systems,
    \item \textbf{Holographic models}, EoS based on a holographic approach to determine homogeneous matter. 
\end{enumerate}
In a second step, different treatments of inhomogeneous matter are distinguished:
\begin{enumerate}
    \item \textbf{NSE} models, using a nuclear statistical equilibrium approach to describe inhomogeneous matter at nonzero temperature. These category concerns mainly the general purpose EoS models, e.g. the models by Hempel \& Schaffner-Bielich. Cold neutron star EoS obtained from the lowest temperature entry of a general purpose without a dedicated description of the crust are ranged in this category, too.
    \item \textbf{SNA} models, using the single nucleus approximation to describe inhomogeneous matter at nonzero temperature. These concern the general purpose EoS models, e.g. the famous EoS by Lattimer \& Swesty or Shen et al.
\item \textbf{Unified EoS} For ``Cold neutron star EoS'' this keyword 
  indicates models where the same nuclear interaction has been used to
  describe the neutron star crust and core, all other models have a
  crust model matched to a core EoS and enter the category
  \item\textbf{Non-unified models (crust-core matched)}, see the description of the
  particular EoS table for details.
  \item\textbf{Thomas-Fermi calculations} concern the NS crust models and indicate the method to determine the crust properties.
  \end{enumerate}

Let us mention that a very convenient way to find a particular table is the bibliography search, where a function exists to search for tables related to publications of one particular author (see section \ref{sec:web}).

\section{Handling and transforming EoS data}
\label{sec:data}

The handling of EoS data is considerably simplified by using the 
software provided on the CompOSE web page. In the most recent version 
of the \texttt{compose}\index{compose}
code, the generation of the necessary parameter files 
to define output quantities and format is guided via terminal input.
There are only few steps required in order to generate
customized data tables from the original sets of EoS tables.

All quantities describing the thermodynamic, compositional and
microscopic properties of the matter are found for arbitrary values of the
table parameters with an interpolation\index{interpolation} scheme
that is described in detail in appendix~\ref{sec:interpol}.
It is possible to create tables with different mesh settings and ranges.
Of course, the ranges should be chosen only within the ranges of 
the basis table for
each model. If they are outside these ranges, error
messages\index{error message} will indicate the problems.

\subsection{Downloading and compiling}

There are four \textsc{Fortran90}
files that are needed in every case to
create the executable \texttt{compose}\index{compose} program.
The file \texttt{composemodules.f90}\index{composemodules.f90}
contains all the 
necessary modules and the file \texttt{compose.f90}\index{compose.f90} 
all required 
(sub-)routines and functions in order to read, interpolate and 
write the EoS data in ASCII format. 
The file \texttt{out\_to\_json.f90}\index{out\_to\_json.f90}
is only relevant 
for the program version used on the CompOSE webpage.
All necessary files are most simply obtained by downloading the
file \texttt{code.zip}.

Using the present version
of the program \texttt{compose}\index{compose}, 
it is not necessary to provide input files beyond those that were 
downloaded from the CompOSE website. The input files that define 
the selected variables and
tabulation scheme are created automatically. Their structure is specified
in appendix \ref{app:files}.

If output of the EoS data 
in HDF5-format is needed, the 
files \texttt{hdf5compose.f90}\index{hdf5compose.f90} 
and \texttt{hdf5writecompose.c}\index{hdf5writecompose.c} are required
in addition to \texttt{compose.f90}\index{compose.f90}
and \texttt{composemodules.f90}\index{composemodules.f90}. 

All the files can be downloaded from the corresponding CompOSE 
web page\index{web page} and they 
have to be compiled with an 
appropriate \textsc{Fortran90}\index{compiler!Fortran90} 
or \textsc{C}\index{compiler!C} compiler.
The program was written using the GNU compilers\index{compiler!GNU} 
\texttt{gfortran} and \texttt{gcc}
and the use of these compilers is encouraged. A sample  
\texttt{Makefile}\index{Makefile} is available 
on the CompOSE web page. It contains in line 26 a switch
to select the compilation without (\texttt{HDF = 0}, default setting)
and with (\texttt{HDF = 1}) the option for the HDF5 output.
For using the HDF5 routines, 
the HDF5 library has to be installed. Compiling the
program files with the provided \texttt{Makefile} generates an
executable with the name \texttt{compose}.

\subsection{Direct use of \texttt{compose}}
\label{sec:direct_use}

In this subsection, the details are described how customized
EoS tables from the EoS models on the
CompOSE web site can be obtained by
running the program \texttt{compose}\index{compose}. 

The operation proceeds in the following steps:

\begin{enumerate}
\item\textbf{Selection of EoS}
\index{equation of state!selection}

In a first step the user selects the EoS that is appropriate for
her/his application from the list given on the CompOSE web pages.
Often, there are different types of tables available for a single
EoS model as described in subsection \ref{sec:tab_structure}.
The tabulated data can depend on one, two or
three of the independent parameters $T$, $n_{b}$ and $Y_{q}$.
There is an option to replace the temperature $T$ by the
entropy per baryon $\mathcal{S}$ as parameter when running the standard
version of \texttt{compose}\index{compose}.
There are at least four different files needed for the successful 
application of the program \texttt{compose}\index{compose}
as specified in subsection \ref{ssec:identtables}: three files
containing information on the parameter grid and at least one
file with the thermodynamic data.

The downloaded parameter and EoS files are always identified with 
the same name and a file name extension that specifies the type
of stored quantities. The program \texttt{compose}\index{compose}
uses these generic names for the input files. Thus the downloaded files
are called
``\texttt{eos.t}''\index{eos.t},
``\texttt{eos.nb}''\index{eos.nb},
``\texttt{eos.yq}''\index{eos.yq}, 
 ``\texttt{eos.thermo}''\index{eos.thermo},
 ``\texttt{eos.compo}''\index{eos.compo} (if available), 
``\texttt{eos.micro}''\index{eos.micro} (if available), and 
``\texttt{eos.init}''\index{eos.init} (if available).

For some EoS, additional files are available: \texttt{eos.mr}\index{eos.mr}, 
the properties of a cold $\beta$-equilibrated spherically symmetric compact star, i.e. mass and radius of a spherically symmetric star, together with optionally tidal deformability and corresponding central density, obtained with the particular EoS model; \texttt{eos.thermo.ns} \index{eos.thermo.ns} and \texttt{eos.nb.ns}\index{eos.nb.ns}, baryon number densities and thermodynamic quantities for cold $\beta$-equilibrated matter for direct use within \textsc{LORENE}\index{LORENE} \cite{LORENE} via the eoscompose class to compute, e.g., initial data for a binary neutron star. The latter files exist only for general purpose EoS tables and have been extracted from the lowest temperature entry of the corresponding table, i.e., in general for a nonzero (but very small) temperature. \texttt{eos.thermo.ns} contains as additional quantity the electron fraction $Y_e$ obtained in $\beta$-equilibrium and the enthalpy density.

Note that for some of the general purpose EoS tables, the information in the file \texttt{eos.thermo.ns} is not sufficient for building compact stars up to the maximum mass. The reason is that for densities exceeding a certain value, $Y_{q}$ corresponding to $\beta$-equilibrated matter becomes lower than the lowest value considered in the table, typically $Y_q = 0.01$ and thus at high densities no $\beta$-equilibrium solution is found. In this case, for some of these models, corresponding cold Neutron Star Matter tables are provided containing the complementary information.

If the user does not run the program \texttt{compose}\index{compose}
in the standard version (see \ref{item:running}) he or she
has to supply two files
(in ASCII format) in addition to the downloaded files.
They specify the parameter values and the quantities that will be
stored in the customized EoS output table.
See appendix \ref{app:files} for details of their structure.
The input files \texttt{eos.parameters}\index{eos.parameters} and 
\texttt{eos.quantities}\index{eos.quantities} are created automatically
when the standard version of the program  \texttt{compose}\index{compose}
is used. Thus it is recommended to run the standard version, in particular
for first-time users of CompOSE.

\item\textbf{Running \texttt{compose}}
\label{item:running}

The user has to run the \texttt{compose}\index{compose} code three times:
\begin{enumerate}
\item to select the output quantities
\item to define the tabulation parameters
\item to generate
the file \texttt{eos.table}\index{eos.table} with the customized EoS table. 
\end{enumerate}
In each case,
the program explicitly asks questions and gives information for possible
answers. The files \texttt{eos.quantities}\index{eos.quantities} and
\texttt{eos.parameters}\index{eos.parameters}
are created automatically. 
See the ``quick guide for users''\index{'quick guide for users} for examples.

In the case that there is no file \texttt{eos.init}\index{eos.init} provided
on the webpage for a particular EoS, it is created automatically.
During the execution of the \texttt{compose}\index{compose} program,
the relevant input data
files are read: first the files with the parameter grids
\texttt{eos.t}\index{eos.t}, \texttt{eos.nb}\index{eos.nb} 
and \texttt{eos.yq}\index{eos.yq}, second the EoS data files
\texttt{eos.thermo}\index{eos.thermo},
\texttt{eos.compo}\index{eos.compo} (if available), 
and \texttt{eos.micro}\index{eos.micro} (if available). 
The read data are analyzed, checked for
consistency and the information is stored in the file
\texttt{eos.init}\index{eos.init}. 

Running the \texttt{compose}\index{compose} code for task 1, a file
\texttt{eos.quantities}\index{eos.quantities}
 is created. When task 2 is selected, a file 
\texttt{eos.parameters}\index{eos.parameters}
is generated. The structure of the files \texttt{eos.init}\index{eos.init},
\texttt{eos.quantities}\index{eos.quantities}, and
\texttt{eos.parameters}\index{eos.parameters}
 is described in detail
in appendex \ref{app:files}.

Each row of the output file \texttt{eos.table} contains 
the three parameter values of the user-chosen grid
and then the selected quantities. Note that for each
quadruple index\index{index!quadruple} $I_{i}$ 
four quantities 
($Y^{\textrm{av}}_{I_{i}}$, $A^{\textrm{av}}_{I_{i}}$, 
$Z^{\textrm{av}}_{I_{i}}$, and $N^{\textrm{av}}_{I_{i}}$)
are given.

\end{enumerate}

\subsection{HDF5 table}
\index{table!HDF5}
The organization of the data in case of the HFD5 table is different
as compared to the ASCII table. All data are stored in a single
data file that is denominated \texttt{eoscompose.h5}\index{eoscompose.h5}.
The names of the data sets are given by the
identifiers\index{identifier!HDF5} in tables
\ref{tab:hdf5} and \ref{tab:hdf5micro}. Note, too, 
that the values of the different quantities at all grid points 
are stored together in one data set. The data sets can 
in addition contain a group of quantities, e.g., 
the $N_{\mathrm{thermo}}$ thermodynamic quantities, 
see tables~\ref{tab:hdf5} and \ref{tab:hdf5micro} for details. 

\subsection{Additional output}
\index{data sheet}

During the execution of task three of the program \texttt{compose}, a file
\texttt{eos.report}\index{eos.report}
is generated that contains information on the
EoS derived from the tabulated data. The information is used for the
creation of the data sheet\index{data sheet} that accompanies each EoS
on the CompOSE web pages\index{web page}. If possible, a second file
\texttt{eos.beta}\index{eos.beta} is produced with the EoS of
$\beta$-equilibrated
matter for the lowest available temperature for all grid points in baryonic
density that are available for the specific EoS and where a solution is found.
Each row consists of four quantities, the density $n_{b}$, the charge
fraction $Y_{q}$, the (free) energy density $f$ (including rest mass
contributions) and the pressure $p$.
These data are used to derive characteristic neutron star parameters
for the data sheet. Finally, a third file \texttt{eos.info.json}
is created that is used for the interactive web version of the program.

\subsection{Error messages}
\index{error message}

During the execution of the subroutines, errors can occur due to
several reasons, e.g.\ parameters out of range etc. In this case,
the execution is stopped and a corresponding error message is generated.

\begin{table}[htb]
\begin{center}
\caption{\label{tab:ident_thermo1}%
Thermodynamic quantities\index{quantity!thermodynamic} 
in the data tables with their units.}
{\small 
\begin{tabular}{cccc}
\toprule
index & quantity/  & unit & description \\ 
$J$& expression & & \\
\toprule
1 & $p$\index{$p$} & MeV~fm${}^{-3}$ & total pressure \\ 
\midrule
2 & $\mathcal{S}=s/n_{b}$\index{$\mathcal{S}$} 
& dimensionless & total entropy per baryon \\ 
\midrule
 3 & $\mu_{b}-m_{n}$\index{$\mu_{b}$} & MeV & baryon chemical potential \\ 
                    &     & & with respect to neutron mass \\
\midrule
4 & $\mu_{q}$\index{$\mu_{q}$} & MeV & 
 {\small charge chemical potential} \\ 
\midrule
5 & $\mu_{l}$\index{$\mu_{l}$} & MeV & lepton chemical potential \\ 
\midrule
6 & $\mathcal{F}/m_{n} -1$\index{$\mathcal{F}$} & dimensionless 
       & scaled   \\  
        & & &  free energy per baryon\\ 
\midrule
7 & $\varepsilon =$\index{$\varepsilon$} & dimensionless 
       & scaled  \\  
       &  $\mathcal{E}/m_{n} -1$\index{$\mathcal{E}$} 
& & internal energy per baryon   \\ 
\midrule
8 & $\mathcal{H}/m_{n}-1$\index{$\mathcal{F}$} & dimensionless 
       & scaled  \\
       & & & enthalpy per baryon \\ 
\midrule
9 & $\mathcal{G}/m_{n}-1$\index{$\mathcal{G}$} & dimensionless 
       & scaled  \\
        & & & free enthalpy per baryon \\ 
\midrule
10 & $\left. \frac{\partial p}{\partial n_{b}} \right|_{\mathcal{E}}$ & MeV 
       & {\small partial derivative of pressure with}\\ 
        & & & {\small respect to baryon number density}\\ 
\midrule
11 & $\left. \frac{\partial p}{\partial
    \mathcal{E}}\right|_{n_{b}}$ & fm$^{-3}$
       & {\small partial derivative of pressure with}\\ 
        & & & {\small respect to internal 
                energy per baryon} \\
\midrule
12 & $c_{s}^{2}$\index{$c_{s}$} & dimensionless 
    & speed of sound squared \\ 
\midrule
13 & $c_{V}$\index{$c_{V}$} & dimensionless & specific heat capacity \\
            &               & & at constant volume \\ 
\midrule
14 & $c_{p}$\index{$c_{p}$} & dimensionless & specific heat capacity \\
            &               & & at constant pressure \\ 
\midrule
15 & $\Gamma = c_{p}/c_{V}$\index{$\Gamma$} & dimensionless & adiabatic index \\
\midrule
16 & $\alpha_{p}$\index{$\alpha_{p}$} & MeV${}^{-1}$ & expansion coefficient \\
                    &              & & at constant pressure \\
\midrule
17 & $\beta_{V}$\index{$\beta_{V}$} & fm${}^{-3}$ & tension coefficient \\
                   &            & & at constant volume \\ 
\midrule
18 & $\kappa_{T}$\index{$\kappa_{T}$} 
& fm${}^{3}$/MeV & isothermal compressibility\\ 
\midrule
19 & $\kappa_{S}$\index{$\kappa_{S}$} 
& fm${}^{3}$/MeV & adiabatic compressibility \\ 
\bottomrule
\end{tabular}
}
\end{center}
\end{table}

\begin{table}[htb]
\begin{center}
\caption{\label{tab:ident_thermo2}%
Additional thermodynamic quantities\index{quantity!thermodynamic} 
in the data tables with their units. (Quantities 25 - 28 are not yet available with the
\texttt{CompOSE} code.)}
{\small 
\begin{tabular}{cccc}
\toprule
index & quantity/  & unit & description \\ 
$J$& expression & & \\
\toprule
20 & $\mathcal{F}$\index{$\mathcal{F}$} 
& MeV & free energy per baryon \\ 
\midrule
21 & $\mathcal{E}$\index{$\mathcal{E}$} 
& MeV & energy$^{\ast}$ per baryon \\ 
\midrule
22 & $\mathcal{H}$\index{$\mathcal{H}$} 
& MeV & enthalpy per baryon \\ 
\midrule
23 & $\mathcal{G}$\index{$\mathcal{G}$} 
& MeV & free enthalpy per baryon \\ 
\midrule 
24 & $e=\mathcal{E}n_{b}$\index{$e$}
& MeV~fm${}^{-3}$ & energy density \\
\midrule
25 & $\zeta$\index{$\zeta$}
& MeV~fm${}^{-2}$ & bulk viscosity \\
\midrule
26 & $\eta$\index{$\eta$}
& MeV~fm${}^{-2}$ & shear viscosity \\
\midrule
27 & $\kappa$\index{$\kappa$}
& fm${}^{-2}$ & thermal conductivity \\
\midrule
28 & $\sigma$\index{$\sigma$}
& fm${}^{-1}$ & electrical conductivity \\
\bottomrule
\end{tabular}
\\[1ex]
${}^{\ast}$ This corresponds
to the "internal energy" in the thermodynamic language\\
including the rest mass contribution.
}
\end{center}
\end{table}

\begin{table}[th]
\begin{center}
\caption{\label{tab:f_derivatives}%
Values and derivatives of the free energy\index{energy!free}
per baryon which are stored in the data tables
with their units.}
{\small
\begin{tabular}{cccc}
\toprule
index $J$ & quantity/ & unit & description \\ 
 & expression & & \\
 \toprule
 1 & $\mathcal{F}$\index{$\mathcal{F}$} 
 & [MeV] & free energy per baryon \\ 
 \midrule
 2 & $\frac{\partial \mathcal{F}}{\partial T}$\index{$\mathcal{F}$} 
 & dimensionless & first temperature derivative\\
 & & & of free energy per baryon \\ 
 \midrule
 3 & $\frac{\partial^{2} \mathcal{F}}{\partial T^{2}}$\index{$\frac{\partial^{2} \mathcal{F}}{\partial T^{2}}$} 
 & [MeV$^{-1}$] & second temperature derivative\\
 & & & of free energy per baryon \\ 
 \midrule
 4 & $\frac{\partial^{2} \mathcal{F}}{\partial T \partial n_{b}}$\index{$\frac{\partial^{2} \mathcal{F}}{\partial T \partial n_{b}}$} 
 & [fm$^{3}$] & mixed second derivative\\
 & & & of free energy per baryon \\ 
 \midrule
 5 & $\frac{\partial^{2} \mathcal{F}}{\partial T \partial Y_{q}}$\index{$\frac{\partial^{2} \mathcal{F}}{\partial T \partial Y_{q}}$} 
 & dimensionless & mixed second derivative\\
 & & & of free energy per baryon \\ 
 \midrule
 6 & $\frac{\partial \mathcal{F}}{\partial n_{b}}$\index{$\frac{\partial \mathcal{F}}{\partial n_{b}}$} 
 & [MeV~fm$^{3}$] & first baryon density derivative\\
 & & & of free energy per baryon \\ 
 \midrule
 7 & $\frac{\partial^{2} \mathcal{F}}{\partial n_{b}^{2}}$\index{$\frac{\partial^{2} \mathcal{F}}{\partial n_{b}^{2}}$} 
 & [MeV~fm$^{6}$] & second baryon density derivative\\
 & & & of free energy per baryon \\ 
 \midrule
 8 & $\frac{\partial^{2} \mathcal{F}}{\partial n_{b} \partial Y_{q}}$\index{$\frac{\partial^{2} \mathcal{F}}{\partial n_{b} \partial Y_{q}}$} 
 & [MeV~fm$^{3}$] & mixed second derivative\\
 & & & of free energy per baryon \\ 
 \midrule
 9 & $\frac{\partial \mathcal{F}}{\partial Y_{q}}$\index{$\frac{\partial \mathcal{F}}{\partial Y_{q}}$} 
 & [MeV] & first hadronic charge fraction derivative\\
 & & & of free energy per baryon \\ 
 \midrule
 10 & $\frac{\partial^{2} \mathcal{F}}{\partial Y_{q}^{2}}$\index{$\frac{\partial^{2} \mathcal{F}}{\partial Y_{q}^{2}}$} 
 & [MeV] & second hadronic charge fraction derivative\\
 & & & of free energy per baryon \\ 
\bottomrule
\end{tabular}
}
\end{center}
\end{table}


\begin{table}[th]
\begin{center}
\caption{\label{tab:ident_compo}%
Quantities containing information on the composition\index{quantity!compositional}
which are stored in the data tables
with their units.
The particle index $I_{i}$\index{$I_{i}$} has been
defined in tables \ref{tab:partindex_fermions_sep},
\ref{tab:partindex_fermions_net}, \ref{tab:partindex_bosons}
and \ref{tab:corrindex}. 
The symbols \% denotes the group of nuclei.} 
{\small
\begin{tabular}{cccc}
\toprule
index $J$ & quantity/  & unit & description \\ 
 & expression  & & \\
\toprule
 & $I_{\textrm{phase}}$\index{$I_{\textrm{phase}}$} 
& dimensionless & phase index \\
\midrule
 & $Y_{I_{i}}$\index{$Y_{i}$} & dimensionless & fraction of particle $I_{i}$ \\
\midrule
$1$ & $Y^{\textrm{av}}_{\%}$\index{$X^{\textrm{av}}$} 
& dimensionless & combined fraction  \\
   & & & of group \% of nuclei \\
\midrule
$2$ & $A^{\textrm{av}}_{\%}$\index{$A^{\textrm{av}}$} 
& dimensionless & average mass number  \\
   & & & of group \% of nuclei \\
\midrule
$3$ & $Z^{\textrm{av}}_{\%}$\index{$Z^{\textrm{av}}$} 
& dimensionless & average charge number \\
  & & & of group \% of nuclei \\
\midrule
$4$ & $N^{\textrm{av}}_{\%}$\index{$N^{\textrm{av}}$}
& dimensionless & average neutron number \\
  & & & of group \% of nuclei \\
\bottomrule
\end{tabular}
}
\end{center}
\end{table}

\begin{table}[htb]
\begin{center}
\caption{\label{tab:ident_micro}%
Microscopic quantities\index{quantity!microscopic}
which are stored in the data tables
with their unit.}
{\small
\begin{tabular}{cccc}
\toprule
index $J$ & quantity/ & unit & description \\ 
 & expression & & \\
\toprule
40 & $m^{L}_{I_{i}}/m_{I_{i}}$\index{$m^{L}_{I_{i}}$} 
& dimensionless & effective Landau mass \\ 
 & & & with respect to \\
 & & & the particle mass \\
\midrule
41 & $m^{D}_{I_{i}}/m_{I_{i}}$\index{$m^{D}_{I_{i}}$}
& dimensionless & effective Dirac mass \\ 
 & & & with respect to \\
 & & & the particle mass \\
\midrule
50 & $U_{I_{i}}$\index{$U_{I_{i}}$} & MeV & nonrelativistic  \\ 
 & & & single-particle potential \\
\midrule
51 & $V_{I_{i}}$\index{$V_{I_{i}}$} & MeV & relativistic  \\ 
  & & & vector self-energy \\
\midrule
52 & $S_{I_{i}}$\index{$S_{I_{i}}$} & MeV & relativistic  \\ 
  & & & scalar self-energy \\
\midrule
60 & $\Delta_{I_{i}}$\index{$\Delta_{I_{i}}$} & MeV & gap \\
\bottomrule
\end{tabular}
}
\end{center}
\end{table}

\begin{table}[htb]
\begin{center}
\caption{\label{tab:ident_err}%
Error quantities which are stored in the data tables
with their unit.}
{\small
\begin{tabular}{cccc}
\toprule
index $J$ & quantity/  & unit & description \\ 
 & expression & & \\
\toprule
1 & $\Delta \mathcal{F}$\index{$\Delta \mathcal{F}$} 
& MeV & absolute error estimate \\ 
   & & & of free energy per baryon \\
\midrule
 2 & $\Delta \mathcal{F}/\mathcal{F}$ & [dimensionless]
 & relative error estimate \\ 
  & & & of free energy per baryon \\
\midrule
 3 & $\Delta \mathcal{E}$\index{$\Delta \mathcal{E}$} 
& MeV & absolute error estimate \\ 
  & & & of internal energy per baryon \\
\midrule
 4 & $\Delta \mathcal{E}/\mathcal{E}$ & [dimensionless]
 & relative error estimate \\ 
  & & & of internal energy per baryon \\
\midrule
 5 & $\Delta \left(\frac{p}{n_{b}}\right)$ & MeV 
 & absolute error estimate \\ 
  & & & of pressure-to-density ratio \\
\midrule
 6 & $\Delta \left(\frac{p}{n_{b}}\right)/
 \left(\frac{p}{n_{b}}\right)$ & [dimensionless] 
 & relative error estimate \\ 
  & & & of pressure-to-density ratio \\
\midrule
 7 & $\Delta \mathcal{S} $\index{$\Delta \mathcal{S}$} & [dimensionless]
 & absolute error estimate \\ 
  & & & of entropy per baryon \\
\midrule
 8 & $\Delta \mathcal{S}/\mathcal{S}$ & [dimensionless] 
 & relative error estimate \\ 
  & & & of entropy per baryon \\
\bottomrule
\end{tabular}
}
\end{center}
\end{table}

\newpage

\begin{table}[th]
\begin{center}
  \caption{\label{tab:hdf5}%
    Thermodynamic quantities 
    which are stored in the HDF5 data file together with their
    units and the name\index{identifier!HDF5} of the corresponding
    data set. If there are two lines in the 
    second column, the first one always
    corresponds to $I_{\mathrm{tab}} = 0$ and the second one to
    $I_{\mathrm{tab}} \neq 0$. 
}
{\small 
\begin{tabular}{cccc}
\toprule
name of  & quantity/  & unit & description \\ 
data set &  expression
 & & \\
\toprule
t & $T$\index{$T$} 
     & MeV & temperature \\ 
\midrule
pointst  &  $\begin{array}{c} N_{\textrm{data}}\index{$N_{\textrm{data}}$} \\ 
                               N_{T}\index{$N_{T}$} \end{array}$
       & dimensionless & number of points in $T$\\ 
\midrule
nb  & $n_{b}$\index{$n_{b}$}  
       & fm$^{-3}$ & baryon number density\\
\midrule 
pointsnb  &  $\begin{array}{c} N_{\textrm{data}}  \\ 
                               N_{n_{b}}\index{$N_{n_{b}}$} \end{array}$
       & dimensionless & number of points in $n_{b}$\\ 
\midrule
yq    & $Y_{q}$\index{$Y_{q}$} & dimensionless & charge
fraction of strongly \\ 
 & & & interacting particles \\
\midrule
pointsyq  &  $\begin{array}{l} N_{\textrm{data}}  \\ 
                               N_{Y_{q}}\index{$N_{Y_{q}}$}  \end{array}$
       & dimensionless & number of points in $Y_{q}$\\ 
\midrule
thermo    & 
  $Q(N_{\textrm{data}},1,1,N_{\textrm{thermo}})$
  & varying & array of
thermodynamical \\ &$Q(N_{n_{b}}, N_{T},N_{Y_{q}},N_{\textrm{thermo}})$\index{$Q$}
&& quantities, see tables~\ref{tab:ident_thermo1}, \ref{tab:ident_thermo2} \\
\midrule
pointsthermo  &  $N_{\textrm{thermo}}$\index{$N_{\textrm{thermo}}$} 
       & dimensionless & number of thermodynamic \\ 
 &&& quantities, see tables~\ref{tab:ident_thermo1}, \ref{tab:ident_thermo2}.\\ 
\midrule
index\_thermo  &  $J(N_{\textrm{thermo}})$\index{$J$}
       & dimensionless & index identifying the \\ &&& thermodynamic
       quantities, \\ &&& see tables~\ref{tab:ident_thermo1}, \ref{tab:ident_thermo2}.\\ 
\midrule
thermo\_add  & $q(N_{\textrm{data}},1,1,N_{\textrm{add}})$
& varying & array of additional \\ 
 &   $q(N_{n_{b}},N_{T},N_{Y_{q}},N_{\textrm{add}})$\index{$q$} 
 & & quantities $q_{\%}$ \\ &&& in file \texttt{eos.thermo}. \\
\midrule
pointsadd  &  $N_{\textrm{add}}$\index{$N_{\textrm{add}}$}
       & dimensionless & number of additional \\ &&& thermodynamic
       quantities \\ &&& in file \texttt{eos.thermo}.\\ 
\midrule
index\_thermo\_add  & $I_{\textrm{add}}(N_{\textrm{add}})$\index{$I_{\textrm{add}}$} & dimensionless &
index identifying the \\ &&& additional quantities \\ & & & $q_{\%}$ from  file \texttt{eos.thermo},\\ &&& 
see eq.\ (\ref{eq:eos.thermo}) \\
\bottomrule
\end{tabular} 
}
\end{center}
\end{table}

\begin{table}[th]
\begin{center}
  \caption{\label{tab:hdf5micro}%
    Compositional and microscopic quantities 
    which are stored in the HDF5 data file together with their
    units and the name\index{identifier!HDF5} of the corresponding
    data set.
    If there are two lines in the 
    second column, the first one always
    corresponds to $I_{\mathrm{tab}} = 0$ and the second one to
    $I_{\mathrm{tab}} \neq 0$.
  }
  {\small 
\begin{tabular}{cccc}
\toprule
name of  & quantity/  & unit & description \\ 
data set &  expression
 & & \\
\toprule
yi    & $\begin{array}{c}
  Y_{I}(N_{\textrm{data}},1,1,N_{\textrm{p}})\index{$Y_{I}$} \\ 
  Y_{I}(N_{n_{b}}, N_{T},N_{Y_{q}},N_{\textrm{p}})
  \end{array} $& dimensionless & array of
particle fractions \\ 
\midrule
pointspairs  &  $N_{\textrm{p}}$\index{$N_{\textrm{p}}$} 
       & dimensionless & number of pairs \\ & & & for compositional data \\  
\midrule
index\_yi  &  $I_{i}(N_{\textrm{p}})$\index{$I_{i}$} 
       & dimensionless & index identifying \\ &&& the particle $i$,\\
       &&& see tables~\ref{tab:partindex_fermions_sep},
       \ref{tab:partindex_fermions_net}, \ref{tab:partindex_bosons}\\ 
\midrule
yav    & $
  Y^{\textrm{av}}(N_{\textrm{data}},1,1,N_{\textrm{q}})$\index{$Y^{\textrm{av}}$}
&dimensionless & array of combined fractions \\
&  $Y^{\textrm{av}}(N_{n_{b}}, N_{T},N_{Y_{q}},N_{\textrm{q}})$
&& of groups of nuclei  \\ 
\midrule
aav    & $
  A^{\textrm{av}}(N_{\textrm{data}},1,1,N_{\textrm{q}})$\index{$A^{\textrm{av}}$}
&dimensionless & average mass number \\
& $ A^{\textrm{av}}(N_{n_{b}}, N_{T},N_{Y_{q}},N_{\textrm{q}}) $
&& of groups of nuclei  \\ 
\midrule
zav    & $
  Z^{\textrm{av}}(N_{\textrm{data}},1,1,N_{\textrm{q}}) $\index{$Z^{\textrm{av}}$}
&dimensionless & average charge number\\
&$  Z^{\textrm{av}}(N_{n_{b}}, N_{T},N_{Y_{q}},N_{\textrm{q}})$
&& of groups of nuclei  \\ 
\midrule
nav    & $
  N^{\textrm{av}}(N_{\textrm{data}},1,1,N_{\textrm{q}}) $\index{$Z^{\textrm{av}}$}
& dimensionless & average neutron number\\
&$  N^{\textrm{av}}(N_{n_{b}}, N_{T},N_{Y_{q}},N_{\textrm{q}})$
& & of groups of nuclei  \\ 
\midrule
pointsav  &  $N_{\textrm{q}}$\index{$N_{\textrm{q}}$} 
       & dimensionless & number of quadruples  \\ &&&for compositional data \\  
\midrule
index\_av  &  $I_{\textrm{av}}(N_{\textrm{q}})$\index{$I_{\textrm{av}}$} 
       & dimensionless & index identifying \\ 
       &&&the group $\%$ of nuclei, \\
       &&& see tables~\ref{tab:partindex_fermions_sep},
       \ref{tab:partindex_fermions_net}, \ref{tab:partindex_bosons}.\\ 
\midrule
micro    & $\begin{array}{c}
  q_{\textrm{mic}}(N_{\textrm{data}},1,1,N_{\textrm{mic}})\index{$q_{\textrm{mic}}$}  \\ 
  q_{\textrm{mic}}(N_{n_{b}}, N_{T},N_{Y_{q}},N_{\textrm{mic}}) 
 \end{array} $&varying & 
 $\begin{array}{c}
 \textrm{array of microscopic}\\
 \textrm{quantities}
 \end{array}$\\ 
\midrule
pointsmicro  &  $N_{\textrm{mic}}$\index{$N_{\textrm{mic}}$} 
       & dimensionless & number of microscopic \\
       & & & quantities\\  
\midrule
index\_micro  &  $K(N_{\textrm{mic}})$\index{$K$}
       & dimensionless & index identifying the\\ & & &
       microscopic quantities, \\
       & & & see subsection~\ref{ssec:micro}\\ 
\midrule
error    & $\begin{array}{c}
  q_{\textrm{err}}(N_{\textrm{data}},1,1,N_{\textrm{err}})\index{$q_{\textrm{err}}$}  \\ 
  q_{\textrm{err}}(N_{n_{b}}, N_{T},N_{Y_{q}},N_{\textrm{err} }) \end{array} 
 $&varying & array of error quantities \\ 
\midrule
pointserr  &  $N_{\textrm{err}}$\index{$N_{\textrm{err}}$}
       & dimensionless & number of microscopic quantities\\  
\midrule
index\_err  &  $J_{\textrm{err}}(N_{\textrm{err}})$\index{$J_{\textrm{err}}$} 
       & dimensionless & index identifying the error\\ & & &
       quantities, see table~\ref{tab:ident_err}.\\ 
\bottomrule
\end{tabular} 
}
\end{center}
\end{table}


\section{Web tools}
\label{sec:web}

The web tools are constructed as an interface to the \textsc{CompOSE} software, i.e. all features of the code described in the previous section are available. The obtained parameter files \texttt{eos.parameters}, \texttt{eos.quantities} as well as the output table \texttt{eos.table} can be downloaded. In addition, it is possible to visualise the results, chosing the axis from the different computed quantities on a linear or a logarthmic scale.

\part{Appendix}
\appendix

\chapter{Technical details}
\label{app:techdetails}

The basic EoS tables only give a selected set of
thermodynamic quantities\index{quantity!thermodynamic} 
at the grid points\index{grid point} 
that are identified with the index triple\index{index!triple}
($i_{T}\index{$i_{T}$},i_{n_{b}}\index{$i_{n_{b}}$},i_{Y_{q}}$\index{$i_{Y_{q}}$}).
 From the stored quantities\index{quantity!stored} 
$Q_{i}$\index{$Q_{i}$}, $i=1,\dots,6$, see section
\ref{ssec:tabthermoquant}, further relevant thermodynamic quantities
can be derived by a smooth interpolation for all possible values
of the parameters\index{parameter} $T$, $n_{b}$ and $Y_{q}$ within the tabulated
ranges. For these quantities, thermodynamic 
consistency\index{consistency!thermodynamic} should be
respected as far as possible. When the thermodynamic
quantities are interpolated separately, however, this condition
is usually not exactly fulfilled. On the other hand, a separate 
interpolation often leads to smoother dependencies of the
thermodynamic quantitities on the parameters avoiding unphysical
oscillations. In addition, different ways of determining a single
quantity gives the opportunity to estimate the error in the
interpolation.
Thus, in the present code \texttt{compose.f90}\index{compose.f90} the strategy of a
direct interpolation\index{interpolation} of individual quantities is followed.

\section{Interpolation}
\label{sec:interpol}

The interpolation\index{interpolation} scheme for a  
quantity in the thermodynamic, compositional and microscopic
EoS tables
generally proceeds in several steps.
In the following, the procedure will be explained for a
three-dimensional general
purpose EoS\index{equation of state!general purpose} table\index{table!three-dimensional}. 
The interpolation scheme follows
the method proposed in Ref.\ \cite{Swe96}
using polynomials of sufficiently high order.


In the sequel, the generic symbol $Q$\index{$Q$} will be used for any tabulated
quantity\index{quantity!tabulated}. 
Its values are given at the grid points\index{grid point} 
that are specified
by a triple of indices\index{index!triple} 
$(i_{T},i_{n_{b}},i_{Y_{q}})=(i,j,k)$, see subsection
\ref{ssec:para}, corresponding to temperature, baryon number density and
charge fraction of strongly interacting particles. 
Thus all values $Q(T(i),n_{b}(j),Y_{q}(k))$\index{$Q$} are known at the grid points. 
In order to calculate the
quantity $Q$ at given values of $T$, $n_{b}$, and $Y_{q}$, first the values
of the indices
$i = i_{T}\index{$i$}\index{$i_{T}$}$, 
$j = i_{n_{b}}\index{$j$}\index{$i_{n_{b}}$}$ and 
$k  = i_{Y_{q}}\index{$k$}\index{$i_{Y_{q}}$}$ are determined
such that
\begin{eqnarray}
 T(i)        \leq & T\index{$T$} & <  T (i+1) \: , \\
 n_{b}(j) \leq & n_{b}\index{$n_{b}$} & <  n_{b}(j+1) \: , \\
 Y_{q}(k) \leq & Y_{q}\index{$Y_{q}$} & <  Y_{q}(k+1) \: .
\end{eqnarray}
Then the interpolation parameters\index{parameter!interpolation}
\begin{eqnarray}
 \xi\index{$\xi$} & = & \frac{T-T(i)}{T(i+1)-T(i)} \\
 \eta\index{$\eta$} & = &
 \frac{n_{b}-n_{b}(j)}{n_{b}(j+1)-n_{b}(j)} \\
 \zeta\index{$\zeta$} & = &
 \frac{Y_{q}-Y_{q}(k)}{Y_{q}(k+1)-Y_{q}(k)} 
\end{eqnarray}
are introduced with
\begin{equation}
 0 \leq \xi < 1 \qquad
 0 \leq \eta < 1 \qquad
 0 \leq \zeta < 1 \: .
\end{equation}
For the interpolation of a quantity $Q$ at given $(T,n_{b},Y_{q})$
we need the tabulated values at least at the eight corners of the
cube\index{cube} with grid points 
$(i,j,k)$,
$(i+1,j,k)$,
$(i,j+1,k)$,
$(i,j,k+1)$,
$(i+1,j+1,k)$,
$(i+1,j,k+1)$,
$(i,j+1,k+1)$,
$(i+1,j+1,k+1)$.

The interpolation proceeds in two steps: first, an interpolation
in the variable $Y_{q}$ such that the three-dimensional grid\index{grid!three-dimensional}
is mapped to a two-dimensional grid\index{grid!two-dimensional} 
with four corners of each square;
second, a two-dimensional
interpolation\index{interpolation!two-dimensional} 
in the variables $T$ and $n_{b}$.

For the interpolation in $Y_{q}$ in the three-dimensional
cube\index{cube} as defined above, four separate one-dimensional 
interpolations\index{interpolation!one-dimensional} along the
lines that connect the grid points 
$(i,j,k)$ and $(i,j,k+1)$,
$(i+1,j,k)$ and $(i+1,j,k+1)$,
$(i,j+1,k)$ and $(i,j+1,k+1)$,
$(i+1,j+1,k)$ and $(i+1,j+1,k+1)$
have to be performed.
After this first step, the interpolation proceeds in the parameters $T$ and
$n_{b}$ by a two-dimensional\index{interpolation!two-dimensional} 
scheme as discussed in subsection 
\ref{subsec:2dim}.

The interpolation in the variables $T$, $n_{b}$ and $Y_{q}$ can
be of different order\index{interpolation!order} in general, defined by the value $I$ of the
variables \texttt{ipl\_t}, \texttt{ipl\_n} and \texttt{ipl\_y}, see section
\ref{sec:data}. In the following, the case of highest order ($I=3$) is
considered first.

\subsection{Interpolation in one dimension}
\index{interpolation!one-dimensional}

The order $I=3$ of the interpolation requires
that the function values, its first and second derivatives are
continuous at the two corner points of each line, i.e.\ six values have to be
reproduced by the interpolation polynomial. Hence, a polynomial of
at least fifth degree has to be used. 
The six independent coefficients $q_{n}$ of a single
polynomial
\begin{equation}
\label{eq:poly}
 Q(T(i),n_{b}(j),Y_{q}) = \sum_{n=0}^{5} q_{n} \zeta^{n}
\end{equation}
can determined from the function values and derivatives\index{derivative} at the corner
points directly. One finds
\begin{equation}
 q_{0}\index{$q_{i}$} =  Q_{ijk}^{(0)} \qquad
 q_{1} = Q_{ijk}^{(1)} \qquad
 q_{2} = \frac{1}{2} Q_{ijk}^{(2)}
\end{equation}
with
\begin{equation}
 Q^{(n)}_{ijk}\index{$Q^{(n)}_{ijk}$}  = 
 \left[Y_{q}(k+1)-Y_{q}(k)\right]^{n}\left. \frac{\partial^{n}Q}{\partial
   Y_{q}^{n}}\right|_{T(i),n_{b}(j),Y_{q}(k)}   
\end{equation}
for $n=0,1,2$, and
\begin{equation}
 Q_{ijk}^{(0)} = \left. \frac{\partial^{0}Q}{\partial
     Y_{q}^{0}}\right|_{T(i),n_{b}(j),Y_{q}(k)}
 = Q(T(i),n_{b}(j),Y_{q}(k))
\end{equation}
in particular. The remaining three coefficients are given by
\begin{eqnarray}
 q_{3} & = & 10 A -4 B + \frac{1}{2} C \\ 
 q_{4} & = & -15 A + 7 B - C \\
 q_{5} & = & 6 A - 3 B + \frac{1}{2}C
\end{eqnarray}
with
\begin{eqnarray}
 A\index{$A$} & = &  Q_{ijk+1}^{(0)}- Q_{ijk}^{(0)} - Q_{ijk}^{(1)} - \frac{1}{2}
 Q_{ijk}^{(2)} \\
 B\index{$B$} & = & Q_{ijk+1}^{(1)} - Q_{ijk}^{(1)} - Q_{ijk}^{(2)} \\
 C\index{$C$} & = & Q_{ijk+1}^{(2)} - Q_{ijk}^{(2)} \: .
\end{eqnarray}

Alternatively, the approach of Ref.\
\cite{Swe96} can be followed.
The value of the quantity $Q$ at given $\zeta$ is found 
with the help of the quintic basis functions\index{basis function!quintic}
\begin{eqnarray}
 \psi_{0}^{(0)}(z) & = & 1 - 10 z^{3} + 15 z^{4} - 6 z^{5}
 \\
 \psi_{1}^{(0)}(z) & = & z - 6 z^{3} + 8 z^{4} - 3 z^{5}
 \\
 \psi_{2}^{(0)}(z) & = & \frac{1}{2} \left( z^{2} - 3 z^{3} + 3 z^{4} -
   z^{5} \right) 
\end{eqnarray}
that have the properties
\begin{equation}
 \psi_{m}^{(n)}(0)  =  \delta_{nm} \qquad
 \psi_{m}^{(n)}(1)  =  0
\end{equation}
for 
\begin{equation}
 \psi_{m}^{(n)}(z)\index{$\psi_{m}^{(n)}$} = \frac{d^{n}\psi_{m}^{(0)}}{dz^{n}}
\end{equation}
with $n,m \in \{0,1,2\}$.
Then one has
\begin{eqnarray}
\label{eq:quintic_z}
 Q(T(i),n_{b}(j),Y_{q}) & = &
 \sum_{n=0}^{2} \left[ Q^{(n)}_{ijk} \psi_{n}^{(0)}(\zeta) +
 (-1)^{n} Q^{(n)}_{ijk+1} \psi_{n}^{(0)}(1-\zeta)\right]
\end{eqnarray}
with six interpolation coefficients\index{interpolation!coefficient}
that are given by
directly by the function values and derivatives\index{derivative}
at the corner points defined by the indices. 
Derivatives are easily found as
\begin{eqnarray}
 \lefteqn{\frac{\partial^{s}}{\partial Y_{q}^{s}} Q(T(i),n_{b}(j),Y_{q})}
 \\ \nonumber & = &
 \frac{1}{\left[Y_{q}(k+1)-Y_{q}(k)\right]^{s}}
 \sum_{n=0}^{2} \left[ Q^{(n)}_{ijk} \psi_{n}^{(s)}(\zeta) +
 (-1)^{n+s} Q^{(n)}_{ijk+1} \psi_{n}^{(s)}(1-\zeta)\right] \: .
\end{eqnarray}

The numerical determination of the derivatives 
depends on the order of interpolation. For $I=3$, centered five-point
finite difference formulas\index{formula!finite difference} 
are used to calculate the first and second
derivative\index{derivative} of a function. Close to the boundaries of the EoS,
centered difference formulas cannot be applied. In this case,
off-center formulas are used. In the case $I=2$, three-point 
finite difference formulas are employed.
A continuity of the
second derivatives at the corner points is not required. Only the
function and the first derivative are demanded to be continuous,
corresponding to four independent quatities. Hence, a polynomial of
third degree is suffiecient and $q_{4} = q_{5} = 0$ in Eq.\ (\ref{eq:poly}).
Then, the coefficients of the second derivatives have to be defined as
\begin{eqnarray}
 Q_{ijk}^{(2)} & = & 6 \left[ Q_{ijk+1}^{(0)} - Q_{ijk}^{(0)} \right]
 - 2 Q_{ijk+1}^{(1)} - 4 Q_{ijk}^{(1)}
 \\ 
 Q_{ijk+1}^{(2)} & = & - 6 \left[ Q_{ijk+1}^{(0)} - Q_{ijk}^{(0)} \right]
 + 4 Q_{ijk+1}^{(1)} + 2 Q_{ijk}^{(1)} \: .
\end{eqnarray}
For $I=1$, the function has to be continuous at the grid points
but there is no condition on the derivatives. In this case,
one sets 
\begin{equation}
 Q_{ijk}^{(1)} = Q_{ijk+1}^{(1)} = Q_{ijk+1}^{(0)}-Q_{ijk}^{(0)} \qquad
 Q_{ijk}^{(2)} = Q_{ijk+1}^{(2)} = 0 \
\end{equation}
for given function values $Q_{ijk}^{(0)}$ and $Q_{ijk+1}^{(0)}$.
Then the polynomial (\ref{eq:poly}) reduces to a linear function
with $q_{2} = q_{3} = q_{4} = q_{5} = 0$.

\subsection{Interpolation in two dimensions}
\label{subsec:2dim}
\index{interpolation!two-dimensional}

The interpolation in two dimensions with conditions on the
continuity of the function and its derivatives is more complicated
than in the one-dimensional case. For an interpolation
order\index{interpolation!order} $I=3$ in
both variables $\xi$ and $\eta$, there are 
four function values,
eight first derivatives and the twelve second derivatives, hence
24 values in total,
that can be used to determine the coefficients of a polynomial
in $\xi$ and $\eta$.  From this consideration, it would seem to be
sufficient to use a polynomial 
\begin{equation}
 Q(T,n_{b},Y_{q}) = \sum_{m=0}^{4} \sum_{n=0}^{4} q_{mn} \xi^{m} \eta^{n}
\end{equation}
that includes powers up to four in both
variables, resulting in $5 \times 5 = 25$ coefficients $q_{nm}$
in total, i.e.\ one more than
required. However, such a form does not guarantee that the function and the
derivatives are continuous not only at the grid points but also along
the boundaries of the interpolation square. In fact, a continuity of
the function $Q$, the first derivatives\index{derivative} $\partial Q/\partial T$ and
$\partial Q/\partial n_{b}$, the second derivatives 
$\partial^{2}Q/\partial T^{2}$,
$\partial^{2}Q/\partial T \partial n_{b}$ and
$\partial^{2}Q/\partial n_{b}^{2}$,
the third derivatives
$\partial^{3}Q/\partial T^{2} \partial n_{b}$ and
$\partial^{3}Q/\partial T \partial n_{b}^{2}$ and
the fourth derivative
$\partial^{4}Q/\partial T^{2} \partial n_{b}^{2}$ at the corners
has to be demanded, determining $4\times 9 = 36 = 6\times 6$
coefficients of a polynomial 
\begin{equation}
\label{eq:bipoly}
 Q(T,n_{b},Y_{q}) = \sum_{m=0}^{5} \sum_{n=0}^{5} q_{mn} \xi^{m} \eta^{n}
\end{equation}
with degree six in each variable.

Instead of determining the coefficients directly as in the case 
of a one dimensional interpolation, it is more advantageous
to use the biquintic
interpolation\index{interpolation!biquintic} 
scheme as in Ref.\ \cite{Swe96} with
\begin{eqnarray}
\label{eq:biquintic_xy}
 \lefteqn{Q(T,n_{b},Y_{q})}
 \\ \nonumber & = &
 \sum_{m=0}^{2} \sum_{n=0}^{2} 
 \left[ Q^{(mn)}_{ij} \psi_{m}^{(0)}(\xi)\psi_{n}^{(0)}(\eta)
 + (-1)^{m} Q^{(mn)}_{i+1j} \psi_{m}^{(0)}(1-\xi)\psi_{n}^{(0)}(\eta)
 \right. \\ \nonumber & & \left.
 + (-1)^{n} Q^{(mn)}_{ij+1} \psi_{m}^{(0)}(\xi)\psi_{n}^{(0)}(1-\eta)
 + (-1)^{m+n} Q^{(mn)}_{i+1j+1} \psi_{m}^{(0)}(1-\xi)\psi_{n}^{(0)}(1-\eta) \right]
\end{eqnarray}
and $4 \times 9 = 36$ coefficients
\begin{eqnarray}
Q^{(mn)}_{ij}\index{$Q_{ij}^{(mn)}$}
& = & \left[T(i+1)-T(i)\right]^{m}\left[n_{b}(j+1)-n_{b}(j)\right]^{n}
 \left. \frac{\partial^{m+n} Q}{\partial T^{m} \partial n_{b}^{n}} 
 \right|_{T(i),n_{b}(j),Y_{q}}  \\
Q^{(mn)}_{i+1j} & = & 
\left[T(i+1)-T(i)\right]^{m}\left[n_{b}(j+1)-n_{b}(j)\right]^{n}
 \left. \frac{\partial^{m+n} Q}{\partial T^{m} \partial n_{b}^{n}} 
 \right|_{T(i+1),n_{b}(j),Y_{q}}  \\
Q^{(mn)}_{ij+1} & = & 
\left[T(i+1)-T(i)\right]^{m}\left[n_{b}(j+1)-n_{b}(j)\right]^{n}
 \left. \frac{\partial^{m+n} Q}{\partial T^{m} \partial n_{b}^{n}} 
 \right|_{T(i),n_{b}(j+1),Y_{q}}  \\
Q^{(mn)}_{i+1j+1} 
 & = & \left[T(i+1)-T(i)\right]^{m}\left[n_{b}(j+1)-n_{b}(j)\right]^{n}
 \left. \frac{\partial^{m+n} Q}{\partial T^{m} \partial n_{b}^{n}} 
 \right|_{T(i+1),n_{b}(j+1),Y_{q}} 
\end{eqnarray}
where $m,n \in \{0,1,2\}$. Of course, it is possible to express the
coefficients $q_{mn}$\index{$q_{mn}$} of the polynomial (\ref{eq:bipoly}) through the
coefficients $Q_{ij}^{(mn)}$\index{$Q_{ij}^{(mn)}$}, 
as done in the code \texttt{compose.f90}.
The derivatives are obtained in the same way as in
the one-dimensional case for the different interpolations orders $I$.
Derivatives\index{derivative} of the function $Q$ with respect to $Y$ and $n_{b}$ can be
found with the help of the relation
\begin{eqnarray}
 \lefteqn{\frac{\partial^{s+t}}{\partial T^{s}\partial n_{b}^{t}}Q(T,n_{b},Y_{q})}
 \\ \nonumber & = &
 \frac{1}{\left[T(i+1)-T(i)\right]^{s}}
 \frac{1}{\left[n_{b}(j+1)-n_{b}(j) \right]^{t}}
 \\ \nonumber & & \times
 \sum_{m=0}^{2} \sum_{n=0}^{2} 
 \left[ Q^{(mn)}_{ij} \psi_{m}^{(s)}(\xi)\psi_{n}^{(t)}(\eta)
 + (-1)^{m+s} Q^{(mn)}_{i+1j} \psi_{m}^{(s)}(1-\xi)\psi_{n}^{(t)}(\eta)
 \right. \\ \nonumber & & \left.
 + (-1)^{n+t} Q^{(mn)}_{ij+1} \psi_{m}^{(s)}(\xi)\psi_{n}^{(t)}(1-\eta)
 + (-1)^{m+n+s+t} Q^{(mn)}_{i+1j+1}
 \psi_{m}^{(s)}(1-\xi)\psi_{n}^{(t)}(1-\eta) \right] \: .
\end{eqnarray}

Above considerations apply to the interpolation in all three
parameters for a general purpose EoS 
table\index{table!general purpose}. 
For neutron matter\index{matter!neutron}, i.e.\
$Y_{q}=0$ the interpolation in the charge fraction of strongly
interacting particles is
trivial since $\zeta=0$ and the tabulated values of the quantities
can be used directly in the second interpolation step in $T$ and
$n_{b}$. A similar procedure applies to the case of an EoS for
symmetric nuclear matter\index{matter!symmetric} of $\beta$-equilibrium.
In the case of a zero-temperature table, the second
interpolation step reduces from two to one dimension.

\section{Matter in beta equilibrium}
\label{sec:betaequi}
\index{matter!$\beta$ equilibrium}

In the case of matter in $\beta$
equilibrium\index{equilibrium!$\beta$}, 
there is an additional
condition on the chemical potentials\index{chemical potential} that reduces the number of
independent parameters\index{parameter!independent} 
by one. Requiring that the weak
interaction reactions\index{reaction!weak interaction}
\begin{equation}
  p + e^{-} \leftrightarrow n + \nu_{e} \qquad
  p + \bar{\nu}_{e} \leftrightarrow n + e^{+}
\end{equation}
and, if muons are considered in the model,
\begin{equation}
  p + \mu^{-} \leftrightarrow n + \nu_{\mu} \qquad
   p + \bar{\nu}_{\mu} \leftrightarrow n + \mu^{+}
\end{equation}
are in weak equilibrium, the relations
\begin{equation}
 \mu_{p} + \mu_{e} = \mu_{n} + \mu_{\nu_{e}}
\end{equation}
and
\begin{equation}
 \mu_{p} + \mu_{\mu} = \mu_{n} + \mu_{\nu_{\mu}} \: ,
\end{equation}
respectively, should hold. 
The chemical potentials of the
neutrinos\index{potential!chemical!neutrino} are considered to be zero, i.e.\
$\mu_{\nu_{e}}=\mu_{\nu_{\mu}}=0$. 
Because $\mu_{p} = \mu_{n}+\mu_{q}$,
$\mu_{e} = \mu_{le}-\mu_{q}$ and $\mu_{\mu} = \mu_{l\mu}-\mu_{q}$,
the constraint can be formulated as
\begin{equation}
 \mu_{l}\index{$\mu_{l}$} = 0
\end{equation}
with the effective lepton chemical potential\index{potential!chemical!lepton}
$\mu_{l} = \mu_{le} = \mu_{l\mu}$. For given temperature $T$
and baryon number density $n_{b}$ there is a unique charge
fraction of strongly interacting particles
$Y_{q}$ that is found by determining the zero of the function
\begin{equation}
 f(Y_{q}) = \mu_{l}(Y_{q}) \: .
\end{equation}

\chapter{Structure of input files}
\label{app:files}

During the execution of the program \texttt{compose}\index{compose} for tasks 
1 and 2, three files (in ASCII format) 
are generated that are used for task 3 as input
files. The structure of these files is described below. Experienced users of 
\texttt{compose}\index{compose} can modify the files 
\texttt{eos.parameters}\index{eos.parameters} and 
\texttt{eos.quantities}\index{eos.quantities} according to their needs
when producing a new EoS table with modified parameters by running only
task 3. The file \texttt{eos.init}\index{eos.init} has to be left
untouched.

\section{\texttt{eos.init}\index{eos.init}}

There are 11 rows in the file \texttt{eos.init}\index{eos.init}.
The first three numbers in the first row indicate whether the EoS
dependence on the temperature, the baryon density, and the
hadronic charge fraction (index = 1) or not (index = 0). The 
following two integers are presently set to zero and will be used
only for future extensions of the \texttt{compose}\index{compose} code.
The next three rows give the minimum and maximum values for the variables
$T$, $n_{b}$, and $Y_{q}$. In the fifth row the number 
zero appears twice in the present version as dummy floating-point variable.
Row six contained four integers. The first indicates whether leptons
are included in the EoS (1: yes, 0: no). The second number gives the
number of additional quantities in the file \texttt{eos.thermo} and the third
number is the total number of thermodynamic quantities in each row of the
file \texttt{eos.thermo}. The last integer indicates whether a calculation
of an EoS table for the condition of $\beta$ equilibrium is possible (1) or not (0).
Row seven gives the number of pairs and quadruples in the file \texttt{eos.compo}
and the following two lines contain the corresponding indices according to tables
\ref{tab:partindex_fermions_sep}, \ref{tab:partindex_fermions_net},
\ref{tab:partindex_bosons} and \ref{tab:corrindex} for pairs (first line)
and according to the data sheet for quadruples (second line).
In row ten the number of microscopic quantities is given and their indices
are listed in the last row of the file \texttt{eos.init}.

\section{\texttt{eos.quantities}\index{eos.quantities}}

This file contains 18 rows. They determine which
and how quantities will be stored in the final EoS table with the name
\texttt{eos.table}\index{eos.table}.
Rows with odd number (one, three, \dots) are comment lines
beginning with the character \texttt{\#}.
The second and fourth lines define the thermodynamic quantities that will be
stored in the output file \texttt{eos.table}. The first entry 
$N_{\textrm{thermo}}$\index{$N_{\textrm{thermo}}$} in row two specifies
the number of thermodynamic quantities that are selected from
tables \ref{tab:ident_thermo1} and \ref{tab:ident_thermo2}
followed by the number $N_{\textrm{add}}$\index{$N_{\textrm{add}}$}
of quantities that are
selected from the stored quantities of the EoS file 
\texttt{eos.thermo}\index{eos.thermo} in addition to
the seven standard quantities as described in subsection 
\ref{ssec:tabthermoquant}.
The last number $N_{\textrm{deriv}}$\index{$N_{\textrm{deriv}}$} in row two 
is the number of quantities from table \ref{tab:f_derivatives} 
that contains the free energy per baryon and various derivatives of it.
The fourth line of the file \texttt{eos.quantities}\index{eos.quantities}
lists the $N_{\textrm{thermo}}$ indices (if $N_{\textrm{thermo}} > 0$)
from the second column of tables \ref{tab:ident_thermo1} 
and \ref{tab:ident_thermo2}.
that define the available thermodynamic quantities,
the $N_{\textrm{add}}$ indices (if $N_{\textrm{add}} > 0$)
of the additionally stored quantities 
and the $N_{\textrm{deriv}}$ indices (if $N_{\textrm{deriv}} > 0$) from the
first column of table \ref{tab:f_derivatives}.
E.g., the first four lines
\begin{quote}
\texttt{\# number of regular, additional and derivative quantities} \\
\texttt{3 1 1} \\
\texttt{\# indices of regular, additional and derivative quantities} \\
\texttt{6 1 2 1 1}
\end{quote}
of an example file \texttt{eos.quantities}\index{eos.quantities}
denote that the quantities $\mathcal{F}/m_{n}-1$, $p$, $\mathcal{S}$,
$q_{1}$ and $\mathcal{F}$ will appear in the 
outpout file \texttt{eos.table}\index{eos.table}
or \texttt{eoscompose.h5}\index{eoscompose.h5}.
Regular thermodynamic quantities are those that are obtained by a direct
interpolation of the tabulated quantities $\mathcal{Q}_{1}$, \dots,
$\mathcal{Q}_{7}$ stored in the file \texttt{eos.thermo} or those that are
obtained from them by applying the thermodynamic relations of
section \ref{sec:thermo_coeff}.
The indices of the regular thermodynamic quantities can have
values of $1$,
$2$, \dots, $23$ corresponding to the definition in tables
\ref{tab:ident_thermo1} and \ref{tab:ident_thermo2}. 
Additional thermodynamic quantities are those that are denoted 
$q_{1}$, \dots, $q_{N_{\textrm{add}}}$ in the file \texttt{eos.thermo}. Their
number $N_{\textrm{add}}$ and meaning depends on the specific EoS table.
Their values are found by a direct interpolation.
The values for $\mathcal{F}$ and its derivatives are found from
a direct interpolation and subsequent derivatives 
of the interpolating function.
If $N_{\textrm{thermo}}$\index{$N_{\textrm{thermo}}$}, 
$N_{\textrm{add}}$\index{$N_{\textrm{add}}$}  
and $N_{\textrm{deriv}}$\index{$N_{\textrm{deriv}}$} are zero the fourth line of the
file \texttt{eos.quantities}\index{eos.quantities} is empty.

The sixth row of the file \texttt{eos.quantities}\index{eos.quantities} 
contains the numbers
$N_{p}$\index{$N_{p}$} and $N_{q}$\index{$N_{q}$} 
of pairs\index{pair} and quadruples\index{quadruple} that are selected from the
compositional quantities\index{quantity!compositional} 
that are stored in the file \texttt{eos.compo}\index{eos.compo}
as described in subsection \ref{ssec:compo}. In the eigth line
the $N_{p}$ particle indices $I_{1}$, \dots, $I_{N_{p}}$ as defined in
tables \ref{tab:partindex_fermions_sep}, \ref{tab:partindex_fermions_net},
\ref{tab:partindex_bosons} and \ref{tab:corrindex} 
are followed by the $N_{q}$ indices that 
identify a particular group of nuclei, see subsection
\ref{ssec:compo}.
E.g., the four lines
\begin{quote}
\texttt{\# number of pairs and quadruples for composition data} \\
\texttt{3 1} \\
\texttt{\# indices of pairs and quadruples for composition data} \\
\texttt{10 11 0 1}
\end{quote}
denote that the fractions $Y_{n}$, $Y_{p}$, $Y_{e}$ are stored in the
file \texttt{eos.table}\index{eos.table} followed by 
$Y^{\textrm{av}}_{1}$\index{$Y^{\textrm{av}}$},
$A^{\textrm{av}}_{1}$\index{$A^{\textrm{av}}$}, 
$Z^{\textrm{av}}_{1}$\index{$Z^{\textrm{av}}$},
and $N^{\textrm{av}}_{1}$\index{$N^{\textrm{av}}$}
of the first set of nuclei.
If both $N_{p}$\index{$N_{p}$} and $N_{q}$\index{$N_{q}$} 
are zero the eighth line of the
file \texttt{eos.quantities}\index{eos.quantities} is empty.

The tenth line of the file \texttt{eos.quantities} contains the number
$N_{\textrm{mic}}$\index{$N_{\textrm{mic}}$} of
microscopic quantities\index{quantity!microscopic} in the file 
\texttt{eos.micro}\index{eos.micro} to be
stored in the output file \texttt{eos.table}\index{eos.table}. Then line 12
is a list of $N_{\textrm{mic}}$ composite indices\index{index!composite} 
$K_{i}$\index{$K_{i}$} as defined in
subsection \ref{ssec:micro}.
E.g., the four lines
\begin{quote}
\texttt{\# number of microscopic quantities} \\
\texttt{2} \\
\texttt{\# indices of microscopic quantities} \\
\texttt{10050 11050}
\end{quote}
denote that the nonrelativistic single-particle potentials
$U_{n}$ and $U_{p}$ are stored in the file \texttt{eos.table}.
If $N_{\textrm{mic}}$  is zero the twelfth line of the
file \texttt{eos.quantities} is empty.

In addition to the quantities considered 
above, error estimates\index{error!estimate} for the
interpolation\index{interpolation} 
of the thermodynamic
quantities $\mathcal{F}$\index{$\mathcal{F}$}, 
$\mathcal{E}$\index{$\mathcal{E}$}, 
$p/n_{b}$\index{$p$} and $\mathcal{S}$\index{$\mathcal{S}$} are available.
The free energy per baryon $\mathcal{F}$ can be obtained by direct
interpolation or by using the homogeneity\index{homogeneity} 
condition (\ref{eq:f_con})
with the interpolated values for the pressure and chemical potentials.
An estimate for the absolute error\index{error!absolute} 
in $\mathcal{F}$ is then given by
\begin{equation}
 \Delta \mathcal{F}\index{$\Delta \mathcal{F}$} = \mathcal{F} + \frac{p}{n_{b}}
 - \left( \mu_{b} + Y_{q} \mu_{q}\right)
\end{equation}
in case of an EoS without leptons and
\begin{equation}
 \Delta \mathcal{F} = \mathcal{F} + \frac{p}{n_{b}}
 - \left( \mu_{b} + Y_{l} \mu_{l}\right)
\end{equation}
in case of an EoS with leptons and charge neutrality.
The relative error\index{error!relative} is then given by 
$\Delta \mathcal{F}/\mathcal{F}-1$.
The consideration above also apply to the internal 
energy per baryon $\mathcal{E}$.
An estimate for the absolute error in $p/n_{b}$ is obtained by comparing
the directly interpolated pressure $p$ with that derived from the
first derivative of the free energy per baryon, i.e.\
\begin{equation}
 \Delta \frac{p}{n_{b}} = \frac{p}{n_{b}} 
 - n_{b} \left. \frac{\partial \mathcal{F}}{\partial n_{b}}
 \right|_{T,Y_{q}} \: .
\end{equation}
Similarly, an error estimate for the entropy per baryon is calculated
from
\begin{equation}
 \Delta \mathcal{S}\index{$\Delta \mathcal{S}$} = \mathcal{S}
 - \left. \frac{\partial \mathcal{F}}{\partial T}
 \right|_{n_{b},Y_{q}} \: .
\end{equation}
The sixteenth line of the file \texttt{eos.quantities}\index{eos.quantities}
defines the number
$N_{\textrm{err}}$\index{$N_{\textrm{err}}$}
of error quantities to be stored in the output file.
Line 18 is a list of the $N_{\textrm{err}}$ indices $J$\index{$J$}
as defined in table \ref{tab:ident_err}.
E.g., the four lines
\begin{quote}
\texttt{\# number of error quantities} \\
\texttt{2} \\
\texttt{\# indices of error quantities} \\
\texttt{1 2 }
\end{quote}
denote that the estimates
$\Delta \mathcal{F}$ and $\Delta \mathcal{F}/\mathcal{F}$ 
are stored in the file \texttt{eos.table}\index{eos.table}.
If $N_{\textrm{err}}$ is zero the eighteenth line of the
file \texttt{eos.quantities} is empty.

The last row of the file \texttt{eos.quantities}\index{eos.quantities}
defines the format\index{format!output} of
the output file  by a single integer $I$.  
There are two methods that are used for storing the EoS data. The
first ($I=1$) is the simple ASCII format\index{format!ASCII} 
using the file name \texttt{eos.table}\index{eos.table}
that easily allows to read the data without the need for further
codes.
The second method ($I \neq 1$) 
uses the more advanced HDF5 format\index{format!HDF5} (see
\url{http://www.hdfgroup.org/HDF5/}) that is widely used in the
astrophysics community.  In this case, the output file carries the
name \texttt{eoscompose.h5}\index{eoscompose.h5} and
each data set is designated with a
particular identifier\index{identifier} 
\texttt{$ \$ $} of the stored 
quantity $Q$. Tables
\ref{tab:hdf5} and \ref{tab:hdf5micro} 
give a list of
all tabulated quantities and corresponding identifiers 
\texttt{$ \$ $}\index{$ \$ $}.

Note that $\varepsilon=\mathcal{E}/m_{n}-1$\index{$\varepsilon$}
in table~\ref{tab:ident_thermo1}
is \emph{not} the specific internal
energy\index{energy!internal!specific}, 
since the total mass
density\index{density!mass!total} is not given by $m_{n} n_{b}$.
We do not store $\mathcal{E}/m_{n}$\index{$\mathcal{E}$} 
because it is in many regimes
largely dominated by the rest mass contribution, such that it is numerically
difficult to keep trace of small variations in the internal energy in that
case. The same argument applies to the other thermodynamic
quantities $\mathcal{F}$\index{$\mathcal{F}$}, 
$\mathcal{G}$\index{$\mathcal{G}$} and 
$\mathcal{H}$\index{$\mathcal{H}$}.
The present choice of storing $\varepsilon$ is also 
motivated by the fact that the total
mass density is not a conserved quantity\index{quantity!conserved} 
throughout the hydrodynamic
evolution, but only the baryon number density, $n_{b}$. 

The individual chemical
potentials\index{potential!chemical!individual} 
for all the particles (baryons\index{baryon},
mesons\index{meson}, nuclei\index{nucleus}, quarks\index{quark}, \dots)
can be calculated from the three ones given in the tables. Note that we
use the relativistic definition\index{definition!relativistic} 
of the chemical potentials.

Particle fractions\index{fraction!particle} $Y_{i}$\index{$Y_{i}$} 
are stored in the data sets with identifier\index{identifier}
$y\#$ where
$\#$\index{$\#$} stands for the particle index $I_{i}$ as defined in tables
\ref{tab:partindex_fermions_sep}, \ref{tab:partindex_fermions_net},
\ref{tab:partindex_bosons} and \ref{tab:corrindex}. E.g., the data identifier
\texttt{y20} and \texttt{y400} 
contain the particle fractions of
$\Delta^{-}$ and $K^{-}$ particles, respectively.
Similarly, \texttt{yav1} stands for the fraction of the first group of
nuclei.

Combining the examples above, the file \texttt{eos.quantities} assumes the
form
\begin{quote}
\texttt{\# number of regular, additional and derivative quantities} \\
\texttt{3 1 1} \\
\texttt{\# indices of regular, additional and derivative quantities} \\
\texttt{6 1 2 1 1} \\
\texttt{\# number of pairs and quadruples for composition data} \\
\texttt{3 1} \\
\texttt{\# indices of pairs and quadruples for composition data} \\
\texttt{10 11 0 1} \\
\texttt{\# number of microscopic quantities} \\
\texttt{2} \\
\texttt{\# indices of microscopic quantities} \\
\texttt{10050 11050} \\
\texttt{\# number of error quantities} \\
\texttt{2} \\
\texttt{\# indices of error quantities} \\
\texttt{1 2 }
\end{quote}
and, consequently,
each line of the output file \texttt{eos.table} 
will contain the following entries:
{
\begin{eqnarray}
 \nonumber 
T \quad n_{b} \quad Y_{q} & & 
\mathcal{F}/m_{n}-1 \quad
 p \quad \mathcal{S} \quad q_{1}  \quad \mathcal{F} \quad Y_{n} 
\quad Y_{p} \quad Y_{e}
 \quad 
\\
 \nonumber  & & 
 Y^{\textrm{av}}_{1}  \quad A^{\textrm{av}}_{1} \quad
 Z^{\textrm{av}}_{1}  \quad N^{\textrm{av}}_{1} \quad
 U_{n} \quad U_{p} \quad \Delta \mathcal{F} \quad \Delta
 \mathcal{F}/\mathcal{F} 
\end{eqnarray}
}
i.e., in total $19$ quantities.

\section{\texttt{eos.parameters}\index{eos.parameters}}

There are at least 
ten lines in the file \texttt{eos.parameters}\index{eos.parameters}.
Rows one, three, five and seven of the file 
are comment lines indicated by the first character \texttt{\#}.
The second row of this file  
contains three integer numbers
$I_{T}$\index{$I_{T}$}, $I_{n_{b}}$\index{$I_{n_{b}}$} 
and $I_{Y_{q}}$\index{$I_{Y_{q}}$} that
specify the interpolation scheme for the temperature, baryon number
density and charge fraction of strongly interacting particles, 
respectively. Presently, there
is the choice between first order ($I_{x}=1$, $x=T,n_{b},Y_{q}$),
second order ($I_{x}=2$) 
and third order ($I_{x}=3$) interpolation\index{interpolation} available with
continuity of all quantities $Q(x)$, all quantities $Q(x)$ and their first
derivative $\partial Q/\partial x$ or all quantities $Q(x)$, their first and 
second derivatives $\partial Q/\partial x$ and $\partial^{2}
Q/\partial x^{2}$, respectively. See appendix~\ref{sec:interpol} for
details on the interpolation scheme.

The fourth row of the file \texttt{eos.parameters} 
contains two integers $I_{\beta}$\index{$I_{\beta}$}
and $I_{\textrm{entr}}$\index{$I_{\textrm{entr}}$}.
The first determines whether the
EoS in the output file \texttt{eos.table} will be generated for matter
in $\beta$ equilibrium ($I_{\beta} = 1$) or not ($I_{\beta} \neq 1$).
Of course, this option is only effective for EoS tables that include
electrons (and muons) and depend on the parameter $Y_{q}$, e.g.\
three-dimensional\index{table!three-dimensional} 
general purpose EoS tables and two-dimensional\index{table!two-dimensional}
zero-temperature EoS tables.
The second integer determines whether 
the EoS in the output file \texttt{eos.table} will be generated 
for given entropy per baryon $\mathcal{S}$
($I_{\textrm{entr}} = 1$) or not ($I_{\textrm{entr}} \neq 1$). If yes, the input
parameters that usually define the temperature values will be interpreted
as defining the entropy per baryon.
This option can only be used for EoS that depend
on the temperature. However it is not guaranteed that there
are solutions, found by interpoalation, for the given parameter values.

In the sixth row of the file, the tabulation scheme for the parameters
is defined by the integer $I_{\textrm{tab}}$\index{$I_{\textrm{tab}}$}. 
For $I_{\textrm{tab}} = 0$ the the parameter values of $T$ (or $\mathcal{S}$),
$n_{b}$ and $Y_{q}$
are listed explicitly in the file 
\texttt{eos.parameters}\index{eos.parameters} as follows:
The eighth row contains the number of data points 
$N_{\textrm{data}}$\index{$N_{\textrm{data}}$} to be generated and
each line of the next $N_{\textrm{data}}$ rows
contains the three parameter values in the form
\begin{equation}
 T \:\: n_{b} \:\: Y_{q}
\end{equation}
or
\begin{equation}
  \mathcal{S} \:\: n_{b} \:\: Y_{q}
\end{equation}
that define the grid points\index{grid point}.
The EoS data that will be stored in the final EoS table will appear
in the same sequence of the parameters as given in the file 
\texttt{eos.parameters}\index{eos.parameters}.
For {$I_{\beta} =1$}, $Y_{q}$ can be set to any value.
For two- and one-dimensional EoS 
tables\index{table!two-dimensional}\index{table!one-dimensional}, 
the parameters that are not
used can be set to arbitrary values, e.q. for a zero-temperature EoS
the first entry in each row can have any finite value.
An example file \texttt{eos.parameters} for the tabulatfor a zero-temperature EoS
the first entry in each row can have any finite value.
An example file \texttt{eos.parameters} for the tabulation
  scheme with $I_{\textrm{tab}}=0$ is given by
\begin{quote}
\texttt{\# interpolation rules for T, n\_b and Y\_q} \\
\texttt{3 3 3} \\
\texttt{\# calculation of beta equilibrium 
(1:~yes, else:~no) and for given \mbox{}\hfill\mbox{} entropy (1:~yes,~else:~no)}\\
\texttt{0 0} \\
\texttt{\# tabulation scheme of parameter values (see manual)} \\
\texttt{0} \\
\texttt{\# parameter values depending on tabulation scheme} \\
\texttt{5} \\
\texttt{0.1 0.1 0.5} \\
\texttt{0.1 0.2 0.5} \\
\texttt{0.1 0.3 0.5} \\
\texttt{0.1 0.4 0.5} \\ 
\texttt{0.1 0.5 0.5} \\
\end{quote}
for generating a table using third order interpolations in $T$, $n_{b}$
and $Y_{q}$. Five data points are calculated for constant $T=0.1$~MeV and
$Y_{q} = 0.5$ at densities of $n_{b} = 0.1$, $0.2$, $0.3$, $0.4$, and $0.5$~fm$^{-3}$.

For $I_{\textrm{tab}} \neq 0$ the parameter values are generated
in a similar way as suggested in subsection \ref{ssec:para}.
In this case, only four additional
rows follow the comment line below the row with 
the specification of $I_{\textrm{tab}}$.
Each column of these rows (in the order temperature, density, charge
fraction) contains the four quantities 
\begin{equation}
 \begin{array}{c}
 p_{\textrm{min}}\index{$p_{\textrm{min}}$} \\
 p_{\textrm{max}}\index{$p_{\textrm{max}}$} \\
 N_{p}\index{$N_{p}$} \\
I_{p}\index{$I_{p}$}
\end{array}
\end{equation}
where $p$\index{$p$} stands for the parameters $T$, $n_{b}$ or $Y_{q}$, 
respectively.
The minimum and maximum values of the parameter $p$ in the EoS table
are denoted by $p_{\textrm{min}}$ and $p_{\textrm{max}}$, respectively. $N_{p}>0$
is the number of data points in the mesh of the parameter $p$ and
$I_{p}$ defines the discretization scheme. If $I_{p} = 0$, the
individual points are given by a linear interpolation
\begin{equation}
  p_{i}\index{$p_{i}$} = p_{\textrm{min}} +
  (p_{\textrm{max}} - p_{\textrm{min}}) \frac{i-1}{N_{p}-1}
\end{equation}
for $i=1,\dots,N_{p}$ with $N_{p}>1$ and 
\begin{equation}
 p_{1} = p_{\textrm{min}}
\end{equation}
for $N_{p}=1$. For $I_{p}\neq 0$, a logarithmic scaling is used with
\begin{equation}
 p_{i} =p_{\textrm{min}} \left( \frac{p_{\textrm{max}}}{p_{\textrm{min}}}\right)^{\frac{i-1}{N_{p}-1}}
\end{equation}
for $i=1,\dots,N_{p}$ with $N_{p}>1$ and 
\begin{equation}
 p_{1} = p_{\textrm{min}}
\end{equation}
for $N_{p}=1$. 
For this tabulation scheme an example file \texttt{eos.parameters}
has the form
\begin{quote}
\texttt{\# interpolation rules for T, n\_b and Y\_q} \\
\texttt{3 3 3} \\
\texttt{\# calculation of beta equilibrium 
(1:~yes, else:~no) and for given \mbox{}\hfill\mbox{} entropy (1:~yes,~else:~no)}\\
\texttt{0 0} \\
\texttt{\# tabulation scheme of parameter values (see manual)} \\
\texttt{1} \\
\texttt{\# parameter values depending on tabulation scheme} \\
\texttt{5.0 0.01 0.3} \\
\texttt{5.0 1.00 0.3} \\
\texttt{ 1  201  1} \\
\texttt{ 0  1  0} 
\end{quote}
Here, $T=5$~MeV and $Y_{q}=0.3$ are kept constant and
the densities $n_{b}$ cover the range from $0.01$~fm$^{-3}$ to
$1.0$~fm$^{-3}$ in $200$ intervals with a logarithmic scaling.

In case the option $I_{\beta}=1$\index{$N_{\beta}$} is set in the third
row of the file \texttt{eos.parameters}, the specification of the
parameter grid in $Y_{q}$ is irrelevant.


\chapter{Structure of the \texttt{compose.f90} program }
\label{app:code}


(Remark: This appendix is not yet updated from version 2.00 of the manual.)

Instead of using the code \texttt{compose.f90} as given, the user can
employ the subroutines directly in her/his program. 
The file \texttt{compose.f90}\index{compose.f90} 
contains one main program,
31 subroutines\index{subroutine} and one function. 
The structure of the program with the dependencies of the subroutines
and functions is
depicted in tables \ref{tab:code}, \ref{tab:code2}
and \ref{tab:code3}.
The file \texttt{composemodules.f90}\index{composemodules.f90}
contains two modules\index{module}. 

The user needs to call the following three subroutines 
\begin{quote}
 \texttt{init\_eos\_table(iwr)} \\
 \texttt{define\_eos\_table(iwr)} \\
 \texttt{get\_eos(t,n,y,b,ipl,i\_beta,i\_entr)}
\end{quote}
from the file \texttt{compose.f90} in order to
generate EoS data. The subroutine 
\begin{quote}
  \texttt{init\_terminal(iwr)}
\end{quote}
is called 
only from the main program for running the standard
terminal version. It is usually
not relevent for users that want to use subroutines within their own programs.
The first two mentioned subroutines
depend on a single integer parameter \texttt{iwr}.
For \texttt{iwr = 1} a progress report during the execution of the
subroutine will be written to the terminal. Otherwise, this action will be
suppressed. The subroutine \texttt{get\_eos} has three \texttt{double
  precision} parameters (\texttt{t}, \texttt{n}, \texttt{y}) 
and three \texttt{integer} parameters (\texttt{ipl},
\texttt{i\_beta}, \texttt{i\_entr})
  as arguments, where the first is a vector of dimension three.
Be sure to include all (sub)routines, functions and the two
modules \texttt{eos\_tables} and \texttt{compose\_internal} in your code.

The subroutine \texttt{init\_eos\_table} has to be called only once 
in order to initialize all relevant tables and quantities. The files
\texttt{eos.t}, \texttt{eos.nb}, \texttt{eos.yq} and
\texttt{eos.thermo} have to exist in order that \texttt{init\_eos\_table}
is executed properly. The files \texttt{eos.compo} and \texttt{eos.micro}
are optional.

The subroutine\index{subroutine} \texttt{define\_eos\_table} 
defines the quantities that
are interpolated and that will be stored in the EoS
table. It reads the file \texttt{eos.quantities}. The subroutine needs to
be called only once. Instead of calling 
\texttt{define\_eos\_table} in the user's program, it is possible that
the relevant parameters are defined directly in the user's code. 
Thus it is possible to change the selection of quantities that
are interpolated during the execution of the user's program.
In the following, the variable names to be specified by the user are
given according to the structure of the input file 
\texttt{eos.quantities}\index{eos.quantities}:
\begin{quote}
 \texttt{\# number of regular, additional and derivative quantities} \\
 \texttt{n\_qty} \quad \texttt{n\_add} \quad \texttt{n\_df} \\
 \texttt{\# indices of regular, additional and derivative quantities} \\
 \texttt{idx\_qty(1)} \dots 
 \texttt{idx\_qty(n\_qty)} \:
 \texttt{idx\_add(1)} \dots
 \texttt{idx\_add(n\_add)} \:
 \texttt{idx\_df(1)} \dots 
 \texttt{idx\_df(n\_df)} \\
 \texttt{\# number of pairs and quadruples for composition data} \\
 \texttt{n\_p} \quad \texttt{n\_q} \\
 \texttt{\# indices of pairs and quadruples for composition data} \\
 \texttt{idx\_p(1)} \quad \dots \quad
 \texttt{idx\_p(n\_p)} \quad 
 \texttt{idx\_q(1)} \quad \dots \quad
 \texttt{idx\_q(n\_q)} \\
 \texttt{\# number of microscopic quantities} \\
 \texttt{n\_m} \\
 \texttt{\# indices of microscopic quantities} \\
 \texttt{idx\_m(1)} \quad \dots \quad
 \texttt{idx\_m(n\_m)} \\
 \texttt{\# number of error quantities} \\
 \texttt{n\_err} \\
 \texttt{\# indices of error quantities} \\
 \texttt{idx\_err(1)} \quad \dots \quad
 \texttt{idx\_err(n\_err)} \\
 \texttt{\# format of output file} \\
 \texttt{iout}
\end{quote}
Note that all variables are \texttt{integers}. The appearing vectors
with fixed size
are defined in the module\index{module} \texttt{compose\_internal}:
\begin{quote}
 \texttt{idx\_qty(dim\_qtyt)} \quad \mbox{with} \quad \texttt{dim\_qtyt = 23} \\
 \texttt{idx\_err(dim\_qtye)} \quad \mbox{with} \quad \texttt{dim\_qtye = 8}
 \: .
\end{quote}
where the dimensions are defined in the module \texttt{eos\_tables}.

Finally, the subroutine\index{subroutine} 
\texttt{get\_eos} can be called as often as needed
with the appropriate arguments. The \texttt{double precision} variables
\texttt{t}, \texttt{n} and \texttt{y} define the temperature $T$ [MeV], baryon
number density $n_{b}$ [fm${}^{-3}$] and the charge fraction
of strongly interacting particles
$Y_{q}$ [dimensionless], respectively, where the EoS is evaluated.
The \texttt{integer} indices \texttt{ipl\_t}, \texttt{ipl\_n} and
\texttt{ipl\_y} define the interpolation rule in $T$, $n_{b}$ and
$Y_{q}$, respectively. For \texttt{ipl\_t = 1} all selected quantities
are interpolated linearly such that the interpolated values agree
with the tabulated values at the grid points of the EoS table.
For \texttt{ipl\_t = 2} the interpolated values of a quantity and
its first derivative with respect to temperature agree with those
at the grid points. For \texttt{ipl\_t = 3} a continuity of also the
second derivatives at the grid points are demanded. Similarly, the
interpolation rules for $n_{b}$ and $Y_{q}$ are determined. If the
index is outside the range $[1,3]$ it is set to $3$.
See appendix~\ref{sec:interpol}  for details of the interpolation scheme.

The results of the interpolation are stored in seven 
\texttt{double precision} vectors/arrays 
with dimensions defined in the module\index{module}
\texttt{eos\_tables} or defined according to the EoS input files:
\begin{quote}
  \texttt{eos\_thermo(dim\_qtyt)}  \quad \mbox{with} \quad \texttt{dim\_qtyt = 23}\\
  \texttt{eos\_thermo\_add(dim\_qty)} \\
  \texttt{eos\_compo\_p(dim\_qtyp)} \\
  \texttt{eos\_compo\_q(dim\_qtyq,3)}  \\  
  \texttt{eos\_micro(dim\_qtym)}  \\
  \texttt{eos\_err(dim\_qtye)} \quad \mbox{with} \quad \texttt{dim\_qtye =
    8} \\
  \texttt{eos\_df(dim\_df)} \quad \mbox{with} \quad \texttt{dim\_df =
    10} \: .
\end{quote}
The vector index in \texttt{eos\_thermo} corresponds to the index $J$
given in tables~\ref{tab:ident_thermo1} and \ref{tab:ident_thermo2}. 
The vector index in
\texttt{eos\_thermo\_add} is just the index $i=1,\dots,N_{\textrm{add}}$ of the 
additional quantities stored in each row of the file \texttt{eos.thermo}.
The (first) index of the vectors/arrays \texttt{eos\_compo\_p}, 
\texttt{eos\_compo\_q} and \texttt{eos\_compo\_m} corresponds to the index
of the vectors \texttt{idx\_p}, \texttt{idx\_q} and \texttt{idx\_m} that are
defined in the file \texttt{eos.quantities} or by the user before
the subroutine \texttt{get\_eos} is called. The second index
of the array \texttt{eos\_compo\_q} is correlated with the index $J$ as
given in table \ref{tab:ident_compo}.
The vector index in \texttt{eos\_err} corresponds to the index $J$
given in table~\ref{tab:ident_err}.
The vector index in \texttt{eos\_df} corresponds to the index $J$
given in table~\ref{tab:f_derivatives}.

\begin{table}[ht]
\begin{center}
\caption{\label{tab:code}%
Modules\index{module}, (sub)routines\index{subroutine} 
and functions in the file \texttt{compose.f90}\index{compose.f90} with their
dependencies (continued on next page).}
{\small 
\begin{tabular}{llll}
\toprule
  & uses \texttt{MODULE} & uses \texttt{SUBROUTINE} & uses \texttt{FUNCTION} \\
\toprule
 \texttt{PROGRAM} & & & \\
\toprule
 \texttt{compose} & & \texttt{run\_terminal} & \\
 & & \texttt{init\_eos\_table} & \\
 & & \texttt{define\_eos\_table} & \\
 & & \texttt{get\_eos\_table} & \\
\toprule
 \texttt{SUBROUTINE} & & & \\
\toprule
 \texttt{init\_eos\_table} & & \texttt{read\_eos\_tables\_tnyb} \\
 & & \texttt{read\_eos\_table\_thermo} & \\
 & & \texttt{read\_eos\_table\_compo} & \\
 & & \texttt{read\_eos\_table\_micro} & \\
 & & \texttt{get\_diff\_rules} & \\
 & & \texttt{init\_ipl\_rule} & \\
 & & \texttt{get\_eos\_report} & \\
\midrule
\texttt{define\_eos\_table} & \texttt{compose\_internal} &
       \texttt{write\_errors} & \\
\midrule
 \texttt{get\_eos\_table} & \texttt{compose\_internal} & \texttt{get\_eos} & \\
 & \texttt{m\_out\_to\_json} & \texttt{write\_errors} & \\
 &  & \texttt{out\_to\_json\_write} & \\
\midrule
 \texttt{get\_eos} & \texttt{compose\_internal} & \texttt{get\_eos\_beta} & \\
 & & \texttt{get\_eos\_sub} & \\
 & & \texttt{write\_errors} & \\
\midrule
 \texttt{get\_eos\_beta} & \texttt{compose\_internal} & \texttt{get\_eos\_sub} & \\
\midrule
\texttt{get\_eos\_sub} & \texttt{compose\_internal} &
       \texttt{get\_eos\_grid\_para} & \\
       & & \texttt{eos\_interpol} & \\
\midrule
 \texttt{get\_eos\_grid\_para} & \texttt{compose\_internal} & & \\
\midrule
 \texttt{eos\_interpol} & \texttt{compose\_internal} & 
 \texttt{eos\_interpol\_d1} & \\ 
 & & \texttt{eos\_interpol\_d2} & \\ 
 & & \texttt{eos\_interpol\_d3} & \\ 
\midrule
 \texttt{eos\_interpol\_d1} & texttt{compose\_internal} & 
 \texttt{get\_interpol\_x} & \texttt{get\_ipl\_rule} \\ 
 & & \texttt{write\_errors} &  \\
\midrule
 \texttt{eos\_interpol\_d2} & \texttt{compose\_internal} & 
 \texttt{get\_idx\_arg2} & \texttt{get\_ipl\_rule} \\ 
 & & \texttt{get\_diff\_rules2} & \\
 & & \texttt{get\_derivatives} & \\
 & & \texttt{get\_coefficients} & \\
 & & \texttt{get\_interpol\_xy} & \\
 & & \texttt{write\_errors} &  \\
\bottomrule
\end{tabular} 
}
\end{center}
\end{table}

\begin{table}[ht]
\begin{center}
\caption{\label{tab:code2}%
Modules\index{module}, (sub)routines\index{subroutine} 
and functions in the file \texttt{compose.f90}\index{compose.f90} with their
dependencies (continued from previous page).}
{\small 
\begin{tabular}{llll}
\toprule
  & uses \texttt{MODULE} & uses \texttt{SUBROUTINE} & uses \texttt{FUNCTION} \\
\toprule
 \texttt{SUBROUTINE} & & & \\
\toprule
 \texttt{eos\_interpol\_d3} & \texttt{compose\_internal} & 
 \texttt{get\_interpol\_x} & \texttt{get\_ipl\_rule} \\ 
 & & \texttt{get\_idx\_arg2} & \\
 & & \texttt{get\_diff\_rules2} & \\
 & & \texttt{get\_derivatives} & \\
 & & \texttt{get\_coefficients} & \\
 & & \texttt{get\_interpol\_xy} & \\
 & & \texttt{write\_errors} &  \\
\midrule
 \texttt{get\_idx\_arg2} & \texttt{compose\_internal} & & \\
\midrule
 \texttt{get\_diff\_rules} & \texttt{compose\_internal} & & \\
\midrule
\texttt{get\_diff\_rules2} & \texttt{compose\_internal} & &
\texttt{get\_ipl\_rule} \\
\midrule
 \texttt{get\_interpol\_x} & \texttt{compose\_internal} & & \\
\midrule
 \texttt{get\_interpol\_xy} & \texttt{compose\_internal} & & \\
\midrule
 \texttt{get\_derivatives} & \texttt{compose\_internal} & & \\
\midrule
 \texttt{get\_coefficients} & \texttt{compose\_internal} & & \\
\midrule
 \texttt{run\_terminal} & \texttt{compose\_internal}
 & \texttt{init\_eos\_table\_term} & \\
 & & \texttt{init\_quant} & \\
 & & \texttt{init\_para} & \\
 & & \texttt{get\_eos\_table\_term} & \\
\midrule
 \texttt{init\_eos\_table\_term} & \texttt{compose\_internal}
 & \texttt{read\_eos\_tables\_tnyb} & \\
 & & \texttt{read\_eos\_table\_thermo} & \\
 & & \texttt{read\_eos\_table\_compo} & \\
 & & \texttt{read\_eos\_table\_micro} & \\
\midrule
 \texttt{get\_eos\_table\_term} & \texttt{compose\_internal}
 & \texttt{read\_eos\_tables\_tnyb} & \\
 & & \texttt{read\_eos\_table\_thermo} & \\
 & & \texttt{read\_eos\_table\_compo} & \\
 & & \texttt{read\_eos\_table\_micro} & \\
 & & \texttt{get\_diff\_rules} & \\
 & & \texttt{get\_ipl\_rule} & \\
 & & \texttt{get\_eos\_report} & \\
 & & \texttt{define\_eos\_table} & \\
 & & \texttt{get\_eos\_table} & \\
\midrule
 \texttt{init\_quant} & \texttt{compose\_internal} & & \\
\midrule
 \texttt{init\_para} & \texttt{compose\_internal} & & \\
\midrule
 \texttt{init\_ipl\_rule} & \texttt{compose\_internal} & & \\
\bottomrule
\end{tabular} 
}
\end{center}
\end{table}

\begin{table}[ht]
\begin{center}
\caption{\label{tab:code3}%
Modules\index{module}, (sub)routines\index{subroutine} 
and functions in the file \texttt{compose.f90}\index{compose.f90} with their
dependencies (continued from previous page).}
{\small 
\begin{tabular}{llll}
\toprule
  & uses \texttt{MODULE} & uses \texttt{SUBROUTINE} & uses \texttt{FUNCTION} \\
\toprule
 \texttt{FUNCTION} & & & \\
\toprule
 \texttt{get\_eos\_report} & \texttt{compose\_internal} & \texttt{get\_eos\_nmp} & \\
 & & \texttt{get\_eos} & \\
 & & \texttt{write\_errors} & \\
\midrule
 \texttt{get\_eos\_nmp} & \texttt{compose\_internal} & \texttt{get\_eos\_sub} & \\
\midrule
 \texttt{write\_errors} & \texttt{compose\_internal} & & \\
\midrule
\texttt{read\_eos\_tables\_tnyb} & \texttt{compose\_internal} &
       \texttt{write\_errors} & \\
\midrule
\texttt{read\_eos\_table\_thermo} & \texttt{compose\_internal} &
       \texttt{write\_errors} & \\
\midrule
\texttt{read\_eos\_table\_compo} & \texttt{compose\_internal} &
       \texttt{write\_errors} & \\
\midrule
\texttt{read\_eos\_table\_micro} & \texttt{compose\_internal} &
       \texttt{write\_errors} & \\
\midrule
 \texttt{get\_ipl\_rule} & \texttt{compose\_internal} & & \\
\bottomrule
\end{tabular} 
}
\end{center}
\end{table}

\chapter{Organization of the CompOSE team}
\label{app:team}
\index{team!organization}
CompOSE is developed in close collaboration with the communities 
the service is meant for.
These are mainly physicists who develop equations of state or
simulate astrophysical phenomena numerically on the computer.
Direct interaction with the users is an essential part in the concept of the
project.




The following persons are engaged in the preparation of
the web site\index{web site}, the manual\index{manual} 
and the tools, numerical codes and tables.
\setkomafont{labelinglabel}{\textbf}
\begin{labeling}{FirstName LongFamilyName}
\item[Debarati Chatterjee] IUCAA, Pune (India),
\item[Veronica Dexheimer] Kent State University (Ohio, USA),
\item[Chikako Ishizuka]Tokyo Institute of Technology (Japan),
\item[Thomas Kl\"ahn]California State University Long Beach (California, USA),
\item[Marco Mancini] Tours University (France),
\item[Jérôme Novak] LUTH, Observatoire de Paris (France), 
\item[Micaela Oertel] LUTH, Observatoire de Paris (France), 
\item[Helena Pais] Coimbra University (Portugal)
\item[Constança Providencia] Coimbra University (Portugal), 
\item[Adriana Raduta] NIPNE, Bucharest (Romania)
\item[Mathieu Servillat] LUTH, Observatoire de Paris (France),
\item[Laura Tolos] ICE (CSIC-IEEC), Barcelona (Spain),
\item[Stefan Typel]Technische Universit\"{a}t Darmstadt 
and GSI Helmholtz\-zentrum f\"{u}r Schwerionen\-forschung, Darmstadt (Germany). 
\end{labeling}

\chapter{Citing CompOSE}\label{citation}

If you make use of the tables provided in CompOSE, you will be guided on the CompOSE web pages to the scientific publications where the particular EoS models have been described in detail. Please cite them when using the tables for scientific purposes
together with a reference to the CompOSE website (https://compose.obspm.fr) and/or the original CompOSE publications: 

\begin{verbatim}

@article{Oertel:2016bki,
    author = "Oertel, M. and Hempel, M. and Klaehn, T. and Typel, S.",
    title = "{Equations of state for supernovae and compact stars}",
    eprint = "1610.03361",
    archivePrefix = "arXiv",
    primaryClass = "astro-ph.HE",
    doi = "10.1103/RevModPhys.89.015007",
    journal = "Rev. Mod. Phys.",
    volume = "89",
    number = "1",
    pages = "015007",
    year = "2017"
}

@article{Typel:2013rza,
    author = "Typel, S. and Oertel, M. and Klaehn, T.",
    title = "{CompOSE CompStar online supernova equations of state
              harmonising the concert of nuclear physics and astrophysics
              compose.obspm.fr}",
    eprint = "1307.5715",
    archivePrefix = "arXiv",
    primaryClass = "astro-ph.SR",
    doi = "10.1134/S1063779615040061",
    journal = "Phys. Part. Nucl.",
    volume = "46",
    number = "4",
    pages = "633--664",
    year = "2015"
}

\end{verbatim}

\chapter{Acknowledgments}
\label{app:ack}


CompOSE would not be possible without the financial and organisatorial
support from a large number of institutions and individual contributors.
We gratefully acknowledge support by CompStar, NewCompStar, and PHAROS, Research Networking Programs
of the European Science Foundation (ESF) and the birthplace of
the CompOSE project, by a grant from the Polish Ministry for
Science and Higher Education (MNiSW) supporting the ``CompStar''-activity,
by the Instytut Fizyki Teoretycznej of the Uniwersytet Wroc\l{}awski,
the National Science Centre Poland
(Narodowe Centrum Nauki, NCN) within the ``Maestro''-programme under
contract No. DEC-2011/02/A/ST2/00306, by the
``hadronphysics3'' network within the seventh framework program of the
European Union,
by the GSI Helmholtzzentrum f\"{u}r Schwerionenforschung GmbH,
by the Helmholtz International
Center for FAIR within the framework of the LOEWE program launched
by the state of Hesse via the Technical University Darmstadt,
by the Helmholtz Association (HGF) through the Nuclear Astrophysics
Virtual Institute (VH-VI-417),
by the ExtreMe Matter Institute EMMI in the framework
of the Helmholtz Alliance `Cosmic Matter in the Laboratory',
by the DFG cluster of excellence ``Origin and Structure of
the Universe'',
by the DFG through grant SFB~1245,
by the SN2NS project ANR-10-BLAN-0503 and
by the MPNS COST action MP1304 ``Exploring fundamental physics with
compact stars (NewCompStar)'' within the EU framework programme.
We thank Jean-Yves Giot for creating the first version of the web site.


\newpage
\addcontentsline{toc}{chapter}{Index}
\printindex


\begin{thebibliography}{99}

\bibitem{manual:1}
  S. Typel, M. Oertel, and T. Kl\"{a}hn,
  ``CompOSE CompStar online supernova equations of state harmonising the concert of nuclear physics and astrophysics compose.obspm.fr,''
  Physics of Particles and Nuclei \textbf{46}, 633 (2015).


\bibitem{Oertel:2016bki}
  M.~Oertel, M.~Hempel, T.~Kl\"{a}hn and S.~Typel,
  ``Equations of state for supernovae and compact stars,''
  Rev.\ Mod.\ Phys.\  \textbf{89}, 015007 (2017). 


\bibitem{CODATA2018}
 E. Tiesinga, D.B. Newell, P.J. Mohr, and B.N. Taylor,
 ``CODATA Recommended Values of the Fundamental Physical Constants: 2018´´,
 NIST SP961 (May 2019)

  

\bibitem{Ame2016a}
  W.J. Huang, G. Audi, Meng Wang, F.G. Kondev, W.J. Huang, S. Naimi, and Xing Xu,
  ``The AME2016 atomic mass evaluation (I). Evaluation of input data; and adjustment procedures,''
  Chinese Physics C \textbf{41}, 030002 (2017);
  
\bibitem{Ame2016b}  
  Meng Wang, G. Audi, F.G. Kondev, W.J. Huang, S. Naimi, and Xing Xu,
  Chinese Physics C
  ``The AME2016 atomic mass evaluation (II). Tables, graphs and references,''
  Chinese Physics C41, 030003 (2017).
  
\bibitem{Huang:2021nwk}
W.~J.~Huang, M.~Wang, F.~G.~Kondev, G.~Audi and S.~Naimi,
``The AME 2020 atomic mass evaluation (I). Evaluation of input data, and adjustment procedures,''
Chin. Phys. C \textbf{45} (2021) no.3, 030002

\bibitem{Wang:2021xhn}
M.~Wang, W.~J.~Huang, F.~G.~Kondev, G.~Audi and S.~Naimi,
``The AME 2020 atomic mass evaluation (II). Tables, graphs and references,''
Chin. Phys. C \textbf{45} (2021) no.3, 030003

  
\bibitem{Nubase2016}
  G.~Audi, F.~G.~Kondev, M.~Wang, W.~J.~Huang and S.~Naimi,
  ``The NUBASE2016 evaluation of nuclear properties,''
  Chinese Physics C \textbf{41}, 030001 (2017).
  
\bibitem{Kondev:2021lzi}
F.~G.~Kondev, M.~Wang, W.~J.~Huang, S.~Naimi and G.~Audi,
``The NUBASE2020 evaluation of nuclear physics properties,''
Chin. Phys. C \textbf{45} (2021) no.3, 030001

 
\bibitem{ParticleDataGroup:2020ssz}
P.~A.~Zyla \textit{et al.} [Particle Data Group],
``Review of Particle Physics,''
PTEP \textbf{2020} (2020) no.8, 083C01

\bibitem{Tolman:1939}  
  R.~C.~Tolman,
  ``Static solutions of Einstein's field equations for spheres of fluid´´
  Phys. Rev. \textbf{55} (1939) 364.

\bibitem{Oppenheimer:1939}
  J.~R.~Oppenheimer and G. M. Volkoff,
  ``On massive neutron cores´´,
  Phys. Rev. \textbf{55} (1939) 374.

\bibitem{Hinderer:2007mb}
T.~Hinderer,
``Tidal Love numbers of neutron stars,''
Astrophys. J. \textbf{677} (2008), 1216-1220.

\bibitem{Hinderer:2009ca}
T.~Hinderer, B.~D.~Lackey, R.~N.~Lang and J.~S.~Read,
``Tidal deformability of neutron stars with realistic equations of state and their gravitational wave signatures in binary inspiral,''
Phys. Rev. D \textbf{81} (2010), 123016.

\bibitem{Stergioulas:1994ea}
N.~Stergioulas and J.~L.~Friedman,
``Comparing models of rapidly rotating relativistic stars constructed by two numerical methods,''
Astrophys. J. \textbf{444} (1995), 306

\bibitem{Nozawa:1998ak}
T.~Nozawa, N.~Stergioulas, E.~Gourgoulhon and Y.~Eriguchi,
``Construction of highly accurate models of rotating neutron stars: Comparison of three different numerical schemes,''
Astron. Astrophys. Suppl. Ser. \textbf{132} (1998), 431

\bibitem{RNS}
RNS - Rapidly Rotating Neutron Star, \texttt{http://www.gravity.phys.uwm.edu/rns}

\bibitem{LORENE}
LORENE - Langage Objet pour la RElativit\'{e} Num\'{e}riquE,
\texttt{https://lorene.obspm.fr/}


\bibitem{Lattimer:1991ib}
J.~M.~Lattimer, M.~Prakash, C.~J.~Pethick and P.~Haensel,
``Direct URCA process in neutron stars,''
Phys. Rev. Lett. \textbf{66} (1991), 2701-2704

\bibitem{Alvarez-Castillo:2016yma}
D.~E.~Alvarez-Castillo and D.~Blaschke,
``Interplay between Symmetry Energy and Excluded Volume Corrections under the Direct Urca Cooling Constraint in Neutron Stars,''
PoS \textbf{MPCS2015} (2016), 026

\bibitem{Strickland:2014pga}
M.~Strickland,
``Anisotropic Hydrodynamics: Three lectures,''
Acta Phys. Polon. B \textbf{45} (2014) no.12, 2355-2394

\bibitem{Dutra:2012mb}
  M.~Dutra, O.~Lourenco, J.~S.~Sa Martins, A.~Delfino, J.~R.~Stone and P.~D.~Stevenson,
  ``Skyrme Interaction and Nuclear Matter Constraints,''
  Phys.\ Rev.\ C \textbf{85}, 035201 (2012).

\bibitem{Sellahewa:2014nia}
  R.~Sellahewa and A.~Rios,
  ``Isovector properties of the Gogny interaction,''
  Phys.\ Rev.\ C \textbf{90}, 054327 (2014).

\bibitem{Dutra:2014qga}
  M.~Dutra, O.~Louren\c{c}o, S.~S.~Avancini, B.~V.~Carlson, A.~Delfino, D.~P.~Menezes, C.~Provid\^encia, S.~Typel and J.~R.~Stone,
  ``Relativistic Mean-Field Hadronic Models under Nuclear Matter Constraints,''
  Phys.\ Rev.\ C \textbf{90}, 055203 (2014).

\bibitem{Stone:2014wza}
  J.~R.~Stone, N.~J.~Stone and S.~A.~Moszkowski,
  ``Incompressibility in finite nuclei and nuclear matter,''
  Phys.\ Rev.\ C \textbf{89}, 044316 (2014).

\bibitem{Swe96}
  F.~D.~Swesty, 
    ``Thermodynamically Consistent Interpolation for Equation of State Tables,''
  J.\ Comp.\ Phys.\ \textbf{127}, 118 (1996).
  


\end{thebibliography}
\end{document}